\documentclass{CUP-JNL-DTM}%
\usepackage{graphicx}
\usepackage{multicol,multirow}
\usepackage{amsmath,amssymb,amsfonts}
\usepackage{mathrsfs}
\usepackage{amsthm}
\usepackage{rotating}
\usepackage{appendix}
\usepackage[numbers]{natbib}
\usepackage{ifpdf}
\usepackage[T1]{fontenc}
\usepackage{newtxtext}
\usepackage{newtxmath}
\usepackage{textcomp}
\usepackage{xcolor}
\usepackage[colorlinks,allcolors=blue]{hyperref}

\theoremstyle{definition}

\numberwithin{equation}{section}

\usepackage{tabularx}
\usepackage{subcaption}
\usepackage{colortbl}
\usepackage{bm}
\usepackage{hyperref}

\newcommand{\Revision}{\textcolor{black}}
\newcommand{\RevisionNew}{\textcolor{black}}

\jname{Data/Math}
\articletype{Research article}
\jyear{2025}

\begin{document}

\newcommand{\HW}{\textcolor{blue}}
\newcommand{\JPF}{\textcolor{green}}
\newcommand{\CW}{\textcolor{orange}}

\newcommand{\lrap}[1]{``#1''}

\newcommand{\Caption}[1]{ \begin{singlespace} \em{\caption{#1}} \end{singlespace} }

\long\def\symbolfootnote[#1]#2{\begingroup \def\thefootnote{\fnsymbol{footnote}}\footnote[#1]{#2} \endgroup} 

\renewcommand{\vec}[1]{ \ensuremath{ \mathbf{ #1 } } }
\newcommand{\ten}[1]{ \ensuremath {\mathbf{#1} } }

\newcommand{\gvec}[1]{ \ensuremath{ \boldsymbol{ #1 } } }
\newcommand{\gten}[1]{ \ensuremath {\boldsymbol{#1} } }


\newcommand{\mn}{_{\mbox{\tiny{N}}}}
\newcommand{\mt}{_{\mbox{\tiny{T}}}}

\newcommand{\dint}{\mbox{d}}
\newcommand{\de}{\,\mbox{det}\,}
\newcommand{\gra}{\,\mbox{grad}\,}
\newcommand{\Gra}{\,\mbox{Grad}\,}
\newcommand{\tra}{\,\mbox{tr}\,}
\newcommand{\Div}{\mbox{Div}\,}
\renewcommand{\div}{\mbox{div}\,}
\newcommand{\sign}{\mbox{sign}}

\newcommand{\ncdot}{\hspace{-0.14cm}\cdot}

\newcommand{\hata}{\ensuremath{ \widehat{a} }}
\newcommand{\hatb}{\ensuremath{ \widehat{b} }}
\newcommand{\hatc}{\ensuremath{ \widehat{c} }}
\newcommand{\hatd}{\ensuremath{ \widehat{d} }}
\newcommand{\hate}{\ensuremath{ \widehat{e} }}
\newcommand{\hatf}{\ensuremath{ \widehat{f} }}
\newcommand{\hatg}{\ensuremath{ \widehat{g} }}
\newcommand{\hath}{\ensuremath{ \widehat{h} }}
\newcommand{\hati}{\ensuremath{ \widehat{i} }}
\newcommand{\hatj}{\ensuremath{ \widehat{j} }}
\newcommand{\hatk}{\ensuremath{ \widehat{k} }}
\newcommand{\hatl}{\ensuremath{ \widehat{l} }}
\newcommand{\hatm}{\ensuremath{ \widehat{m} }}
\newcommand{\hatn}{\ensuremath{ \widehat{n} }}
\newcommand{\hato}{\ensuremath{ \widehat{o} }}
\newcommand{\hatp}{\ensuremath{ \widehat{p} }}
\newcommand{\hatq}{\ensuremath{ \widehat{q} }}
\newcommand{\hatr}{\ensuremath{ \widehat{r} }}
\newcommand{\hats}{\ensuremath{ \widehat{s} }}
\newcommand{\hatt}{\ensuremath{ \widehat{t} }}
\newcommand{\hatu}{\ensuremath{ \widehat{u} }}
\newcommand{\hatv}{\ensuremath{ \widehat{v} }}
\newcommand{\hatw}{\ensuremath{ \widehat{w} }}
\newcommand{\hatx}{\ensuremath{ \widehat{x} }}
\newcommand{\haty}{\ensuremath{ \widehat{y} }}
\newcommand{\hatz}{\ensuremath{ \widehat{z} }}

\newcommand{\hatA}{\ensuremath{ \widehat{A} }}
\newcommand{\hatB}{\ensuremath{ \widehat{B} }}
\newcommand{\hatC}{\ensuremath{ \widehat{C} }}
\newcommand{\hatD}{\ensuremath{ \widehat{D} }}
\newcommand{\hatE}{\ensuremath{ \widehat{E} }}
\newcommand{\hatF}{\ensuremath{ \widehat{F} }}
\newcommand{\hatG}{\ensuremath{ \widehat{G} }}
\newcommand{\hatH}{\ensuremath{ \widehat{H} }}
\newcommand{\hatI}{\ensuremath{ \widehat{I} }}
\newcommand{\hatJ}{\ensuremath{ \widehat{J} }}
\newcommand{\hatK}{\ensuremath{ \widehat{K} }}
\newcommand{\hatL}{\ensuremath{ \widehat{L} }}
\newcommand{\hatM}{\ensuremath{ \widehat{M} }}
\newcommand{\hatN}{\ensuremath{ \widehat{N} }}
\newcommand{\hatO}{\ensuremath{ \widehat{O} }}
\newcommand{\hatP}{\ensuremath{ \widehat{P} }}
\newcommand{\hatQ}{\ensuremath{ \widehat{Q} }}
\newcommand{\hatR}{\ensuremath{ \widehat{R} }}
\newcommand{\hatS}{\ensuremath{ \widehat{S} }}
\newcommand{\hatT}{\ensuremath{ \widehat{T} }}
\newcommand{\hatU}{\ensuremath{ \widehat{U} }}
\newcommand{\hatV}{\ensuremath{ \widehat{V} }}
\newcommand{\hatW}{\ensuremath{ \widehat{W} }}
\newcommand{\hatX}{\ensuremath{ \widehat{X} }}
\newcommand{\hatY}{\ensuremath{ \widehat{Y} }}
\newcommand{\hatZ}{\ensuremath{ \widehat{Z }}}

\newcommand{\tila}{\ensuremath{ \widetilde{a} }}
\newcommand{\tilb}{\ensuremath{ \widetilde{b} }}
\newcommand{\tilc}{\ensuremath{ \widetilde{c} }}
\newcommand{\tild}{\ensuremath{ \widetilde{d} }}
\newcommand{\tile}{\ensuremath{ \widetilde{e} }}
\newcommand{\tilf}{\ensuremath{ \widetilde{f} }}
\newcommand{\tilg}{\ensuremath{ \widetilde{g} }}
\newcommand{\tilh}{\ensuremath{ \widetilde{h} }}
\newcommand{\tili}{\ensuremath{ \widetilde{i} }}
\newcommand{\tilj}{\ensuremath{ \widetilde{j} }}
\newcommand{\tilk}{\ensuremath{ \widetilde{k} }}
\newcommand{\till}{\ensuremath{ \widetilde{l} }}
\newcommand{\tilm}{\ensuremath{ \widetilde{m} }}
\newcommand{\tiln}{\ensuremath{ \widetilde{n} }}
\newcommand{\tilo}{\ensuremath{ \widetilde{o} }}
\newcommand{\tilp}{\ensuremath{ \widetilde{p} }}
\newcommand{\tilq}{\ensuremath{ \widetilde{q} }}
\newcommand{\tilr}{\ensuremath{ \widetilde{r} }}
\newcommand{\tils}{\ensuremath{ \widetilde{s} }}
\newcommand{\tilt}{\ensuremath{ \widetilde{t} }}
\newcommand{\tilu}{\ensuremath{ \widetilde{u} }}
\newcommand{\tilv}{\ensuremath{ \widetilde{v} }}
\newcommand{\tilw}{\ensuremath{ \widetilde{w} }}
\newcommand{\tilx}{\ensuremath{ \widetilde{x} }}
\newcommand{\tily}{\ensuremath{ \widetilde{y} }}
\newcommand{\tilz}{\ensuremath{ \widetilde{z} }}

\newcommand{\tilA}{\ensuremath{ \widetilde{A} }}
\newcommand{\tilB}{\ensuremath{ \widetilde{B} }}
\newcommand{\tilC}{\ensuremath{ \widetilde{C} }}
\newcommand{\tilD}{\ensuremath{ \widetilde{D} }}
\newcommand{\tilE}{\ensuremath{ \widetilde{E} }}
\newcommand{\tilF}{\ensuremath{ \widetilde{F} }}
\newcommand{\tilG}{\ensuremath{ \widetilde{G} }}
\newcommand{\tilH}{\ensuremath{ \widetilde{H} }}
\newcommand{\tilI}{\ensuremath{ \widetilde{I} }}
\newcommand{\tilJ}{\ensuremath{ \widetilde{J} }}
\newcommand{\tilK}{\ensuremath{ \widetilde{K} }}
\newcommand{\tilL}{\ensuremath{ \widetilde{L} }}
\newcommand{\tilM}{\ensuremath{ \widetilde{M} }}
\newcommand{\tilN}{\ensuremath{ \widetilde{N} }}
\newcommand{\tilO}{\ensuremath{ \widetilde{O} }}
\newcommand{\tilP}{\ensuremath{ \widetilde{P} }}
\newcommand{\tilQ}{\ensuremath{ \widetilde{Q} }}
\newcommand{\tilR}{\ensuremath{ \widetilde{R} }}
\newcommand{\tilS}{\ensuremath{ \widetilde{S} }}
\newcommand{\tilT}{\ensuremath{ \widetilde{T} }}
\newcommand{\tilU}{\ensuremath{ \widetilde{U} }}
\newcommand{\tilV}{\ensuremath{ \widetilde{V} }}
\newcommand{\tilW}{\ensuremath{ \widetilde{W} }}
\newcommand{\tilX}{\ensuremath{ \widetilde{X} }}
\newcommand{\tilY}{\ensuremath{ \widetilde{Y} }}
\newcommand{\tilZ}{\ensuremath{ \widetilde{Z }}}

\newcommand{\calA}{\ensuremath{ \mathcal{A} }}
\newcommand{\calB}{\ensuremath{ \mathcal{B} }}
\newcommand{\calC}{\ensuremath{ \mathcal{C} }}
\newcommand{\calD}{\ensuremath{ \mathcal{D} }}
\newcommand{\calE}{\ensuremath{ \mathcal{E} }}
\newcommand{\calF}{\ensuremath{ \mathcal{F} }}
\newcommand{\calG}{\ensuremath{ \mathcal{G} }}
\newcommand{\calH}{\ensuremath{ \mathcal{H} }}
\newcommand{\calI}{\ensuremath{ \mathcal{I} }}
\newcommand{\calJ}{\ensuremath{ \mathcal{J} }}
\newcommand{\calK}{\ensuremath{ \mathcal{K} }}
\newcommand{\calL}{\ensuremath{ \mathcal{L} }}
\newcommand{\calM}{\ensuremath{ \mathcal{M} }}
\newcommand{\calN}{\ensuremath{ \mathcal{N} }}
\newcommand{\calO}{\ensuremath{ \mathcal{O} }}
\newcommand{\calP}{\ensuremath{ \mathcal{P} }}
\newcommand{\calQ}{\ensuremath{ \mathcal{Q} }}
\newcommand{\calR}{\ensuremath{ \mathcal{R} }}
\newcommand{\calS}{\ensuremath{ \mathcal{S} }}
\newcommand{\calT}{\ensuremath{ \mathcal{T} }}
\newcommand{\calU}{\ensuremath{ \mathcal{U} }}
\newcommand{\calV}{\ensuremath{ \mathcal{V} }}
\newcommand{\calW}{\ensuremath{ \mathcal{W} }}
\newcommand{\calX}{\ensuremath{ \mathcal{X} }}
\newcommand{\calY}{\ensuremath{ \mathcal{Y} }}
\newcommand{\calZ}{\ensuremath{ \mathcal{Z} }}

\newcommand{\bbA}{\ensuremath{ \mathbb{A} }}
\newcommand{\bbB}{\ensuremath{ \mathbb{B} }}
\newcommand{\bbC}{\ensuremath{ \mathbb{C} }}
\newcommand{\bbD}{\ensuremath{ \mathbb{D} }}
\newcommand{\bbE}{\ensuremath{ \mathbb{E} }}
\newcommand{\bbF}{\ensuremath{ \mathbb{F} }}
\newcommand{\bbG}{\ensuremath{ \mathbb{G} }}
\newcommand{\bbH}{\ensuremath{ \mathbb{H} }}
\newcommand{\bbI}{\ensuremath{ \mathbb{I} }}
\newcommand{\bbJ}{\ensuremath{ \mathbb{J} }}
\newcommand{\bbK}{\ensuremath{ \mathbb{K} }}
\newcommand{\bbL}{\ensuremath{ \mathbb{L} }}
\newcommand{\bbM}{\ensuremath{ \mathbb{M} }}
\newcommand{\bbN}{\ensuremath{ \mathbb{N} }}
\newcommand{\bbO}{\ensuremath{ \mathbb{O} }}
\newcommand{\bbP}{\ensuremath{ \mathbb{P} }}
\newcommand{\bbQ}{\ensuremath{ \mathbb{Q} }}
\newcommand{\bbR}{\ensuremath{ \mathbb{R} }}
\newcommand{\bbS}{\ensuremath{ \mathbb{S} }}
\newcommand{\bbT}{\ensuremath{ \mathbb{T} }}
\newcommand{\bbU}{\ensuremath{ \mathbb{U} }}
\newcommand{\bbV}{\ensuremath{ \mathbb{V} }}
\newcommand{\bbW}{\ensuremath{ \mathbb{W} }}
\newcommand{\bbX}{\ensuremath{ \mathbb{X} }}
\newcommand{\bbY}{\ensuremath{ \mathbb{Y} }}
\newcommand{\bbZ}{\ensuremath{ \mathbb{Z} }}

\newcommand{\tenA}{\ensuremath{ \ten{A} }}
\newcommand{\tenB}{\ensuremath{ \ten{B} }}
\newcommand{\tenC}{\ensuremath{ \ten{C} }}
\newcommand{\tenD}{\ensuremath{ \ten{D} }}
\newcommand{\tenE}{\ensuremath{ \ten{E} }}
\newcommand{\tenF}{\ensuremath{ \ten{F} }}
\newcommand{\tenG}{\ensuremath{ \ten{G} }}
\newcommand{\tenH}{\ensuremath{ \ten{H} }}
\newcommand{\tenI}{\ensuremath{ \ten{I} }}
\newcommand{\tenJ}{\ensuremath{ \ten{J} }}
\newcommand{\tenK}{\ensuremath{ \ten{K} }}
\newcommand{\tenL}{\ensuremath{ \ten{L} }}
\newcommand{\tenM}{\ensuremath{ \ten{M} }}
\newcommand{\tenN}{\ensuremath{ \ten{N} }}
\newcommand{\tenO}{\ensuremath{ \ten{O} }}
\newcommand{\tenP}{\ensuremath{ \ten{P} }}
\newcommand{\tenQ}{\ensuremath{ \ten{Q} }}
\newcommand{\tenR}{\ensuremath{ \ten{R} }}
\newcommand{\tenS}{\ensuremath{ \ten{S} }}
\newcommand{\tenT}{\ensuremath{ \ten{T} }}
\newcommand{\tenU}{\ensuremath{ \ten{U} }}
\newcommand{\tenV}{\ensuremath{ \ten{V} }}
\newcommand{\tenW}{\ensuremath{ \ten{W} }}
\newcommand{\tenX}{\ensuremath{ \ten{X} }}
\newcommand{\tenY}{\ensuremath{ \ten{Y} }}
\newcommand{\tenZ}{\ensuremath{ \ten{Z} }}

\newcommand{\tena}{\ensuremath{ \ten{a} }}
\newcommand{\tenb}{\ensuremath{ \ten{b} }}
\newcommand{\tenc}{\ensuremath{ \ten{c} }}
\newcommand{\tend}{\ensuremath{ \ten{d} }}
\newcommand{\tene}{\ensuremath{ \ten{e} }}
\newcommand{\tenf}{\ensuremath{ \ten{f} }}
\newcommand{\teng}{\ensuremath{ \ten{g} }}
\newcommand{\tenh}{\ensuremath{ \ten{h} }}
\newcommand{\teni}{\ensuremath{ \ten{i} }}
\newcommand{\tenj}{\ensuremath{ \ten{j} }}
\newcommand{\tenk}{\ensuremath{ \ten{k} }}
\newcommand{\tenl}{\ensuremath{ \ten{l} }}
\newcommand{\tenm}{\ensuremath{ \ten{m} }}
\newcommand{\tenn}{\ensuremath{ \ten{n} }}
\newcommand{\teno}{\ensuremath{ \ten{o} }}
\newcommand{\tenp}{\ensuremath{ \ten{p} }}
\newcommand{\tenq}{\ensuremath{ \ten{q} }}
\newcommand{\tenr}{\ensuremath{ \ten{r} }}
\newcommand{\tens}{\ensuremath{ \ten{s} }}
\newcommand{\tent}{\ensuremath{ \ten{t} }}
\newcommand{\tenu}{\ensuremath{ \ten{u} }}
\newcommand{\tenv}{\ensuremath{ \ten{v} }}
\newcommand{\tenw}{\ensuremath{ \ten{w} }}
\newcommand{\tenx}{\ensuremath{ \ten{x} }}
\newcommand{\teny}{\ensuremath{ \ten{y} }}
\newcommand{\tenz}{\ensuremath{ \ten{z} }}

\newcommand{\ttena}{\dot{\tena}}
\newcommand{\ttenb}{\dot{\tenb}}
\newcommand{\ttenc}{\dot{\tenc}}
\newcommand{\ttend}{\dot{\tend}}
\newcommand{\ttene}{\dot{\tene}}
\newcommand{\ttenf}{\dot{\tenf}}

\newcommand{\ttenE}{\dot{\tenE}}
\newcommand{\ttenF}{\dot{\tenF}}
\newcommand{\ttenS}{\dot{\tenS}}
\newcommand{\ttenT}{\dot{\tenT}}
\newcommand{\ttenU}{\dot{\tenU}}
\newcommand{\ttenV}{\dot{\tenV}}
\newcommand{\ttenW}{\dot{\tenW}}
\newcommand{\ttenX}{\dot{\tenX}}
\newcommand{\ttenY}{\dot{\tenY}}
\newcommand{\ttenZ}{\dot{\tenZ}}

\newcommand{\ltenA}{\stackrel{\triangle}{\tenA}}
\newcommand{\ltenD}{\stackrel{\triangle}{\tenD}}
\newcommand{\ltenM}{\stackrel{\triangle}{\tenM}}

\newcommand{\btena}{\bar{\tena}}
\newcommand{\btenb}{\bar{\tenb}}
\newcommand{\btenc}{\bar{\tenc}}
\newcommand{\btend}{\bar{\tend}}
\newcommand{\btene}{\bar{\tene}}
\newcommand{\btenf}{\bar{\tenf}}
\newcommand{\bteng}{\bar{\teng}}

\newcommand{\btenA}{\bar{\tenA}}
\newcommand{\btenB}{\bar{\tenB}}
\newcommand{\btenC}{\bar{\tenC}}
\newcommand{\btenD}{\bar{\tenD}}
\newcommand{\btenE}{\bar{\tenE}}
\newcommand{\btenF}{\bar{\tenF}}
\newcommand{\btenK}{\bar{\tenK}}
\newcommand{\btenL}{\bar{\tenL}}
\newcommand{\btenM}{\bar{\tenM}}
\newcommand{\btenN}{\bar{\tenN}}
\newcommand{\btenO}{\bar{\tenO}}
\newcommand{\btenP}{\bar{\tenP}}
\newcommand{\btenQ}{\bar{\tenQ}}
\newcommand{\btenR}{\bar{\tenR}}
\newcommand{\btenS}{\bar{\tenS}}
\newcommand{\btenT}{\bar{\tenT}}


\newcommand{\lbtenE}{\Delta\bar{\tenE}}

\newcommand{\hattena}{\ensuremath{ \widehat{\ten{a}} }}
\newcommand{\hattenb}{\ensuremath{ \widehat{\ten{b}} }}
\newcommand{\hattenc}{\ensuremath{ \widehat{\ten{c}} }}
\newcommand{\hattend}{\ensuremath{ \widehat{\ten{d}} }}
\newcommand{\hattene}{\ensuremath{ \widehat{\ten{e}} }}
\newcommand{\hattenf}{\ensuremath{ \widehat{\ten{f}} }}
\newcommand{\hatteng}{\ensuremath{ \widehat{\ten{g}} }}
\newcommand{\hattenh}{\ensuremath{ \widehat{\ten{h}} }}
\newcommand{\hatteni}{\ensuremath{ \widehat{\ten{i}} }}
\newcommand{\hattenj}{\ensuremath{ \widehat{\ten{j}} }}
\newcommand{\hattenk}{\ensuremath{ \widehat{\ten{k}} }}
\newcommand{\hattenl}{\ensuremath{ \widehat{\ten{l}} }}
\newcommand{\hattenm}{\ensuremath{ \widehat{\ten{m}} }}
\newcommand{\hattenn}{\ensuremath{ \widehat{\ten{n}} }}
\newcommand{\hatteno}{\ensuremath{ \widehat{\ten{o}} }}
\newcommand{\hattenp}{\ensuremath{ \widehat{\ten{p}} }}
\newcommand{\hattenq}{\ensuremath{ \widehat{\ten{q}} }}
\newcommand{\hattenr}{\ensuremath{ \widehat{\ten{r}} }}
\newcommand{\hattens}{\ensuremath{ \widehat{\ten{s}} }}
\newcommand{\hattent}{\ensuremath{ \widehat{\ten{t}} }}
\newcommand{\hattenu}{\ensuremath{ \widehat{\ten{u}} }}
\newcommand{\hattenv}{\ensuremath{ \widehat{\ten{v}} }}
\newcommand{\hattenw}{\ensuremath{ \widehat{\ten{w}} }}
\newcommand{\hattenx}{\ensuremath{ \widehat{\ten{x}} }}
\newcommand{\hatteny}{\ensuremath{ \widehat{\ten{y}} }}
\newcommand{\hattenz}{\ensuremath{ \widehat{\ten{z}} }}

\newcommand{\hattenA}{\ensuremath{ \widehat{\ten{A}} }}
\newcommand{\hattenB}{\ensuremath{ \widehat{\ten{B}} }}
\newcommand{\hattenC}{\ensuremath{ \widehat{\ten{C}} }}
\newcommand{\hattenD}{\ensuremath{ \widehat{\ten{D}} }}
\newcommand{\hattenE}{\ensuremath{ \widehat{\ten{E}} }}
\newcommand{\hattenF}{\ensuremath{ \widehat{\ten{F}} }}
\newcommand{\hattenG}{\ensuremath{ \widehat{\ten{G}} }}
\newcommand{\hattenH}{\ensuremath{ \widehat{\ten{H}} }}
\newcommand{\hattenI}{\ensuremath{ \widehat{\ten{I}} }}
\newcommand{\hattenJ}{\ensuremath{ \widehat{\ten{J}} }}
\newcommand{\hattenK}{\ensuremath{ \widehat{\ten{K}} }}
\newcommand{\hattenL}{\ensuremath{ \widehat{\ten{L}} }}
\newcommand{\hattenM}{\ensuremath{ \widehat{\ten{M}} }}
\newcommand{\hattenN}{\ensuremath{ \widehat{\ten{N}} }}
\newcommand{\hattenO}{\ensuremath{ \widehat{\ten{O}} }}
\newcommand{\hattenP}{\ensuremath{ \widehat{\ten{P}} }}
\newcommand{\hattenQ}{\ensuremath{ \widehat{\ten{Q}} }}
\newcommand{\hattenR}{\ensuremath{ \widehat{\ten{R}} }}
\newcommand{\hattenS}{\ensuremath{ \widehat{\ten{S}} }}
\newcommand{\hattenT}{\ensuremath{ \widehat{\ten{T}} }}
\newcommand{\hattenU}{\ensuremath{ \widehat{\ten{U}} }}
\newcommand{\hattenV}{\ensuremath{ \widehat{\ten{V}} }}
\newcommand{\hattenW}{\ensuremath{ \widehat{\ten{W}} }}
\newcommand{\hattenX}{\ensuremath{ \widehat{\ten{X}} }}
\newcommand{\hattenY}{\ensuremath{ \widehat{\ten{Y}} }}
\newcommand{\hattenZ}{\ensuremath{ \widehat{\ten{Z}} }}

\newcommand{\tiltena}{\ensuremath{ \widetilde{\ten{a}} }}
\newcommand{\tiltenb}{\ensuremath{ \widetilde{\ten{b}} }}
\newcommand{\tiltenc}{\ensuremath{ \widetilde{\ten{c}} }}
\newcommand{\tiltend}{\ensuremath{ \widetilde{\ten{d}} }}
\newcommand{\tiltene}{\ensuremath{ \widetilde{\ten{e}} }}
\newcommand{\tiltenf}{\ensuremath{ \widetilde{\ten{f}} }}
\newcommand{\tilteng}{\ensuremath{ \widetilde{\ten{g}} }}
\newcommand{\tiltenh}{\ensuremath{ \widetilde{\ten{h}} }}
\newcommand{\tilteni}{\ensuremath{ \widetilde{\ten{i}} }}
\newcommand{\tiltenj}{\ensuremath{ \widetilde{\ten{j}} }}
\newcommand{\tiltenk}{\ensuremath{ \widetilde{\ten{k}} }}
\newcommand{\tiltenl}{\ensuremath{ \widetilde{\ten{l}} }}
\newcommand{\tiltenm}{\ensuremath{ \widetilde{\ten{m}} }}
\newcommand{\tiltenn}{\ensuremath{ \widetilde{\ten{n}} }}
\newcommand{\tilteno}{\ensuremath{ \widetilde{\ten{o}} }}
\newcommand{\tiltenp}{\ensuremath{ \widetilde{\ten{p}} }}
\newcommand{\tiltenq}{\ensuremath{ \widetilde{\ten{q}} }}
\newcommand{\tiltenr}{\ensuremath{ \widetilde{\ten{r}} }}
\newcommand{\tiltens}{\ensuremath{ \widetilde{\ten{s}} }}
\newcommand{\tiltent}{\ensuremath{ \widetilde{\ten{t}} }}
\newcommand{\tiltenu}{\ensuremath{ \widetilde{\ten{u}} }}
\newcommand{\tiltenv}{\ensuremath{ \widetilde{\ten{v}} }}
\newcommand{\tiltenw}{\ensuremath{ \widetilde{\ten{w}} }}
\newcommand{\tiltenx}{\ensuremath{ \widetilde{\ten{x}} }}
\newcommand{\tilteny}{\ensuremath{ \widetilde{\ten{y}} }}
\newcommand{\tiltenz}{\ensuremath{ \widetilde{\ten{z}} }}

\newcommand{\tiltenA}{\ensuremath{ \widetilde{\ten{A}} }}
\newcommand{\tiltenB}{\ensuremath{ \widetilde{\ten{B}} }}
\newcommand{\tiltenC}{\ensuremath{ \widetilde{\ten{C}} }}
\newcommand{\tiltenD}{\ensuremath{ \widetilde{\ten{D}} }}
\newcommand{\tiltenE}{\ensuremath{ \widetilde{\ten{E}} }}
\newcommand{\tiltenF}{\ensuremath{ \widetilde{\ten{F}} }}
\newcommand{\tiltenG}{\ensuremath{ \widetilde{\ten{G}} }}
\newcommand{\tiltenH}{\ensuremath{ \widetilde{\ten{H}} }}
\newcommand{\tiltenI}{\ensuremath{ \widetilde{\ten{I}} }}
\newcommand{\tiltenJ}{\ensuremath{ \widetilde{\ten{J}} }}
\newcommand{\tiltenK}{\ensuremath{ \widetilde{\ten{K}} }}
\newcommand{\tiltenL}{\ensuremath{ \widetilde{\ten{L}} }}
\newcommand{\tiltenM}{\ensuremath{ \widetilde{\ten{M}} }}
\newcommand{\tiltenN}{\ensuremath{ \widetilde{\ten{N}} }}
\newcommand{\tiltenO}{\ensuremath{ \widetilde{\ten{O}} }}
\newcommand{\tiltenP}{\ensuremath{ \widetilde{\ten{P}} }}
\newcommand{\tiltenQ}{\ensuremath{ \widetilde{\ten{Q}} }}
\newcommand{\tiltenR}{\ensuremath{ \widetilde{\ten{R}} }}
\newcommand{\tiltenS}{\ensuremath{ \widetilde{\ten{S}} }}
\newcommand{\tiltenT}{\ensuremath{ \widetilde{\ten{T}} }}
\newcommand{\tiltenU}{\ensuremath{ \widetilde{\ten{U}} }}
\newcommand{\tiltenV}{\ensuremath{ \widetilde{\ten{V}} }}
\newcommand{\tiltenW}{\ensuremath{ \widetilde{\ten{W}} }}
\newcommand{\tiltenX}{\ensuremath{ \widetilde{\ten{X}} }}
\newcommand{\tiltenY}{\ensuremath{ \widetilde{\ten{Y}} }}
\newcommand{\tiltenZ}{\ensuremath{ \widetilde{\ten{Z}} }}

\newcommand{\veca}{\ensuremath{ \vec{a} }}
\newcommand{\vecb}{\ensuremath{ \vec{b} }}
\newcommand{\vecc}{\ensuremath{ \vec{c} }}
\newcommand{\vecd}{\ensuremath{ \vec{d} }}
\newcommand{\vece}{\ensuremath{ \vec{e} }}
\newcommand{\vecf}{\ensuremath{ \vec{f} }}
\newcommand{\vecg}{\ensuremath{ \vec{g} }}
\newcommand{\vech}{\ensuremath{ \vec{h} }}
\newcommand{\veci}{\ensuremath{ \vec{i} }}
\newcommand{\vecj}{\ensuremath{ \vec{j} }}
\newcommand{\veck}{\ensuremath{ \vec{k} }}
\newcommand{\vecl}{\ensuremath{ \vec{l} }}
\newcommand{\vecm}{\ensuremath{ \vec{m} }}
\newcommand{\vecn}{\ensuremath{ \vec{n} }}
\newcommand{\veco}{\ensuremath{ \vec{o} }}
\newcommand{\vecp}{\ensuremath{ \vec{p} }}
\newcommand{\vecq}{\ensuremath{ \vec{q} }}
\newcommand{\vecr}{\ensuremath{ \vec{r} }}
\newcommand{\vecs}{\ensuremath{ \vec{s} }}
\newcommand{\vect}{\ensuremath{ \vec{t} }}
\newcommand{\vecu}{\ensuremath{ \vec{u} }}
\newcommand{\vecv}{\ensuremath{ \vec{v} }}
\newcommand{\vecw}{\ensuremath{ \vec{w} }}
\newcommand{\vecx}{\ensuremath{ \vec{x} }}
\newcommand{\vecy}{\ensuremath{ \vec{y} }}
\newcommand{\vecz}{\ensuremath{ \vec{z} }}

\newcommand{\vecA}{\ensuremath{ \vec{A} }}
\newcommand{\vecB}{\ensuremath{ \vec{B} }}
\newcommand{\vecC}{\ensuremath{ \vec{C} }}
\newcommand{\vecD}{\ensuremath{ \vec{D} }}
\newcommand{\vecE}{\ensuremath{ \vec{E} }}
\newcommand{\vecF}{\ensuremath{ \vec{F} }}
\newcommand{\vecG}{\ensuremath{ \vec{G} }}
\newcommand{\vecH}{\ensuremath{ \vec{H} }}
\newcommand{\vecI}{\ensuremath{ \vec{I} }}
\newcommand{\vecJ}{\ensuremath{ \vec{J} }}
\newcommand{\vecK}{\ensuremath{ \vec{K} }}
\newcommand{\vecL}{\ensuremath{ \vec{L} }}
\newcommand{\vecM}{\ensuremath{ \vec{M} }}
\newcommand{\vecN}{\ensuremath{ \vec{N} }}
\newcommand{\vecO}{\ensuremath{ \vec{O} }}
\newcommand{\vecP}{\ensuremath{ \vec{P} }}
\newcommand{\vecQ}{\ensuremath{ \vec{Q} }}
\newcommand{\vecR}{\ensuremath{ \vec{R} }}
\newcommand{\vecS}{\ensuremath{ \vec{S} }}
\newcommand{\vecT}{\ensuremath{ \vec{T} }}
\newcommand{\vecU}{\ensuremath{ \vec{U} }}
\newcommand{\vecV}{\ensuremath{ \vec{V} }}
\newcommand{\vecW}{\ensuremath{ \vec{W} }}
\newcommand{\vecX}{\ensuremath{ \vec{X} }}
\newcommand{\vecY}{\ensuremath{ \vec{Y} }}
\newcommand{\vecZ}{\ensuremath{ \vec{Z} }}

\newcommand{\tveca}{\ensuremath{ \dot{\vec{a}} }}
\newcommand{\tvecb}{\dot{\vecb}}
\newcommand{\tvecc}{\ensuremath{ \vec{c} }}
\newcommand{\tvecd}{\ensuremath{ \vec{d} }}
\newcommand{\tvece}{\ensuremath{ \vec{e} }}
\newcommand{\tvecf}{\ensuremath{ \vec{f} }}
\newcommand{\tvecg}{\ensuremath{ \dot{\vec{g}} }}
\newcommand{\tvech}{\ensuremath{ \vec{h} }}
\newcommand{\tveci}{\ensuremath{ \vec{i} }}
\newcommand{\tvecj}{\ensuremath{ \vec{j} }}
\newcommand{\tveck}{\ensuremath{ \vec{k} }}
\newcommand{\tvecl}{\ensuremath{ \vec{l} }}
\newcommand{\tvecm}{\ensuremath{ \vec{m} }}
\newcommand{\tvecn}{\dot{\ensuremath{ \vec{n} }}}
\newcommand{\tveco}{\dot{\ensuremath{ \vec{o} }}}
\newcommand{\tvecp}{\dot{\ensuremath{ \vec{p} }}}
\newcommand{\tvecq}{\dot{\ensuremath{ \vec{q} }}}
\newcommand{\tvecr}{\ensuremath{ \vec{r} }}
\newcommand{\tvecs}{\ensuremath{ \vec{s} }}
\newcommand{\tvect}{\ensuremath{ \vec{t} }}
\newcommand{\tvecu}{\dot{\ensuremath{ \vec{u} }}}
\newcommand{\tvecv}{\dot{\ensuremath{ \vec{v} }}}
\newcommand{\tvecw}{\dot{\ensuremath{ \vec{w} }}}
\newcommand{\tvecx}{\dot{\ensuremath{ \vec{x} }}}
\newcommand{\tvecy}{\dot{\ensuremath{ \vec{y} }}}
\newcommand{\tvecz}{\dot{\ensuremath{ \vec{z} }}}

\newcommand{\tvecA}{\ensuremath{ \vec{A} }}
\newcommand{\tvecB}{\ensuremath{ \vec{B} }}
\newcommand{\tvecC}{\ensuremath{ \vec{C} }}
\newcommand{\tvecD}{\ensuremath{ \vec{D} }}
\newcommand{\tvecE}{\ensuremath{ \vec{E} }}
\newcommand{\tvecF}{\ensuremath{ \vec{F} }}
\newcommand{\tvecG}{\ensuremath{ \vec{G} }}
\newcommand{\tvecH}{\ensuremath{ \vec{H} }}
\newcommand{\tvecI}{\ensuremath{ \vec{I} }}
\newcommand{\tvecJ}{\ensuremath{ \vec{J} }}
\newcommand{\tvecK}{\ensuremath{ \vec{K} }}
\newcommand{\tvecL}{\ensuremath{ \vec{L} }}
\newcommand{\tvecM}{\ensuremath{ \vec{M} }}
\newcommand{\tvecN}{\ensuremath{ \vec{N} }}
\newcommand{\tvecO}{\ensuremath{ \vec{O} }}
\newcommand{\tvecP}{\ensuremath{ \vec{P} }}
\newcommand{\tvecQ}{\ensuremath{ \vec{Q} }}
\newcommand{\tvecR}{\ensuremath{ \vec{R} }}
\newcommand{\tvecS}{\ensuremath{ \vec{S} }}
\newcommand{\tvecT}{\ensuremath{ \vec{T} }}
\newcommand{\tvecU}{\ensuremath{ \vec{U} }}
\newcommand{\tvecV}{\ensuremath{ \vec{V} }}
\newcommand{\tvecW}{\ensuremath{ \vec{W} }}
\newcommand{\tvecX}{\ensuremath{ \vec{X} }}
\newcommand{\tvecY}{\ensuremath{ \vec{Y} }}
\newcommand{\tvecZ}{\ensuremath{ \vec{Z} }}

\newcommand{\no}{\hspace{-0.14cm}\,}

\newcommand{\lveca}{\Delta\veca}
\newcommand{\lvecb}{\Delta\vecb}
\newcommand{\lvecc}{\Delta\vecc}
\newcommand{\lvecd}{\Delta\vecd}
\newcommand{\lvece}{\Delta\vece}
\newcommand{\lvecf}{\Delta\vecf}
\newcommand{\lvecg}{\Delta\vecg}
\newcommand{\lvech}{\Delta\vech}
\newcommand{\lveci}{\Delta\veci}
\newcommand{\lvecj}{\Delta\vecj}
\newcommand{\lveck}{\Delta\veck}
\newcommand{\lvecl}{\Delta\vecl}
\newcommand{\lvecm}{\Delta\vecm}
\newcommand{\lvecn}{\Delta\vecn}
\newcommand{\lveco}{\Delta\veco}
\newcommand{\lvecp}{\Delta\vecp}
\newcommand{\lvecq}{\Delta\vecq}
\newcommand{\lvecr}{\Delta\vecr}
\newcommand{\lvecs}{\Delta\vecs}
\newcommand{\lvect}{\Delta\vect}
\newcommand{\lvecu}{\Delta\vecu}
\newcommand{\lvecv}{\Delta\vecv}
\newcommand{\lvecw}{\Delta\vecw}
\newcommand{\lvecx}{\Delta\vecx}
\newcommand{\lvecy}{\Delta\vecy}
\newcommand{\lvecz}{\Delta\vecz}

\newcommand{\lvecN}{\Delta\vecN}


\newcommand{\dveca}{\delta\veca}
\newcommand{\dvecb}{\delta\vecb}
\newcommand{\dvecc}{\delta\vecc}
\newcommand{\dvecd}{\delta\vecd}
\newcommand{\dvece}{\delta\vece}
\newcommand{\dvecf}{\delta\vecf}
\newcommand{\dvecg}{\delta\vecg}
\newcommand{\dvech}{\delta\vech}
\newcommand{\dveci}{\delta\veci}
\newcommand{\dvecj}{\delta\vecj}
\newcommand{\dveck}{\delta\veck}
\newcommand{\dvecl}{\delta\vecl}
\newcommand{\dvecm}{\delta\vecm}
\newcommand{\dvecu}{\delta\vecu}
\newcommand{\dvecv}{\delta\vecv}
\newcommand{\dvecw}{\delta\vecw}
\newcommand{\dvecx}{\delta\vecx}
\newcommand{\dvecy}{\delta\vecy}
\newcommand{\dvecz}{\delta\vecz}

\newcommand{\bveca}{\bar{\veca}}
\newcommand{\bvecb}{\bar{\vecb}}
\newcommand{\bvecc}{\bar{\vecc}}
\newcommand{\bvecd}{\bar{\vecd}}
\newcommand{\bvece}{\bar{\vece}}
\newcommand{\bvecf}{\bar{\vecf}}
\newcommand{\bvecg}{\bar{\vecg}}
\newcommand{\bvecm}{\bar{\vecm}}
\newcommand{\bvecn}{\bar{\vecn}}
\newcommand{\bveco}{\bar{\veco}}
\newcommand{\bvecp}{\bar{\vecp}}
\newcommand{\bvecq}{\bar{\vecq}}
\newcommand{\bvecr}{\bar{\vecr}}
\newcommand{\bvecs}{\bar{\vecs}}
\newcommand{\bvect}{\bar{\vect}}
\newcommand{\bvecu}{\bar{\vecu}}
\newcommand{\bvecv}{\bar{\vecv}}
\newcommand{\bvecw}{\bar{\vecw}}
\newcommand{\bvecx}{\bar{\vecx}}
\newcommand{\bvecy}{\bar{\vecy}}
\newcommand{\bvecz}{\bar{\vecz}}

\newcommand{\bvecA}{\bar{\vecA}}
\newcommand{\bvecB}{\bar{\vecB}}
\newcommand{\bvecC}{\bar{\vecC}}
\newcommand{\bvecD}{\bar{\vecD}}
\newcommand{\bvecE}{\bar{\vecE}}
\newcommand{\bvecF}{\bar{\vecF}}
\newcommand{\bvecG}{\bar{\vecG}}
\newcommand{\bvecH}{\bar{\vecH}}
\newcommand{\bvecI}{\bar{\vecI}}
\newcommand{\bvecJ}{\bar{\vecJ}}
\newcommand{\bvecK}{\bar{\vecK}}
\newcommand{\bvecL}{\bar{\vecL}}
\newcommand{\bvecM}{\bar{\vecM}}
\newcommand{\bvecN}{\bar{\vecN}}
\newcommand{\bvecO}{\bar{\vecO}}
\newcommand{\bvecP}{\bar{\vecP}}
\newcommand{\bvecR}{\bar{\vecR}}
\newcommand{\bvecS}{\bar{\vecS}}
\newcommand{\bvecT}{\bar{\vecT}}
\newcommand{\bvecU}{\bar{\vecU}}


\newcommand{\lbveca}{\Delta\bar{\veca}}
\newcommand{\lbvecb}{\Delta\bar{\vecb}}
\newcommand{\lbvecc}{\Delta\bar{\vecc}}
\newcommand{\lbvecd}{\Delta\bar{\vecd}}
\newcommand{\lbvece}{\Delta\bar{\vece}}
\newcommand{\lbvecf}{\Delta\bar{\vecf}}
\newcommand{\lbvecg}{\Delta\bar{\vecg}}
\newcommand{\lbvech}{\Delta\bar{\vech}}
\newcommand{\lbveci}{\Delta\bar{\veci}}
\newcommand{\lbvecj}{\Delta\bar{\vecj}}
\newcommand{\lbvect}{\Delta\bar{\vect}}
\newcommand{\lbvecu}{\Delta\bar{\vecu}}
\newcommand{\lbvecx}{\Delta\bar{\vecx}}


\newcommand{\dbveca}{\delta\bar{\veca}}
\newcommand{\dbvecb}{\delta\bar{\vecb}}
\newcommand{\dbvecc}{\delta\bar{\vecc}}
\newcommand{\dbvecd}{\delta\bar{\vecd}}
\newcommand{\dbvece}{\delta\bar{\vece}}
\newcommand{\dbvecf}{\delta\bar{\vecf}}
\newcommand{\dbvecg}{\delta\bar{\vecg}}
\newcommand{\dbvech}{\delta\bar{\vech}}
\newcommand{\dbveci}{\delta\bar{\veci}}
\newcommand{\dbvecj}{\delta\bar{\vecj}}
\newcommand{\dbvect}{\delta\bar{\vect}}
\newcommand{\dbvecu}{\delta\bar{\vecu}}
\newcommand{\dbvecx}{\delta\bar{\vecx}}

\newcommand{\hatveca}{\ensuremath{ \widehat{\vec{a}} }}
\newcommand{\hatvecb}{\ensuremath{ \widehat{\vec{b}} }}
\newcommand{\hatvecc}{\ensuremath{ \widehat{\vec{c}} }}
\newcommand{\hatvecd}{\ensuremath{ \widehat{\vec{d}} }}
\newcommand{\hatvece}{\ensuremath{ \widehat{\vec{e}} }}
\newcommand{\hatvecf}{\ensuremath{ \widehat{\vec{f}} }}
\newcommand{\hatvecg}{\ensuremath{ \widehat{\vec{g}} }}
\newcommand{\hatvech}{\ensuremath{ \widehat{\vec{h}} }}
\newcommand{\hatveci}{\ensuremath{ \widehat{\vec{i}} }}
\newcommand{\hatvecj}{\ensuremath{ \widehat{\vec{j}} }}
\newcommand{\hatveck}{\ensuremath{ \widehat{\vec{k}} }}
\newcommand{\hatvecl}{\ensuremath{ \widehat{\vec{l}} }}
\newcommand{\hatvecm}{\ensuremath{ \widehat{\vec{m}} }}
\newcommand{\hatvecn}{\ensuremath{ \widehat{\vec{n}} }}
\newcommand{\hatveco}{\ensuremath{ \widehat{\vec{o}} }}
\newcommand{\hatvecp}{\ensuremath{ \widehat{\vec{p}} }}
\newcommand{\hatvecq}{\ensuremath{ \widehat{\vec{q}} }}
\newcommand{\hatvecr}{\ensuremath{ \widehat{\vec{r}} }}
\newcommand{\hatvecs}{\ensuremath{ \widehat{\vec{s}} }}
\newcommand{\hatvect}{\ensuremath{ \widehat{\vec{t}} }}
\newcommand{\hatvecu}{\ensuremath{ \widehat{\vec{u}} }}
\newcommand{\hatvecv}{\ensuremath{ \widehat{\vec{v}} }}
\newcommand{\hatvecw}{\ensuremath{ \widehat{\vec{w}} }}
\newcommand{\hatvecx}{\ensuremath{ \widehat{\vec{x}} }}
\newcommand{\hatvecy}{\ensuremath{ \widehat{\vec{y}} }}
\newcommand{\hatvecz}{\ensuremath{ \widehat{\vec{z}} }}

\newcommand{\hatvecA}{\ensuremath{ \widehat{\vec{A}} }}
\newcommand{\hatvecB}{\ensuremath{ \widehat{\vec{B}} }}
\newcommand{\hatvecC}{\ensuremath{ \widehat{\vec{C}} }}
\newcommand{\hatvecD}{\ensuremath{ \widehat{\vec{D}} }}
\newcommand{\hatvecE}{\ensuremath{ \widehat{\vec{E}} }}
\newcommand{\hatvecF}{\ensuremath{ \widehat{\vec{F}} }}
\newcommand{\hatvecG}{\ensuremath{ \widehat{\vec{G}} }}
\newcommand{\hatvecH}{\ensuremath{ \widehat{\vec{H}} }}
\newcommand{\hatvecI}{\ensuremath{ \widehat{\vec{I}} }}
\newcommand{\hatvecJ}{\ensuremath{ \widehat{\vec{J}} }}
\newcommand{\hatvecK}{\ensuremath{ \widehat{\vec{K}} }}
\newcommand{\hatvecL}{\ensuremath{ \widehat{\vec{L}} }}
\newcommand{\hatvecM}{\ensuremath{ \widehat{\vec{M}} }}
\newcommand{\hatvecN}{\ensuremath{ \widehat{\vec{N}} }}
\newcommand{\hatvecO}{\ensuremath{ \widehat{\vec{O}} }}
\newcommand{\hatvecP}{\ensuremath{ \widehat{\vec{P}} }}
\newcommand{\hatvecQ}{\ensuremath{ \widehat{\vec{Q}} }}
\newcommand{\hatvecR}{\ensuremath{ \widehat{\vec{R}} }}
\newcommand{\hatvecS}{\ensuremath{ \widehat{\vec{S}} }}
\newcommand{\hatvecT}{\ensuremath{ \widehat{\vec{T}} }}
\newcommand{\hatvecU}{\ensuremath{ \widehat{\vec{U}} }}
\newcommand{\hatvecV}{\ensuremath{ \widehat{\vec{V}} }}
\newcommand{\hatvecW}{\ensuremath{ \widehat{\vec{W}} }}
\newcommand{\hatvecX}{\ensuremath{ \widehat{\vec{X}} }}
\newcommand{\hatvecY}{\ensuremath{ \widehat{\vec{Y}} }}
\newcommand{\hatvecZ}{\ensuremath{ \widehat{\vec{Z}} }}

\newcommand{\tilveca}{\ensuremath{ \widetilde{\vec{a}} }}
\newcommand{\tilvecb}{\ensuremath{ \widetilde{\vec{b}} }}
\newcommand{\tilvecc}{\ensuremath{ \widetilde{\vec{c}} }}
\newcommand{\tilvecd}{\ensuremath{ \widetilde{\vec{d}} }}
\newcommand{\tilvece}{\ensuremath{ \widetilde{\vec{e}} }}
\newcommand{\tilvecf}{\ensuremath{ \widetilde{\vec{f}} }}
\newcommand{\tilvecg}{\ensuremath{ \widetilde{\vec{g}} }}
\newcommand{\tilvech}{\ensuremath{ \widetilde{\vec{h}} }}
\newcommand{\tilveci}{\ensuremath{ \widetilde{\vec{i}} }}
\newcommand{\tilvecj}{\ensuremath{ \widetilde{\vec{j}} }}
\newcommand{\tilveck}{\ensuremath{ \widetilde{\vec{k}} }}
\newcommand{\tilvecl}{\ensuremath{ \widetilde{\vec{l}} }}
\newcommand{\tilvecm}{\ensuremath{ \widetilde{\vec{m}} }}
\newcommand{\tilvecn}{\ensuremath{ \widetilde{\vec{n}} }}
\newcommand{\tilveco}{\ensuremath{ \widetilde{\vec{o}} }}
\newcommand{\tilvecp}{\ensuremath{ \widetilde{\vec{p}} }}
\newcommand{\tilvecq}{\ensuremath{ \widetilde{\vec{q}} }}
\newcommand{\tilvecr}{\ensuremath{ \widetilde{\vec{r}} }}
\newcommand{\tilvecs}{\ensuremath{ \widetilde{\vec{s}} }}
\newcommand{\tilvect}{\ensuremath{ \widetilde{\vec{t}} }}
\newcommand{\tilvecu}{\ensuremath{ \widetilde{\vec{u}} }}
\newcommand{\tilvecv}{\ensuremath{ \widetilde{\vec{v}} }}
\newcommand{\tilvecw}{\ensuremath{ \widetilde{\vec{w}} }}
\newcommand{\tilvecx}{\ensuremath{ \widetilde{\vec{x}} }}
\newcommand{\tilvecy}{\ensuremath{ \widetilde{\vec{y}} }}
\newcommand{\tilvecz}{\ensuremath{ \widetilde{\vec{z}} }}

\newcommand{\tilvecA}{\ensuremath{ \widetilde{\vec{A}} }}
\newcommand{\tilvecB}{\ensuremath{ \widetilde{\vec{B}} }}
\newcommand{\tilvecC}{\ensuremath{ \widetilde{\vec{C}} }}
\newcommand{\tilvecD}{\ensuremath{ \widetilde{\vec{D}} }}
\newcommand{\tilvecE}{\ensuremath{ \widetilde{\vec{E}} }}
\newcommand{\tilvecF}{\ensuremath{ \widetilde{\vec{F}} }}
\newcommand{\tilvecG}{\ensuremath{ \widetilde{\vec{G}} }}
\newcommand{\tilvecH}{\ensuremath{ \widetilde{\vec{H}} }}
\newcommand{\tilvecI}{\ensuremath{ \widetilde{\vec{I}} }}
\newcommand{\tilvecJ}{\ensuremath{ \widetilde{\vec{J}} }}
\newcommand{\tilvecK}{\ensuremath{ \widetilde{\vec{K}} }}
\newcommand{\tilvecL}{\ensuremath{ \widetilde{\vec{L}} }}
\newcommand{\tilvecM}{\ensuremath{ \widetilde{\vec{M}} }}
\newcommand{\tilvecN}{\ensuremath{ \widetilde{\vec{N}} }}
\newcommand{\tilvecO}{\ensuremath{ \widetilde{\vec{O}} }}
\newcommand{\tilvecP}{\ensuremath{ \widetilde{\vec{P}} }}
\newcommand{\tilvecQ}{\ensuremath{ \widetilde{\vec{Q}} }}
\newcommand{\tilvecR}{\ensuremath{ \widetilde{\vec{R}} }}
\newcommand{\tilvecS}{\ensuremath{ \widetilde{\vec{S}} }}
\newcommand{\tilvecT}{\ensuremath{ \widetilde{\vec{T}} }}
\newcommand{\tilvecU}{\ensuremath{ \widetilde{\vec{U}} }}
\newcommand{\tilvecV}{\ensuremath{ \widetilde{\vec{V}} }}
\newcommand{\tilvecW}{\ensuremath{ \widetilde{\vec{W}} }}
\newcommand{\tilvecX}{\ensuremath{ \widetilde{\vec{X}} }}
\newcommand{\tilvecY}{\ensuremath{ \widetilde{\vec{Y}} }}
\newcommand{\tilvecZ}{\ensuremath{ \widetilde{\vec{Z}} }}

\newcommand{\tilveclam}{\ensuremath{ \widetilde{\gvec{\lambda}} }}

\newcommand{\ttvecd}{\ensuremath{ \ddot{\vec{d}} }}
\newcommand{\ttvect}{\ensuremath{ \ddot{\vec{t}} }}
\newcommand{\ttvecu}{\ensuremath{ \ddot{\vec{u}} }}
\newcommand{\ttvecv}{\ensuremath{ \ddot{\vec{v}} }}
\newcommand{\ttvecw}{\ensuremath{ \ddot{\vec{w}} }}
\newcommand{\ttvecx}{\ensuremath{ \ddot{\vec{x}} }}
\newcommand{\ttvecy}{\ensuremath{ \ddot{\vec{y}} }}
\newcommand{\ttvecz}{\ensuremath{ \ddot{\vec{z}} }}

\newcommand{\scaa}{\ensuremath{\mathrm{a}}}
\newcommand{\scab}{\ensuremath{\mathrm{b}}}
\newcommand{\scac}{\ensuremath{\mathrm{c}}}
\newcommand{\scad}{\ensuremath{\mathrm{d}}}
\newcommand{\scae}{\ensuremath{\mathrm{e}}}
\newcommand{\scaf}{\ensuremath{\mathrm{f}}}
\newcommand{\scag}{\ensuremath{\mathrm{g}}}
\newcommand{\scah}{\ensuremath{\mathrm{h}}}
\newcommand{\scai}{\ensuremath{\mathrm{i}}}
\newcommand{\scaj}{\ensuremath{\mathrm{j}}}
\newcommand{\scak}{\ensuremath{\mathrm{k}}}
\newcommand{\scal}{\ensuremath{\mathrm{l}}}
\newcommand{\scam}{\ensuremath{\mathrm{m}}}
\newcommand{\scan}{\ensuremath{\mathrm{n}}}
\newcommand{\scao}{\ensuremath{\mathrm{o}}}
\newcommand{\scap}{\ensuremath{\mathrm{p}}}
\newcommand{\scaq}{\ensuremath{\mathrm{q}}}
\newcommand{\scar}{\ensuremath{\mathrm{r}}}
\newcommand{\scas}{\ensuremath{\mathrm{s}}}
\newcommand{\scat}{\ensuremath{\mathrm{t}}}
\newcommand{\scau}{\ensuremath{\mathrm{u}}}
\newcommand{\scav}{\ensuremath{\mathrm{v}}}
\newcommand{\scaw}{\ensuremath{\mathrm{w}}}
\newcommand{\scax}{\ensuremath{\mathrm{x}}}
\newcommand{\scay}{\ensuremath{\mathrm{y}}}
\newcommand{\scaz}{\ensuremath{\mathrm{z}}}

\newcommand{\scaA}{\ensuremath{\mathrm{A}}}
\newcommand{\scaB}{\ensuremath{\mathrm{B}}}
\newcommand{\scaC}{\ensuremath{\mathrm{C}}}
\newcommand{\scaD}{\ensuremath{\mathrm{D}}}
\newcommand{\scaE}{\ensuremath{\mathrm{E}}}
\newcommand{\scaF}{\ensuremath{\mathrm{F}}}
\newcommand{\scaG}{\ensuremath{\mathrm{G}}}
\newcommand{\scaH}{\ensuremath{\mathrm{H}}}
\newcommand{\scaI}{\ensuremath{\mathrm{I}}}
\newcommand{\scaJ}{\ensuremath{\mathrm{J}}}
\newcommand{\scaK}{\ensuremath{\mathrm{K}}}
\newcommand{\scaL}{\ensuremath{\mathrm{L}}}
\newcommand{\scaM}{\ensuremath{\mathrm{M}}}
\newcommand{\scaN}{\ensuremath{\mathrm{N}}}
\newcommand{\scaO}{\ensuremath{\mathrm{O}}}
\newcommand{\scaP}{\ensuremath{\mathrm{P}}}
\newcommand{\scaQ}{\ensuremath{\mathrm{Q}}}
\newcommand{\scaR}{\ensuremath{\mathrm{R}}}
\newcommand{\scaS}{\ensuremath{\mathrm{S}}}
\newcommand{\scaT}{\ensuremath{\mathrm{T}}}
\newcommand{\scaU}{\ensuremath{\mathrm{U}}}
\newcommand{\scaV}{\ensuremath{\mathrm{V}}}
\newcommand{\scaW}{\ensuremath{\mathrm{W}}}
\newcommand{\scaX}{\ensuremath{\mathrm{X}}}
\newcommand{\scaY}{\ensuremath{\mathrm{Y}}}
\newcommand{\scaZ}{\ensuremath{\mathrm{Z}}}

\newcommand{\bscaa}{\bar{\scaa}}
\newcommand{\bscab}{\bar{\scab}}
\newcommand{\bscac}{\bar{\scac}}
\newcommand{\bscad}{\bar{\scad}}
\newcommand{\bscae}{\bar{\scae}}
\newcommand{\bscaf}{\bar{\scaf}}
\newcommand{\bscag}{\bar{\scag}}
\newcommand{\bscah}{\bar{\scah}}
\newcommand{\bscai}{\bar{\scai}}
\newcommand{\bscaj}{\bar{\scaj}}
\newcommand{\bscak}{\bar{\scak}}
\newcommand{\bscal}{\bar{\scal}}
\newcommand{\bscam}{\bar{\scam}}
\newcommand{\bscan}{\bar{\scan}}
\newcommand{\bscao}{\bar{\scao}}
\newcommand{\bscap}{\bar{\scap}}
\newcommand{\bscaq}{\bar{\scaq}}
\newcommand{\bscar}{\bar{\scar}}
\newcommand{\bscas}{\bar{\scas}}
\newcommand{\bscat}{\bar{\scat}}

\newcommand{\bscaA}{\bar{\scaA}}
\newcommand{\bscaB}{\bar{\scaB}}
\newcommand{\bscaC}{\bar{\scaC}}
\newcommand{\bscaD}{\bar{\scaD}}
\newcommand{\bscaE}{\bar{\scaE}}
\newcommand{\bscaF}{\bar{\scaF}}
\newcommand{\bscaG}{\bar{\scaG}}
\newcommand{\bscaH}{\bar{\scaH}}
\newcommand{\bscaI}{\bar{\scaI}}
\newcommand{\bscaJ}{\bar{\scaJ}}
\newcommand{\bscaK}{\bar{\scaK}}
\newcommand{\bscaL}{\bar{\scaL}}
\newcommand{\bscaM}{\bar{\scaM}}
\newcommand{\bscaN}{\bar{\scaN}}
\newcommand{\bscaO}{\bar{\scaO}}
\newcommand{\bscaP}{\bar{\scaP}}
\newcommand{\bscaQ}{\bar{\scaQ}}
\newcommand{\bscaR}{\bar{\scaR}}
\newcommand{\bscaS}{\bar{\scaS}}
\newcommand{\bscaT}{\bar{\scaT}}

\newcommand{\hatscan}{\ensuremath{\widehat{\scan}}}

\newcommand{\tilscag}{\ensuremath{\widetilde{\scag}}}
\newcommand{\tilscat}{\ensuremath{\widetilde{\scat}}}

\newcommand{\tillam}{\ensuremath{\widetilde{\lambda}}}

\newcommand{\tscag}{\ensuremath{\dot{\scag}}}

\newcommand{\lscaa}{\Delta\scaa}
\newcommand{\lscab}{\Delta\scab}
\newcommand{\lscac}{\Delta\scac}
\newcommand{\lscad}{\Delta\scad}
\newcommand{\lscae}{\Delta\scae}
\newcommand{\lscaf}{\Delta\scaf}
\newcommand{\lscag}{\Delta\scag}
\newcommand{\lscah}{\Delta\scah}
\newcommand{\lscai}{\Delta\scai}
\newcommand{\lscaj}{\Delta\scaj}
\newcommand{\lscak}{\Delta\scak}
\newcommand{\lscal}{\Delta\scal}
\newcommand{\lscam}{\Delta\scam}
\newcommand{\lscan}{\Delta\scan}
\newcommand{\lscao}{\Delta\scao}
\newcommand{\lscap}{\Delta\scap}
\newcommand{\lscaq}{\Delta\scaq}
\newcommand{\lscar}{\Delta\scar}
\newcommand{\lscas}{\Delta\scas}
\newcommand{\lscat}{\Delta\scat}

\newcommand{\lscaA}{\Delta\scaA}
\newcommand{\lscaD}{\Delta\scaD}
\newcommand{\lscaM}{\Delta\scaM}
\newcommand{\lscaN}{\Delta\scaN}

\newcommand{\balpha     }{\bar{\alpha}}
\newcommand{\bbeta      }{\bar{\beta}}
\newcommand{\bgamma     }{\bar{\gamma}}
\newcommand{\bdelta     }{\bar{\delta}}
\newcommand{\bepsilon   }{\bar{\epsilon}}
\newcommand{\bvareps    }{\bar{\varepsilon}}
\newcommand{\blambda    }{\bar{\lambda}}
\newcommand{\bxi        }{\bar{\xi}}
\newcommand{\bsigma     }{\bar{\sigma}}
\newcommand{\bvarsigma  }{\bar{\varsigma}}
\newcommand{\btau       }{\bar{\tau}}

\newcommand{\tileps     }{\widetilde{\epsilon}}
\newcommand{\tillambda  }{\widetilde{\lambda}}
\newcommand{\tilsigma   }{\widetilde{\sigma}}

\newcommand{\vecalpha     }{\ensuremath{ \gvec{\alpha} }}
\newcommand{\vecbeta      }{\ensuremath{ \gvec{\beta} }}
\newcommand{\vecgamma     }{\ensuremath{ \gvec{\gamma} }}
\newcommand{\vecdelta     }{\ensuremath{ \gvec{\delta} }}
\newcommand{\vecepsilon   }{\ensuremath{ \gvec{\epsilon} }}
\newcommand{\vecvarepsilon}{\ensuremath{ \gvec{\varepsilon} }}
\newcommand{\veczeta      }{\ensuremath{ \gvec{\zeta} }}
\newcommand{\veceta       }{\ensuremath{ \gvec{\eta} }}
\newcommand{\vectheta     }{\ensuremath{ \gvec{\theta} }}
\newcommand{\vecvartheta  }{\ensuremath{ \gvec{\vartheta} }}
\newcommand{\veciota      }{\ensuremath{ \gvec{\iota} }}
\newcommand{\veckappa     }{\ensuremath{ \gvec{\kappa} }}
\newcommand{\veclam       }{\ensuremath{ \gvec{\lambda} }}
\newcommand{\vecmu        }{\ensuremath{ \gvec{\mu} }}
\newcommand{\vecnu        }{\ensuremath{ \gvec{\nu} }}
\newcommand{\vecxi        }{\ensuremath{ \gvec{\xi} }}
\newcommand{\vecpi        }{\ensuremath{ \gvec{\pi} }}
\newcommand{\vecvarpi     }{\ensuremath{ \gvec{\varphi} }}
\newcommand{\vecrho       }{\ensuremath{ \gvec{\rho} }}
\newcommand{\vecvarrho    }{\ensuremath{ \gvec{\varrho} }}
\newcommand{\vecsigma     }{\ensuremath{ \gvec{\sigma} }}
\newcommand{\vecvarsigma  }{\ensuremath{ \gvec{\varsigma} }}
\newcommand{\vectau       }{\ensuremath{ \gvec{\tau} }}
\newcommand{\vecupsilon   }{\ensuremath{ \gvec{\upsilon} }}
\newcommand{\vecphi       }{\ensuremath{ \gvec{\phi} }}
\newcommand{\vecvarphi    }{\ensuremath{ \gvec{\varphi} }}
\newcommand{\vecchi       }{\ensuremath{ \gvec{\chi} }}
\newcommand{\vecpsi       }{\ensuremath{ \gvec{\psi} }}
\newcommand{\vecomega     }{\ensuremath{ \gvec{\omega} }}
\newcommand{\vecUpsilon   }{\ensuremath{ \gvec{\Upsilon} }}

\newcommand{\bveceps      }{\ensuremath{ \bar{\gvec{\epsilon}} }}
\newcommand{\bveceta      }{\ensuremath{ \bar{\gvec{\eta}} }}
\newcommand{\bveclam      }{\ensuremath{ \bar{\gvec{\lambda}} }}
\newcommand{\bvecsig      }{\ensuremath{ \bar{\gvec{\sigma}} }}
\newcommand{\bvecvarsigma }{\ensuremath{ \bar{\gvec{\varsigma}} }}
\newcommand{\bvectau      }{\ensuremath{ \bar{\gvec{\tau}} }}
\newcommand{\bvecupsilon  }{\ensuremath{ \bar{\gvec{\upsilon} }}}

\newcommand{\tilveceps    }{\widetilde{\vecepsilon}}
\newcommand{\tilvecsig    }{\widetilde{\vecsigma}}

\newcommand{\tveclam}{\ensuremath{ \dot{\veclam }}}

\newcommand{\lveceps}{\Delta\vecepsilon}
\newcommand{\lveclam}{\Delta\veclam}
\newcommand{\lvecsig}{\Delta\vecsigma}
\newcommand{\lvectau}{\Delta\vectau}
\newcommand{\lvecxi }{\Delta\vecxi}

\newcommand{\dveceps}{\delta\vecepsilon}
\newcommand{\dveclam}{\delta\veclam}
\newcommand{\dvecsig}{\delta\vecsigma}
\newcommand{\dvectau}{\delta\vectau}
\newcommand{\dvecxi }{\delta\vecxi}


\newcommand{\lbveclam}{\Delta\bar{\veclam}}

\newcommand{\tenalpha     }{\ensuremath{ \gten{\alpha} }}
\newcommand{\tenbeta      }{\ensuremath{ \gten{\beta} }}
\newcommand{\tengamma     }{\ensuremath{ \gten{\gamma} }}
\newcommand{\tendelta     }{\ensuremath{ \gten{\delta} }}
\newcommand{\tenepsilon   }{\ensuremath{ \gten{\epsilon} }}
\newcommand{\teneps       }{\ensuremath{ \gten{\varepsilon} }}
\newcommand{\tenzeta      }{\ensuremath{ \gten{\zeta} }}
\newcommand{\teneta       }{\ensuremath{ \gten{\eta} }}
\newcommand{\tentheta     }{\ensuremath{ \gten{\theta} }}
\newcommand{\tenvartheta  }{\ensuremath{ \gten{\vartheta} }}
\newcommand{\teniota      }{\ensuremath{ \gten{\iota} }}
\newcommand{\tenkappa     }{\ensuremath{ \gten{\kappa} }}
\newcommand{\tenlambda    }{\ensuremath{ \gten{\lambda} }}
\newcommand{\tenmu        }{\ensuremath{ \gten{\mu} }}
\newcommand{\tennu        }{\ensuremath{ \gten{\nu} }}
\newcommand{\tenxi        }{\ensuremath{ \gten{\xi} }}
\newcommand{\tenpi        }{\ensuremath{ \gten{\pi} }}
\newcommand{\tenvarpi     }{\ensuremath{ \gten{\varphi} }}
\newcommand{\tenrho       }{\ensuremath{ \gten{\rho} }}
\newcommand{\tenvarrho    }{\ensuremath{ \gten{\varrho} }}
\newcommand{\tensig       }{\ensuremath{ \gten{\sigma} }}
\newcommand{\tenvarsigma  }{\ensuremath{ \gten{\varsigma} }}
\newcommand{\tentau       }{\ensuremath{ \gten{\tau} }}
\newcommand{\tenupsilon   }{\ensuremath{ \gten{\upsilon} }}
\newcommand{\tenphi       }{\ensuremath{ \gten{\phi} }}
\newcommand{\tenvarphi    }{\ensuremath{ \gten{\varphi} }}
\newcommand{\tenchi       }{\ensuremath{ \gten{\chi} }}
\newcommand{\tenpsi       }{\ensuremath{ \gten{\psi} }}
\newcommand{\tenomega     }{\ensuremath{ \gten{\omega} }}

\newcommand{\tenOmega     }{\ensuremath{ \gten{\Omega} }}

\newcommand{\tilteneps    }{\widetilde{\teneps}}
\newcommand{\tiltensig    }{\widetilde{\tensig}}


\newcommand{\bteneps}{\ensuremath{ \bar{\teneps }}}
\newcommand{\btensig}{\ensuremath{ \bar{\tensig }}}


\newcommand{\tteneps}{\ensuremath{ \dot{\teneps }}}
\newcommand{\ttensig}{\ensuremath{ \dot{\tensig }}}


\newcommand{\ltenalpha}{\Delta\tenalpha}
\newcommand{\ltenbeta }{\Delta\tenbeta}
\newcommand{\lteneps  }{\Delta\teneps}
\newcommand{\ltensig  }{\Delta\tensig}

\newcommand{\tgamma}{\ensuremath{ \dot{\gamma} }}
\newcommand{\txi}{\ensuremath{ \dot{\xi} }}
\newcommand{\tlam}{\ensuremath{ \dot{\lambda} }}
\newcommand{\tomega}{\ensuremath{ \dot{\omega} }}

\newcommand{\lgamma}{\Delta\gamma}
\newcommand{\llambda}{\Delta\lambda}
\newcommand{\lxi}{\Delta\xi}
\newcommand{\lsigma}{\stackrel{\triangle}{\sigma}}
\newcommand{\ltau}{\stackrel{\triangle}{\tau}}

\newcommand{\hatalpha     }{\ensuremath{ \widehat{\alpha} }}
\newcommand{\hatbeta      }{\ensuremath{ \widehat{\beta} }}
\newcommand{\hatgamma     }{\ensuremath{ \widehat{\gamma} }}
\newcommand{\hatdelta     }{\ensuremath{ \widehat{\delta} }}
\newcommand{\hatepsilon   }{\ensuremath{ \widehat{\epsilon} }}
\newcommand{\hatvarepsilon}{\ensuremath{ \widehat{\varepsilon} }}
\newcommand{\hatzeta      }{\ensuremath{ \widehat{\zeta} }}
\newcommand{\hateta       }{\ensuremath{ \widehat{\eta} }}
\newcommand{\hattheta     }{\ensuremath{ \widehat{\theta} }}
\newcommand{\hatvartheta  }{\ensuremath{ \widehat{\vartheta} }}
\newcommand{\hatiota      }{\ensuremath{ \widehat{\iota} }}
\newcommand{\hatkappa     }{\ensuremath{ \widehat{\kappa} }}
\newcommand{\hatlambda    }{\ensuremath{ \widehat{\lambda} }}
\newcommand{\hatmu        }{\ensuremath{ \widehat{\mu} }}
\newcommand{\hatnu        }{\ensuremath{ \widehat{\nu} }}
\newcommand{\hatxi        }{\ensuremath{ \widehat{\xi} }}
\newcommand{\hatpi        }{\ensuremath{ \widehat{\pi} }}
\newcommand{\hatvarpi     }{\ensuremath{ \widehat{\varphi} }}
\newcommand{\hatrho       }{\ensuremath{ \widehat{\rho} }}
\newcommand{\hatvarrho    }{\ensuremath{ \widehat{\varrho} }}
\newcommand{\hatsigma     }{\ensuremath{ \widehat{\sigma} }}
\newcommand{\hatvarsigma  }{\ensuremath{ \widehat{\varsigma} }}
\newcommand{\hattau       }{\ensuremath{ \widehat{\tau} }}
\newcommand{\hatupsilon   }{\ensuremath{ \widehat{\upsilon} }}
\newcommand{\hatphi       }{\ensuremath{ \widehat{\phi} }}
\newcommand{\hatvarphi    }{\ensuremath{ \widehat{\varphi} }}
\newcommand{\hatchi       }{\ensuremath{ \widehat{\chi} }}
\newcommand{\hatpsi       }{\ensuremath{ \widehat{\psi} }}
\newcommand{\hatomega     }{\ensuremath{ \widehat{\omega} }}

\newcommand{\hatteneps}{\ensuremath{ \widehat{\teneps} }}
\newcommand{\hattensig}{\ensuremath{ \widehat{\tensig} }}

\newcommand{\ionesi}{\scaI\utensig}
\newcommand{\itwosi}{\scaI\scaI\utensig}
\newcommand{\ithrsi}{\scaI\scaI\scaI\utensig}
\newcommand{\itwos}{\scaI\scaI\utens}
\newcommand{\ithrs}{\scaI\scaI\scaI\utens}

\newcommand{\onetwo}{\frac{1}{2}}
\newcommand{\thrtwo}{\frac{3}{2}}
\newcommand{\onethr}{\frac{1}{3}}
\newcommand{\twothr}{\frac{2}{3}}
\newcommand{\forthr}{\frac{4}{3}}
\newcommand{\onefor}{\frac{1}{4}}
\newcommand{\onesix}{\frac{1}{6}}
\newcommand{\oneeig}{\frac{1}{8}}
\newcommand{\onenin}{\frac{1}{9}}
\newcommand{\onetwe}{\frac{1}{12}}

\newcommand{\tengf}{\teng^{\flat}}
\newcommand{\tengs}{\teng^{\sharp}}

\newcommand{\Lin}{^{Lin}}

\newcommand{\uscan}{_{\mbox{\tiny{N}}}}

\newcommand{\ena}{\ensuremath{^{n+1}}}
\newcommand{\sena}{\ensuremath{^{1\,n+1}}}
\newcommand{\mena}{\ensuremath{^{2\,n+1}}}

\newcommand{\ea}{^{\alpha}}
\newcommand{\eb}{^{\beta}}
\newcommand{\ec}{^{\gamma}}
\newcommand{\ed}{^{\delta}}
\newcommand{\ex}{^{\xi}}

\newcommand{\eat}{^{\alpha T}}
\newcommand{\ebt}{^{\beta T}}
\newcommand{\ect}{^{\gamma T}}
\newcommand{\edt}{^{\delta T}}
\newcommand{\eet}{^{\epsilon T}}

\newcommand{\eaa}{^{\alpha\alpha}}
\newcommand{\eab}{^{\alpha\beta}}
\newcommand{\eac}{^{\alpha\gamma}}
\newcommand{\ead}{^{\alpha\delta}}
\newcommand{\eba}{^{\beta\alpha}}
\newcommand{\ebc}{^{\beta\gamma}}
\newcommand{\ebd}{^{\beta\delta}}
\newcommand{\ecb}{^{\gamma\beta}}
\newcommand{\ecd}{^{\gamma\delta}}
\newcommand{\edb}{^{\delta\beta}}
\newcommand{\ede}{^{\delta\epsilon}}
\newcommand{\eec}{^{\epsilon\gamma}}
\newcommand{\exx}{^{\xi\xi}}

\newcommand{\ua}{_{\alpha}}
\newcommand{\ub}{_{\beta}}
\newcommand{\uc}{_{\gamma}}
\newcommand{\ud}{_{\delta}}
\newcommand{\ue}{_{\epsilon}}
\newcommand{\ux}{_{\xi}}

\newcommand{\uaa}{_{\alpha\alpha}}
\newcommand{\uab}{_{\alpha\beta}}
\newcommand{\uac}{_{\alpha\gamma}}
\newcommand{\uba}{_{\beta\alpha}}
\newcommand{\ubb}{_{\beta\beta}}
\newcommand{\ubc}{_{\beta\gamma}}
\newcommand{\ubd}{_{\beta\delta}}
\newcommand{\ucd}{_{\gamma\delta}}
\newcommand{\ucb}{_{\gamma\beta}}
\newcommand{\ueb}{_{\epsilon\beta}}
\newcommand{\ued}{_{\epsilon\delta}}
\newcommand{\udb}{_{\delta\beta}}
\newcommand{\uta}{_{{\mbox{\tiny{T}}}\alpha}}
\newcommand{\utb}{_{{\mbox{\tiny{T}}}\beta}}
\newcommand{\utc}{_{{\mbox{\tiny{T}}}\gamma}}
\newcommand{\uxx}{_{\xi\xi}}

\newcommand{\uka}{_{,\alpha}}
\newcommand{\ukb}{_{,\beta}}
\newcommand{\ukc}{_{,\gamma}}
\newcommand{\ukx}{_{,\xi}}

\newcommand{\uakb}{_{\alpha,\beta}}
\newcommand{\uakc}{_{\alpha,\gamma}}
\newcommand{\ubkc}{_{\beta,\gamma}}
\newcommand{\ubkd}{_{\beta,\delta}}
\newcommand{\ubke}{_{\beta,\epsilon}}
\newcommand{\uckd}{_{\gamma,\delta}}
\newcommand{\udke}{_{\delta,\epsilon}}

\newcommand{\ukaa}{_{,\alpha\alpha}}
\newcommand{\ukab}{_{,\alpha\beta}}
\newcommand{\ukba}{_{,\beta\alpha}}
\newcommand{\ukbb}{_{,\beta\beta}}
\newcommand{\ukbc}{_{,\beta\gamma}}
\newcommand{\ukxx}{_{,\xi\xi}}
\newcommand{\uxkx}{_{\xi,\xi}}
\newcommand{\uxkxx}{_{\xi,\xi\xi}}

\newcommand{\ubkcd}{_{\beta,\gamma\delta}}

\newcommand{\uga}{_{g\alpha}}
\newcommand{\ugb}{_{g\beta}}
\newcommand{\ugc}{_{g\gamma}}
\newcommand{\ugd}{_{g\delta}}
\newcommand{\ugka}{_{g,\alpha}}
\newcommand{\ugkb}{_{g,\beta}}
\newcommand{\ugkc}{_{g,\gamma}}
\newcommand{\ugkx}{_{g,\xi}}
\newcommand{\ugakb}{_{g\alpha,\beta}}
\newcommand{\ugbkc}{_{g\beta,\gamma}}

\newcommand{\uana}{_{\alpha\,n+1}}
\newcommand{\ubna}{_{\beta\,n+1}}

\newcommand{\ukana}{_{,\alpha\,n+1}}
\newcommand{\ukano}{_{,\alpha\,n}}

\newcommand{\uano}{_{\alpha\,n}}
\newcommand{\ubno}{_{\beta\,n}}

\newcommand{\umN}{_{\mbox{\tiny{N}}}}
\newcommand{\umT}{_{\mbox{\tiny{T}}}}

\newcommand{\utenb}{_{\tenb}}
\newcommand{\utenp}{_{\tenp}}
\newcommand{\utens}{_{\tens}}
\newcommand{\utenC}{_{\tenC}}
\newcommand{\utenE}{_{\tenE}}

\newcommand{\utensna}{_{\tens\,n+1}}

\newcommand{\uteneps}{_{\teneps}}
\newcommand{\utenepse}{_{\teneps^e}}
\newcommand{\utenepsp}{_{\teneps^p}}
\newcommand{\utensig}{_{\tensig}}
\newcommand{\utensigsig}{_{\tensig\tensig}}

\newcommand{\utenepsna}{_{\teneps\,n+1}}
\newcommand{\utensigna}{_{\tensig\,n+1}}

\newcommand{\gvecx}{\grave{\vecx}}


\newcommand{\mska}{_{s,\alpha}}
\newcommand{\mskb}{_{s,\beta}}
\newcommand{\mskc}{_{s,\gamma}}
\newcommand{\mfka}{_{f,\alpha}}
\newcommand{\mfkb}{_{f,\beta}}
\newcommand{\mfkc}{_{f,\gamma}}


\newcommand{\dmska}{_{Ag,\alpha}}
\newcommand{\dmskb}{_{Ag,\beta}}
\newcommand{\dmskc}{_{Ag,\gamma}}
\newcommand{\dmfka}{_{Ag,\alpha}}
\newcommand{\dmfkb}{_{Ag,\beta}}
\newcommand{\dmfkc}{_{Ag,\gamma}}

\newcommand{\gmska}{_{g,\alpha}}
\newcommand{\gmskb}{_{g,\beta}}
\newcommand{\gmskc}{_{g,\gamma}}
\newcommand{\gmfka}{_{g,\alpha}}
\newcommand{\gmfkb}{_{g,\beta}}
\newcommand{\gmfkc}{_{g,\gamma}}

\newcommand{\sumgp}{\sum_{g=1}^{n_{gp}}}
\newcommand{\sumni}{\sum_{I=1}^{n_{I}}}
\newcommand{\sumnj}{\sum_{J=1}^{n_{J}}}
\newcommand{\sumseg}{\sum_{seg}}
\newcommand{\sumel}{\sum_{e=1}^{n_{el}}}

\newcommand{\suma}{\sum_{\alpha=1}^{n_{\alpha}}}

\newcommand{\gng}{\scag_{N\,g}}
\newcommand{\gtg}{g_{T\,g}}
\newcommand{\lng}{\lambda_{N\,g}}
\newcommand{\ltg}{\lambda_{T\,g}}
\newcommand{\ttg}{t_{T\,g}}

\renewcommand{\d}[1]{\text{$\hspace{0.1cm}$d $\hspace{-0.11cm}#1$}}
\newcommand{\del}{\ensuremath{\partial}}
\newcommand{\divx}[1]{\text{$\hspace{0.1cm}$div$\left(#1\right)$}}
\newcommand{\divX}[1]{\text{$\hspace{0.1cm}$Div$\left(#1\right)$}}
\newcommand{\grad}[1]{\ensuremath{ \boldsymbol{\nabla}{#1}}}
\newcommand{\gradx}[1]{\ensuremath{ \boldsymbol{\nabla}_{\vecx}{#1}}}
\newcommand{\gradX}[1]{\ensuremath{ \boldsymbol{\nabla}_{\vecX}{#1}}}
\newcommand{\parder}[2]{\ensuremath{ \frac{\del #1}{\del #2} }}
\newcommand{\tder}[1]{\ensuremath{ \frac{\d{#1}}{\d{} \hspace{0.05cm}{t}} }}
\newcommand{\dx}{\ensuremath{ \d{\vecx} }}
\newcommand{\dX}{\ensuremath{ \d{\vecX} }}
\newcommand{\da}{\ensuremath{ \d{a} }}
\newcommand{\dA}{\ensuremath{ \d{A} }}
\newcommand{\dv}{\ensuremath{ \d{v} }}
\newcommand{\dV}{\ensuremath{ \d{V} }}
\newcommand{\dxis}{\ensuremath{ \d{\xi} }}
\newcommand{\lda}{\stackrel{\triangle}{\da}}

\newcommand{\dr}{\ensuremath{ \d{r} }}
\newcommand{\dphi}{\ensuremath{ \d{\phi} }}
\newcommand{\dz}{\ensuremath{\d{z}}}
\newcommand{\du}{\ensuremath{\d{u}}}
\newcommand{\dy}{\ensuremath{\d{y}}}
\newcommand{\dxscal}{\ensuremath{\d{x}}}

\renewcommand{\cos}[1]{ \text{cos}\hspace{0.0cm}\left( {#1} \right) }
\renewcommand{\sin}[1]{ \text{sin}\hspace{0.0cm}\left( {#1} \right) }
\renewcommand{\ln}[1]{\text{$\hspace{0.1cm}$ln$\left(#1\right)$}}
\renewcommand{\exp}[1]{\ensuremath{ \,\text{exp}{\left( #1 \right)} }}

\newcommand{\define}{\ensuremath{\stackrel{\mathrm{def}}{=}}}
\newcommand{\p}{\ensuremath{ ^{\prime} }}
\newcommand{\pp}{\ensuremath{ ^{\prime \prime} }}
\newcommand{\first}{$\ensuremath{ 1^{\text{st}} }\,$}
\newcommand{\second}{$\ensuremath{ 2^{\text{nd}} }\,$}
\newcommand{\lin}[1]{\ensuremath{ \calL[#1] }}

\newcommand{\tenfour}[1]{ \ensuremath {\boldsymbol{\mathcal{#1}} } }
\newcommand{\tr}[1]{\text{$\hspace{0.1cm}$tr$\left(#1\right)$}}
\newcommand{\dev}{\ensuremath{ ^{\prime} }}
\renewcommand{\det}[1]{\text{$\hspace{0.1cm}$det$\left(#1\right)$}}
\newcommand{\inv}{\ensuremath{ ^{-1} }}

\renewcommand{\it}{\ensuremath{ ^{-T} }}
\newcommand{\sym}{\ensuremath{ ^{\text{sym}} }}
\newcommand{\skw}{\ensuremath{ ^{\text{skw}} }}
\newcommand{\adj}{\ensuremath{ ^{\sharp}  }}
\newcommand{\mg}[1]{\ensuremath{ \left\| #1 \right\| }}
\newcommand{\mgv}[1]{\ensuremath{ \left| #1 \right| }}
\newcommand{\s}{\ensuremath{ ^{2}  }}
\newcommand{\ione}[1]{\ensuremath{ I_{#1} }}
\newcommand{\itwo}[1]{\ensuremath{ I\hspace{-0.1cm}I_{#1} }}
\newcommand{\ithree}[1]{\ensuremath{ I\hspace{-0.1cm}I\hspace{-0.1cm}I_{#1} }}
\newcommand{\ionep}[1]{\ensuremath{ I_{#1}\p }}
\newcommand{\itwop}[1]{\ensuremath{ I\hspace{-0.1cm}I_{#1}\p }}
\newcommand{\ithreep}[1]{\ensuremath{ I\hspace{-0.1cm}I\hspace{-0.1cm}I_{#1}\p }}
\newcommand{\ionepp}[1]{\ensuremath{ I_{#1}\pp }}
\newcommand{\itwopp}[1]{\ensuremath{ I\hspace{-0.1cm}I_{#1}\pp }}
\newcommand{\ithreepp}[1]{\ensuremath{ I\hspace{-0.1cm}I\hspace{-0.1cm}I_{#1}\pp }}

\newcommand{\gus}{\frac{\partial\scag}{\partial\tensig}}
\newcommand{\guss}{\frac{\partial^2\scag}{\partial^2\tensig}}

\newcommand{\percent}{\ensuremath{ \%  }}
\newcommand{\IE}{\ensuremath{I\hspace{-0.12cm}E  }}
\newcommand{\II}{\ensuremath{1\hspace{-0.12cm}1  }}
\newcommand{\tenIE}{\ensuremath{ \ten{\IE}  }}
\newcommand{\tenII}{\ensuremath{ \ten{\II}  }}
\newcommand{\hattenIE}{\ensuremath{ \widehat{\tenIE}  }}
\newcommand{\IC}{\ensuremath{C\hspace{-0.22cm}C  }}
\newcommand{\cc}{\ensuremath{c\hspace{-0.22cm}c  }}
\newcommand{\tenIC}{\ensuremath{ \ten{\IC}  }}
\newcommand{\tencc}{\ensuremath{ \ten{\cc}  }}
\newcommand{\lsup}[1]{\ensuremath{ {}^{[#1]} \hspace{-0.05cm} }}
\newcommand{\lsub}[1]{\ensuremath{ {}_{[#1]} \hspace{-0.05cm} }}
\newcommand{\lsupb}[2]{\ensuremath{ {}^{#1}_{#2} \hspace{-0.05cm} }}
\newcommand{\Phihat}{\ensuremath{ \widehat{\Phi} }}
\newcommand{\Psihat}{\ensuremath{ \widehat{\Psi} }}
\renewcommand{\s}{\ensuremath{ ^*} }
\renewcommand{\k}{\ensuremath{ ^K} }
\newcommand{\kp}{\ensuremath{ ^{K+1}} }
\newcommand{\km}{\ensuremath{ ^{K-1}} }
\newcommand{\tol}[1]{\ensuremath{ \text{TOL}_{#1} } }
\newcommand{\pen}{\ensuremath{ \calK }}
\newcommand{\tenone}{\ensuremath{ \ten{1} }}
\newcommand{\err}{\ensuremath{ \pounds }}
\newcommand{\li}{\ensuremath{\left[\hspace{-0.15cm}\left[ }}
\newcommand{\ri}{\ensuremath{\right]\hspace{-0.15cm}\right] }}
\renewcommand{\t}{\ensuremath{ ^{\scaT} }}
\newcommand{\tria}{\ensuremath{ ^{tr} }}

\newcommand{\iso}[1]{\ensuremath{ \widetilde{#1} }}
\newcommand{\tenFisox}{\ensuremath{ \tenF_{\vecx} } }
\newcommand{\tenFisoX}{\ensuremath{ \tenF_{\vecX} } }
\newcommand{\tenCisox}{\ensuremath{ \tenC_{\vecx} } }
\newcommand{\tenCisoX}{\ensuremath{ \tenC_{\vecX} } }
\newcommand{\Jisox}{\ensuremath{ J_{\vecx} } }
\newcommand{\JisoX}{\ensuremath{ J_{\vecX} } }
\newcommand{\calRiso}{ \ensuremath{ \iso{\calR}}}
\newcommand{\dzeta}{\ensuremath{ \d{\veczeta} } }
\newcommand{\daiso}{\ensuremath{ \d{\iso{a}} }}
\newcommand{\dviso}{\ensuremath{ \d{\iso{v}} }}
\newcommand{\vecniso}{\ensuremath{ \iso{\vecn} } }

\newcommand{\dgn}{\delta \bar{g}_{\mbox{\tiny{N}}}}
\newcommand{\lgn}{\Delta \bar{g}_{\mbox{\tiny{N}}}}
\newcommand{\ldgn}{\Delta\delta \bar{g}_{\mbox{\tiny{N}}}}

\newcommand{\dbxi}{\delta \bar{\xi}}
\newcommand{\lbxi}{\Delta \bar{\xi}}
\newcommand{\ldbxi}{\Delta\delta \bar{\xi}}

\newcommand{\intbco}{\int\limits_{\varphi\left(B^1\right)}}
\newcommand{\intbct}{\int\limits_{\varphi\left(B^2\right)}}
\newcommand{\intbcc}{\int\limits_{\varphi\left(B_c\right)}}
\newcommand{\intpbco}{\int\limits_{\varphi\left(\partial B^1\right)}}
\newcommand{\intpbct}{\int\limits_{\varphi\left(\partial B^2\right)}}
\newcommand{\intpbcon}{\int\limits_{\varphi\left(\partial B^1_n\right)}}
\newcommand{\intpbctn}{\int\limits_{\varphi\left(\partial B^2_n\right)}}
\newcommand{\intpbcc}{\int\limits_{\Gamma_c}}
\newcommand{\intbro}{\int\limits_{B^1}}
\newcommand{\intbrt}{\int\limits_{B^2}}
\newcommand{\intbrc}{\int\limits_{B_c}}

\newcommand{\intprco}{\int\limits_{\partial B^1}}
\newcommand{\intprct}{\int\limits_{\partial B^2}}
\newcommand{\intprcon}{\int\limits_{\partial B^1_n}}
\newcommand{\intprctn}{\int\limits_{\partial B^2_n}}
\newcommand{\intprcc}{\int\limits_{\partial B_c}}

\newcommand{\mnab}{n_{AB}}
\newcommand{\mnac}{n_{AC}}
\newcommand{\mnad}{n_{AD}}
\newcommand{\mnae}{n_{AE}}

\newcommand{\hthr}{\frac{h}{3}}
\newcommand{\hsix}{\frac{h}{6}}


\newcommand{\pna}{_{p\,n+1}}
\newcommand{\pno}{_{p\,n}}
\newcommand{\Ina}{_{I\,n+1}}
\newcommand{\Ino}{_{I\,n}}
\newcommand{\Jna}{_{J\,n+1}}
\newcommand{\Jno}{_{J\,n}}
\newcommand{\sal}{^{1\,\alpha}}
\newcommand{\sbl}{^{1\,\beta}}
\newcommand{\mal}{^{2\,\alpha}}
\newcommand{\xisg}{\left(\xi^1_{g\,n+1}\right)}
\newcommand{\ximg}{\left(\xi^2_{g\,n+1}\right)}
\newcommand{\xisp}{\left(\xi^1_{p\,n+1}\right)}
\newcommand{\ximp}{\left(\xi^2_{p\,n+1}\right)}
\newcommand{\xisgo}{\left(\xi^1_{g\,n}\right)}
\newcommand{\xiso}{\left(\xi^1_{p\,n}\right)}
\newcommand{\ximo}{\left(\xi^2_{p\,n}\right)}
\newcommand{\xise}{\left(\xi^1_g\left(\eta\right)\right)}
\newcommand{\xime}{\left(\xi^2_g\left(\eta\right)\right)}
\newcommand{\xipe}{\left(\xi^1_p\left(\eta\right)\right)}
\newcommand{\xisa}{\left(\xi^1_A\right)}
\newcommand{\xima}{\left(\xi^2_A\right)}
\newcommand{\xisq}{\left(\xi^1_Q\right)}
\newcommand{\ximq}{\left(\xi^2_Q\right)}

\newcommand{\etag}{\left(\eta\right)}
\newcommand{\xig}{\left(\xi_g\right)}

\newcommand{\xisf}{\left(\xi^1_1\right)}
\newcommand{\ximf}{\left(\xi^2_1\right)}
\newcommand{\xiss}{\left(\xi^1_2\right)}
\newcommand{\xims}{\left(\xi^2_2\right)}

\newcommand{\ana}{\ensuremath{ _{\alpha\,n+1} }}
\newcommand{\bna}{\ensuremath{ _{\beta\,n+1} }}

\newcommand{\sna}{\ensuremath{ _{s\,n+1} }}
\newcommand{\nna}{\ensuremath{ _{\mbox{\tiny{N}}\,n+1} }}
\newcommand{\tna}{\ensuremath{ _{\mbox{\tiny{T}}\,n+1} }}
\newcommand{\tnaa}{\ensuremath{_{\mbox{\tiny{T}}\alpha\,n+1\,}}}
\newcommand{\tnab}{\ensuremath{_{\mbox{\tiny{T}}\,n+1\,\beta}}}
\newcommand{\tnax}{\ensuremath{_{\mbox{\tiny{T}}\,n+1\,\xi}}}

\newcommand{\nA}{\ensuremath{ _{\mbox{\tiny{N}} A} }}
\newcommand{\Ng}{\ensuremath{ _{\mbox{\tiny{N}} g} }}
\newcommand{\tA}{\ensuremath{ _{\mbox{\tiny{T}} A} }}
\newcommand{\ta}{\ensuremath{ _{\mbox{\tiny{T}} \alpha} }}
\newcommand{\Tag}{\ensuremath{ _{\mbox{\tiny{T}} \alpha g} }}
\newcommand{\taA}{\ensuremath{ _{\mbox{\tiny{T}}\alpha A} }}
\newcommand{\tbA}{\ensuremath{ _{\mbox{\tiny{T}}\beta A} }}

\newcommand{\tana}{\ensuremath{ _{\mbox{\tiny{T}} A\,n+1} }}

\newcommand{\taana}{\ensuremath{ _{\mbox{\tiny{T}}\alpha A\,n+1} }}
\newcommand{\tbana}{\ensuremath{ _{\mbox{\tiny{T}}\beta A\,n+1} }}

\newcommand{\taano}{\ensuremath{ _{\mbox{\tiny{T}}\alpha A\,n} }}

\newcommand{\iana}{\ensuremath{ _{i A\,n+1} }}
\newcommand{\nana}{\ensuremath{ _{\mbox{\tiny{N}} A\,n+1} }}
\newcommand{\nano}{\ensuremath{ _{\mbox{\tiny{N}} A\,n} }}
\newcommand{\dana}{\ensuremath{ _{\mbox{\tiny{D}} A\,n+1} }}
\newcommand{\tno}{\ensuremath{ _{\mbox{\tiny{T}}\,n} }}
\newcommand{\tano}{\ensuremath{ _{\mbox{\tiny{T}}A\,n} }}
\newcommand{\na}{\ensuremath{ _{n+1} }}
\newcommand{\ina}{\ensuremath{ _{i\,n+1} }}
\newcommand{\jna}{\ensuremath{ _{j\,n+1} }}

\newcommand{\gna}{\ensuremath{ _{g\,n+1} }}
\newcommand{\qna}{\ensuremath{ _{q\,n+1} }}
\newcommand{\gno}{\ensuremath{ _{g\,n} }}

\newcommand{\Ana}{\ensuremath{ _{A\,n+1} }}
\newcommand{\Bna}{\ensuremath{ _{B\,n+1} }}
\newcommand{\Cna}{\ensuremath{ _{C\,n+1} }}
\newcommand{\Ano}{\ensuremath{ _{A\,n} }}
\newcommand{\Bno}{\ensuremath{ _{B\,n} }}
\newcommand{\Cno}{\ensuremath{ _{C\,n} }}

\newcommand{\nBna}{\ensuremath{ _{\mbox{\tiny{N}}B\,n+1} }}
\newcommand{\tBna}{\ensuremath{ _{\mbox{\tiny{T}}B\,n+1} }}

\newcommand{\xina}{\left(\vecxi_{n+1}\right)}
\newcommand{\xino}{\left(\vecxi_{n}\right)}
\newcommand{\xigna}{\left(\vecxi_{g\,n+1}\right)}
\newcommand{\xigno}{\left(\vecxi_{g\,n}\right)}

\newcommand{\agna}{\ensuremath{ _{\alpha\,g\,n+1} }}
\newcommand{\kagna}{\ensuremath{ _{,\alpha\,g\,n+1} }}
\newcommand{\agno}{\ensuremath{ _{\alpha\,g\,n} }}
\newcommand{\kagno}{\ensuremath{ _{,\alpha\,g\,n} }}

\newcommand{\ogna}{\ensuremath{ _{1\,g\,n+1} }}
\newcommand{\kogna}{\ensuremath{ _{,1\,g\,n+1} }}
\newcommand{\ogno}{\ensuremath{ _{1\,g\,n} }}
\newcommand{\kogno}{\ensuremath{ _{,1\,g\,n} }}

\newcommand{\tgna}{\ensuremath{ _{2\,g\,n+1} }}
\newcommand{\ktgna}{\ensuremath{ _{,2\,g\,n+1} }}
\newcommand{\tgno}{\ensuremath{ _{2\,g\,n} }}
\newcommand{\ktgno}{\ensuremath{ _{,2\,g\,n} }}

\newcommand{\aAna}{\ensuremath{ _{\alpha\,A\,n+1} }}
\newcommand{\bAna}{\ensuremath{ _{\beta\,A\,n+1} }}
\newcommand{\aBna}{\ensuremath{ _{\alpha\,B\,n+1} }}

\newcommand{\nnAna}{\ensuremath{ _{33\,A\,n+1} }}
\newcommand{\aaAna}{\ensuremath{ _{\alpha\alpha\,A\,n+1} }}
\newcommand{\abAna}{\ensuremath{ _{12\,A\,n+1} }}
\newcommand{\anAna}{\ensuremath{ _{\alpha3\,A\,n+1} }}

\newcommand{\iAna}{\ensuremath{ _{i\,A\,n+1} }}
\newcommand{\jAna}{\ensuremath{ _{j\,A\,n+1} }}
\newcommand{\kAna}{\ensuremath{ _{k\,A\,n+1} }}
\newcommand{\kBna}{\ensuremath{ _{k\,B\,n+1} }}

\newcommand{\dthetc}{\delta\bar{\vartheta}^1}
\newcommand{\dthets}{\delta\vartheta^2}
\newcommand{\dphic}{\delta\bar{\varphi}^1}
\newcommand{\dphis}{\delta\varphi^2}

\newcommand{\Dthetc}{\Delta\bar{\vartheta}^1}
\newcommand{\Dthets}{\Delta\vartheta^2}
\newcommand{\Dphic}{\Delta\bar{\varphi}^1}
\newcommand{\Dphis}{\Delta\varphi^2}

\newcommand{\dxi}{\delta\bar{\xi}}
\newcommand{\dxia}{\delta\bar{\xi}^{\alpha}}
\newcommand{\dxib}{\delta\bar{\xi}^{\beta}}
\newcommand{\dxic}{\delta\bar{\xi}^{\gamma}}
\newcommand{\dxid}{\delta\bar{\xi}^{\delta}}
\newcommand{\dxie}{\delta\bar{\xi}^{\epsilon}}
\newcommand{\dxif}{\delta\bar{\xi}^{\eta}}

\newcommand{\Dxi}{\Delta\bar{\xi}}
\newcommand{\Dxia}{\Delta\bar{\xi}^{\alpha}}
\newcommand{\Dxib}{\Delta\bar{\xi}^{\beta}}
\newcommand{\Dxic}{\Delta\bar{\xi}^{\gamma}}
\newcommand{\Dxid}{\Delta\bar{\xi}^{\delta}}
\newcommand{\Dxie}{\Delta\bar{\xi}^{\epsilon}}
\newcommand{\Dxif}{\Delta\bar{\xi}^{\eta}}

\newcommand{\Ddxi}{\Delta\delta\bar{\xi}}
\newcommand{\Ddxia}{\Delta\delta\bar{\xi}^{\alpha}}
\newcommand{\Ddxib}{\Delta\delta\bar{\xi}^{\beta}}
\newcommand{\Ddxic}{\Delta\delta\bar{\xi}^{\gamma}}
\newcommand{\Ddxid}{\Delta\delta\bar{\xi}^{\delta}}
\newcommand{\Ddxie}{\Delta\delta\bar{\xi}^{\epsilon}}
\newcommand{\Ddxif}{\Delta\delta\bar{\xi}^{\eta}}

\newcommand{\dvecaa}{\delta\bar{\veca}_{\alpha}^1}
\newcommand{\dvecab}{\delta\bar{\veca}_{\beta}^1}

\newcommand{\dvecaat}{\delta\bar{\veca}^{1\alpha}}
\newcommand{\dvecabt}{\delta\bar{\veca}^{1\beta}}

\newcommand{\Dvecaa}{\Delta\bar{\veca}_{\alpha}^1}
\newcommand{\Dvecab}{\Delta\bar{\veca}_{\beta}^1}
\newcommand{\Dvecac}{\Delta\bar{\veca}_{\gamma}^1}

\newcommand{\Dvecaat}{\Delta\bar{\veca}^{1\alpha}}
\newcommand{\Dvecabt}{\Delta\bar{\veca}^{1\beta}}

\newcommand{\dvecn}{\delta\bar{\vecn}^1}
\newcommand{\Dvecn}{\Delta\bar{\vecn}^1}
\newcommand{\Ddvecn}{\Delta\delta\bar{\vecn}^1}

\newcommand{\dvecua}{\delta\bar{\vecu}_{,\alpha}^2}
\newcommand{\dvecub}{\delta\bar{\vecu}_{,\beta}^2}
\newcommand{\dvecuc}{\delta\bar{\vecu}_{,\gamma}^2}
\newcommand{\dvecud}{\delta\bar{\vecu}_{,\vartheta}^2}
\newcommand{\dvecue}{\delta\bar{\vecu}_{,\theta}^2}

\newcommand{\Dvecua}{\Delta\bar{\vecu}_{,\alpha}^2}
\newcommand{\Dvecub}{\Delta\bar{\vecu}_{,\beta}^2}
\newcommand{\Dvecuc}{\Delta\bar{\vecu}_{,\gamma}^2}
\newcommand{\Dvecud}{\Delta\bar{\vecu}_{,\vartheta}^2}
\newcommand{\Dvecue}{\Delta\bar{\vecu}_{,\theta}^2}

\newcommand{\dvecuab}{\delta\bar{\vecu}_{,\alpha\beta}^2}
\newcommand{\dvecuac}{\delta\bar{\vecu}_{,\alpha\gamma}^2}
\newcommand{\dvecubc}{\delta\bar{\vecu}_{,\beta\gamma}^2}
\newcommand{\dvecuad}{\delta\bar{\vecu}_{,\alpha\vartheta}^2}
\newcommand{\dvecuae}{\delta\bar{\vecu}_{,\alpha\theta}^2}

\newcommand{\Dvecuab}{\Delta\bar{\vecu}_{,\alpha\beta}^2}
\newcommand{\Dvecuac}{\Delta\bar{\vecu}_{,\alpha\gamma}^2}
\newcommand{\Dvecubc}{\Delta\bar{\vecu}_{,\beta\gamma}^2}
\newcommand{\Dvecuad}{\Delta\bar{\vecu}_{,\alpha\vartheta}^2}
\newcommand{\Dvecuae}{\Delta\bar{\vecu}_{,\alpha\theta}^2}

\newcommand{\dvecus}{\delta\vecu^1}
\newcommand{\Dvecus}{\Delta\vecu^1}
\newcommand{\vecxa}{\bar{\vecx}_{,\alpha}^2}
\newcommand{\vecxb}{\bar{\vecx}_{,\beta}^2}
\newcommand{\vecxab}{\bar{\vecx}_{,\alpha\beta}^2}

\newcommand{\gthet}{g_{\vartheta}}
\newcommand{\dgthet}{\delta g_{\vartheta}}
\newcommand{\Dgthet}{\Delta g_{\vartheta}}

\newcommand{\thetg}{\theta_G}
\newcommand{\dthetg}{\delta\theta_G}
\newcommand{\Dthetg}{\Delta\theta_G}

\newcommand{\gphi}{g_{\varphi}}
\newcommand{\dgphi}{\delta g_{\varphi}}
\newcommand{\Dgphi}{\Delta g_{\varphi}}

\newcommand{\tta}{t_{T\alpha}}
\newcommand{\ttb}{t_{T\beta}}
\newcommand{\ttc}{t_{T\gamma}}

\newcommand{\ttx}{t_{T\xi}}

\newcommand{\ttat}{t_{T}^{\alpha}}

\newcommand{\Dtta}{\Delta t_{T\alpha}}
\newcommand{\Dttb}{\Delta t_{T\beta}}
\newcommand{\Dttc}{\Delta t_{T\gamma}}

\newcommand{\ttra}{t_{t\alpha}^{tr}}
\newcommand{\ttrb}{t_{t\beta}^{tr}}
\newcommand{\ttrc}{t_{t\gamma}^{tr}}

\newcommand{\ttrat}{t_{T}^{trial\alpha}}
\newcommand{\ttrbt}{t_{T}^{trial\beta}}
\newcommand{\ttrct}{t_{T}^{trial\gamma}}
\newcommand{\ttrdt}{t_{T}^{trial\vartheta}}

\newcommand{\Dttra}{\Delta t_{T\alpha}^{trial}}
\newcommand{\Dttrb}{\Delta t_{T\beta}^{trial}}
\newcommand{\Dttrc}{\Delta t_{T\gamma}^{trial}}

\begin{Frontmatter}

\title[Article Title]{Gaussian Processes enabled model calibration in the context of deep geological disposal}

\author[1]{Lennart Paul}
\author[2]{Jorge-Humberto Urrea-Quintero}
\author[1]{Umer Fiaz}
\author[4]{Ali Hussein}
\author[3]{Hazem Yaghi}
\author[1]{Joachim Stahlmann}
\author[3]{Ulrich Römer}
\author[2]{Henning Wessels}

\address[1]{\orgdiv{Institute of Geomechanics and Geotechnical Engineering}, \orgname{Technische Universität Braunschweig}, \orgaddress{\city{Braunschweig}, \postcode{38106}, \country{Germany}} \email{lennart.paul@tu-braunschweig.de}}

\address[2]{\orgdiv{Institute of Applied Mechanics, Division Data-Driven Modeling of Mechanical Systems}, \orgname{Technische Universität Braunschweig}, \orgaddress{\city{Braunschweig}, \postcode{38106}, \country{Germany}}}

\address[3]{\orgdiv{Institute for Acoustics and Dynamics}, \orgname{Technische Universität Braunschweig}, \orgaddress{\city{Braunschweig}, \postcode{38106}, \country{Germany}}}

\address[4]{\orgname{Bundesgesellschaft für Endlagerung mbH (BGE)}, \orgaddress{\city{Peine}, \postcode{31224}, \country{Germany}}}

\authormark{Paul et al.}





\abstract{\Revision{Deep geological repositories are critical for the long-term storage of hazardous materials, where understanding the mechanical behavior of emplacement drifts is essential for safety assurance. This study presents a surrogate modeling approach for the mechanical response of emplacement drifts in rock salt formations, utilizing Gaussian Processes (GPs). The surrogate model serves as an efficient substitute for high-fidelity mechanical simulations in many-query scenarios, including time-dependent sensitivity analyses and calibration tasks. By significantly reducing computational demands, this approach facilitates faster design iterations and enhances the interpretation of monitoring data.
The findings indicate that only a few key parameters are sufficient to accurately reflect in-situ conditions in complex rock salt models. Identifying these parameters is crucial for ensuring the reliability and safety of deep geological disposal systems.}}

\end{Frontmatter}

\section*{Impact Statement}
This study provides key contributions in processing real-world geomechanical monitoring data from deep geological cavities, applying GP-based global sensitivity analysis using time-dependent Sobol' indices, and achieving efficient calibration of geomechanical models. The structured approach demonstrates that only a few critical material parameters need calibration to accurately reflect in-situ monitoring data within complex constitutive models of rock salt. This finding is particularly significant for safety-critical applications such as deep geological disposal, where precise modeling is essential for long-term safety and stability. The results emphasize the efficiency and accuracy of the GP-based surrogate model in simplifying the calibration process while maintaining high fidelity to real-world conditions.

\section{Introduction}\label{sec:intro}

Deep repository material models are complex geological models that account for the mechanics of soil and rock, hydrological properties, thermal effects, and chemical interactions \cite{Pitz2023BenchmarkingTH2M, Claret2024eurad}. They are highly parametrized and the associated numerical analyses are computationally expensive \cite{Wojnarowicz2024OptRepo, Kurgyis2024UncertaintiesRepo}. Their complexity has prevented their widespread adoption for many-query tasks, that is, simulations to explore multiple scenarios and the effects of uncertainties on the deep repository prognosis \cite{Kurgyis2024UncertaintiesRepo}.
In this regard, we propose using Gaussian Processes (GPs) as surrogates for high-fidelity geological models. GPs can approximate the outputs of complex simulations with much lower computational cost, enabling efficient calibration, design and validation \cite{Myren2021comparisonGPsNNs,Radaideh2020DeepGPs, Sung2024RevModelCalibration}. Speeding up the simulations with the use of a GP-based surrogate will not only allow for validated prognoses for the repository through the assimilation of data but also help in making informed decisions and enhancing the overall robustness and adaptability of the repository management process.

\subsection{Constitutive models for deep geological repositories}
Rock salt formations are one of the potential host rocks in Germany considered for secure long-term nuclear waste storage due to their distinctive mechanical and hydraulic properties \citep{StandAG2017, Bollingfehr2017DesignSalt}. Ensuring the natural integrity of geological barriers is crucial for safety, therefore reliable numerical calculations that depend on advanced material models are essential \RevisionNew{\cite{Bollingfehr2017DesignSalt}}. The thermo-mechanical behavior of rock salt is typically assessed by means of laboratory tests, in particular, short-term triaxial compression tests and long-term creep tests \citep{Langer1985, Wittke2014, Fecker2018}. \RevisionNew{In order to include all the characteristics of rock salt, various constitutive models have been developed to effectively simulate the hydraulic, thermal, and mechanical behavior of rock salt; refer, for example, to \cite{Saltmech2007, ARMA2013, SaltMech2022}}. These constitutive models are able to capture for instance multiple creep phases, healing, and dependency on, e.g., stress condition, time, temperature, and humidity. Several different approaches to capturing the complex behavior of rock salt were compared and briefly described in past research projects \citep{Vergleich2010, Vergleich2016, Vergleich2022}. Among them is the constitutive model \textit{TUBSsalt}, which was developed by the Institute for Geomechanics and Geotechnics (IGG, TU Braunschweig) and presented for the first time in \cite{TUBSsalt2015}. It has been shown in \cite{SaltMech2022}, \cite{Vergleich2016} and \cite{Vergleich2022} that \textit{TUBSsalt}, as well as other constitutive models, accurately capture the thermal and mechanical characteristics of rock salt. The application of such complex constitutive models requires expert knowledge and ideally data to infer the numerous model parameters. Therefore, this constitutive model is the focus of the presented calibration process.

\subsection{Gaussian processes as efficient surrogates for computationally expensive models}
GPs are non-parametric probabilistic models that use Bayesian inference to make predictions and learn from data. They are particularly useful for modeling complex input-output data relationships and for making predictions in situations where the data is noisy or incomplete
\citep{Kennedy2000PredOut, Kennedy2001BayesianCal,gu2018scaledGPS,teckentrup2020GPsConvergence,myren2021comparisonGPs_NNs,gramacy2020surrogates}. 
One of the key advantages of GPs over concurrent regression ansatzes, such as neural networks, is their ability to effectively handle uncertainty \citep{schulz2018GPstutorial} and their applicability in a low-data regime. GPs can generate probabilistic predictions of the model output, considering the uncertainty in both the input data and the model itself. 
This makes GPs a good alternative for calibrating complex and non-linear computational models (see, e.g., \cite{wu2018invUQGPs,mahdaviara2021PermeabilityGPs,li2023GPsDams,veasna2023GPsPlasticity} and \cite{Sung2024RevModelCalibration} for a recent review on computer model calibration). Additionally, global sensitivity analysis for computationally expensive models, which assesses the influence of input parameters on model output, becomes computationally feasible with GP-based surrogates \citep{Oakley2004ProbSA, marrel2009calculations, Srivastava2017SA}. 
GPs have been applied to sophisticated computer models in various fields, including nuclear physics \citep{Kejzlar2020NuclearPhysics}, environmental science \citep{Cheng2021intersatellite}, and digital twining \citep{Thelen2022DigitalTwinI, Thelen2023DigitalTwinII}.

In \cite{sacks1989designs}, the concept of using GPs as surrogates to predict model outputs at untried locations within the parameter space was introduced. The use of GPs for calibrating computer models was then pioneered by \cite{Kennedy2000PredOut} and \cite{Kennedy2001BayesianCal}. Subsequent work by \cite{Bayarri2007CompModels} and \cite{Higdon2004DataCompModels} refined this framework, with Higdon adopting a fully Bayesian approach. Extending these methodologies to multivariate data \citep{Higdon2008HDOut}, and further advancements by others, such as \cite{Bayarri2007FuncOut}, have enhanced the calibration of models with multivariate outputs. \cite{Santner2003ExpDesign}, and additional studies, including those by \cite{Fang2005CompExpDesign}, \cite{loeppky2009SampleSize}, and \cite{Baker2022StoCompModelsg}, provide in-depth discussions on the optimal design of computer experiments, focusing on the strategic layout of model evaluations in parameter space.
Beyond calibration, GP emulators facilitate understanding variability in model outputs when parameters are uncertain, known as uncertainty analysis. The tutorial by \cite{OHagan2006BayesianCompCode} offers an accessible introduction to uncertainty analysis using GP emulators. 

\vspace{0.5cm}

\Revision{Unlike much of the existing literature on surrogate modeling for complex mechanical systems, which often relies on synthetic data, our study goes beyond the implementation of a surrogate modeling approach and exploits it to enable the calibration of the \textit{TUBSsalt} constitutive model using $14$ yr of real-world monitoring data collected in an open drift located in the northern main drift of Gorleben, Germany. The primary contribution of this paper is to demonstrate the applicability and effectiveness of a GP-based data-driven methodology in addressing a real-world problem. Specifically:}
\begin{itemize}
    \item We develop a GP-based surrogate model that approximates the deformation behavior of a drift in rock salt formations and verify its accuracy in closely replicating the high-fidelity model's behavior.
    \item We perform a time-dependent sensitivity analysis using Sobol' indices, utilizing the surrogate model to extract clear insights from the complex geomechanical constitutive model.
    \item We calibrate the model parameters using real in-situ monitoring data from the Gorleben site, benefiting from the efficiency of the GP-based surrogate model.
\end{itemize}

The remainder of this article is structured as follows: \autoref{sec:drift_model} introduces the mechanical aspects of modeling a deep geological drift in a rock salt formation. The formulation of the calibration process from experimental and monitoring data as well as the global sensitivity analysis is presented in \autoref{sec:calibration}. GPs as surrogate models are described in \autoref{sec:GPs}. \autoref{sec:results} contains the results of the training and validation of as well as calibration with the surrogate model. The paper concludes with a summary and outlook in \autoref{sec:conclusion}.

\section{Mechanical modeling and numerical solution of a drift in the deep geological formation of rock salt}\label{sec:drift_model}

This section presents the geomechanical model for a drift located on the north-western flank of the Gorleben salt dome in Germany. The latter has been explored with regard to its suitability as a location for a repository for high-level radioactive waste for decades.
First, the drift location is described, and the assumptions for the computational model are outlined. Then, the kinematics, governing equations, and the constitutive model \textit{TUBSsalt} for rock salt are presented. Finally, we outline the numerical solution of the boundary value problem. 

\subsection{Detailed site description and problem geometry}\label{sec:model_assumptions}

The mechanical model considered in this work is based on the cross-section of a drift located on the north-western flank of the Gorleben salt dome, a former salt exploration mine in Germany. Monitoring data were collected and provided by the German Federal Company for Radioactive Waste Disposal (BGE mbH).



The considered measurement cross-section is located in the northern main drift of Gorleben with a depth of $840$ m below the top edge of the ground. Excavation in the area of the measurement location was finished on  October 19, 1999 without a recut being carried out afterwards. The monitoring data consists of time series of convergence measurements, which indicate the change in distance between opposing fixed points inside the rock, from which the deformation rate of the drift contour is derived. At each measuring location, the horizontal and vertical distances are recorded periodically as a standard procedure. Consequently, the computational model adopted in this work represents a similar open emplacement drift of a deep geological repository based on the location where the monitoring data were obtained. Since this phase does not involve storing highly radioactive and heat-generating waste, the problem considered is purely mechanical. 

Depending on crystallinity and the presence of secondary aggregates, the rock salt can be divided into homogeneous areas based on their viscous behavior, specifically the steady-state (secondary) creep rate \citep{Hunsche2003Gorleben}. \RevisionNew{A vertical geological cross-section of the Gorleben salt dome and its homogeneous salt areas can be found in \cite{BGR2008Gorleben}}. Since the measurement location is homogeneously surrounded by a salt formation known as \textit{Streifensalz} \textit{z2HS2}, interactions between different homogeneous areas are not taken into account. Therefore, the entire numerical model can be constructed under the assumption of a uniform rock salt material.

\begin{figure}[h]
    \centering
    \includegraphics[width=0.90\textwidth]{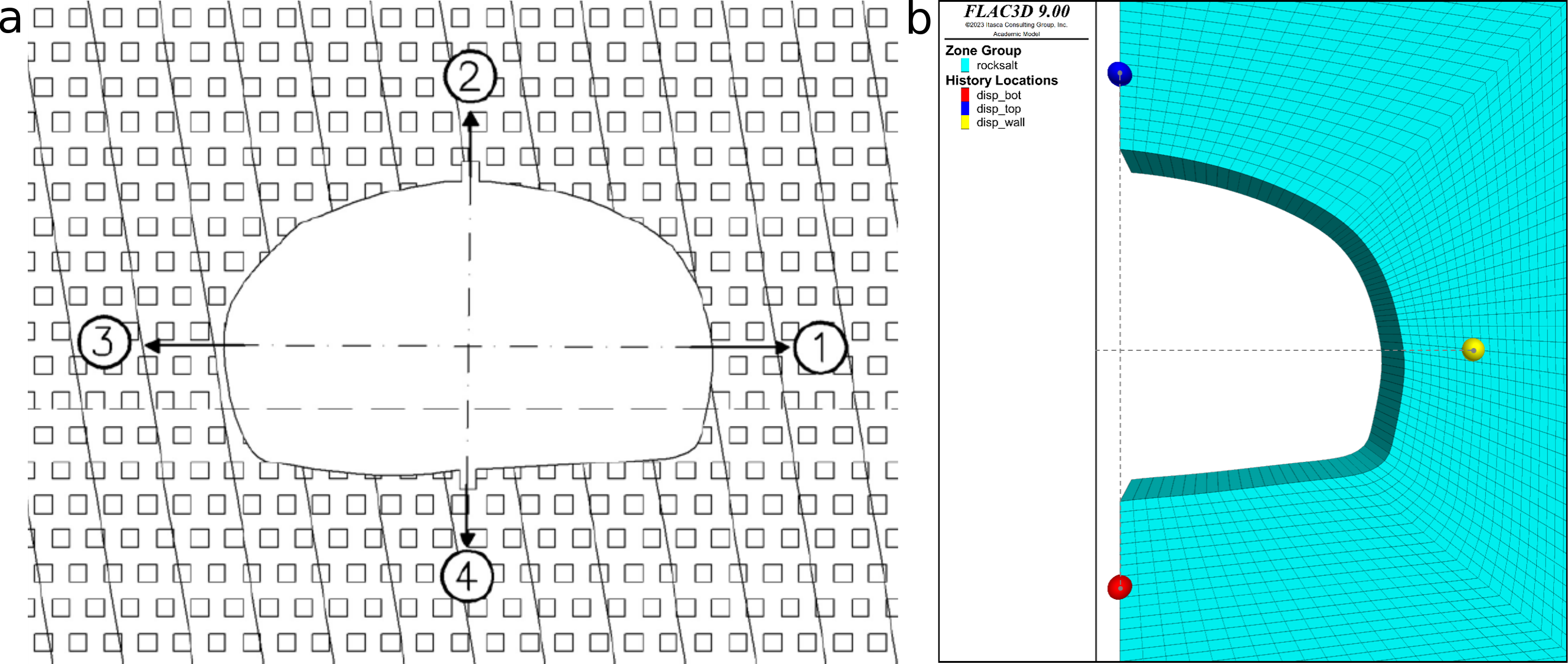}
    \caption{\textbf{Cross-sectional representation of the drift area.} \textbf{a.} Cross-section of the measurement locations from the Gorleben site, \Revision{provided by the BGE mbH.}
    \textbf{b.} Computational model of the drift in FLAC3D, showing the history locations used for evaluating displacements and the mesh used in the numerical solution.}
    \label{fig:high_fidelity_model_simulation}
\end{figure}

\autoref{fig:high_fidelity_model_simulation}\textbf{a} depicts the cross-section of the measurement location together with the vertical (2-4) and horizontal (1-3) measurement distances. 
\autoref{fig:high_fidelity_model_simulation}\textbf{b} corresponds to the computational model of the drift in FLAC3D with history locations for the evaluation of the displacements.

\subsection{Continuum mechanics}\label{sec:mechanics}

Modeling in continuum mechanics involves three ingredients: \textbf{i.} balance equations, \textbf{ii.} kinematics and \textbf{iii.} a constitutive model.
This section introduces these three ingredients of continuum mechanics for the deep repository under the assumptions presented in the previous Section. 
\Revision{We follow the notation for general continuum mechanical equations introduced by \cite{Anand2020CMM_solids_Book, FLAC3D2023}. The constitutive equations for \textit{TUBSsalt} are presented as they were introduced in \cite{TUBSsalt2015, Vergleich2022TUBS}.}

\subsubsection{Balance equations and kinematics}\label{sec:kinematics}

In the current framework, we are focused only on the changes in mechanical quantities due to the excavation of the drift, while keeping the temperature constant, as there is no significant temperature development during the initial phase of the repository operation. Therefore, only the balance of linear momentum needs to be considered, which in its strong form and current configuration is given by
\begin{equation}\label{eq:momentum}
\operatorname{div} \tensig + \rho \, \tenb = \rho \, \frac{\mbox{d} \vecv}{\mbox{d}t} \text{ in } \Omega,
\end{equation}
where $\Omega$ denotes the computational domain. Further, $\tensig$ denotes the Cauchy stress, $\rho$ the density, $\tenb$ acceleration caused by external body forces (here: gravitation), $\vecv$ the velocity vector and $\mbox{d}\vecv/\mbox{d}t$ the acceleration.
Dirichlet and Neumann-type boundary conditions are defined at boundaries ~$\Gamma_D$ and $\Gamma_N$, respectively, as
\begin{equation}
\begin{aligned}
    \vecv &= \bar{\vecv} \text{ on } \Gamma_D \\
    \tensig \cdot \tenn &= \bar{\tent}  \text{ on } \Gamma_N, 
\end{aligned}
\end{equation}
where $\bar{\vecv}$ and $\bar{\tent}$ denote a prescribed velocity and traction and $\vecn$ the surface normal vector. Due to the distinct creep mechanisms of rock salt, a time-dependent problem needs to be considered. The mechanical initial conditions are defined as
\begin{equation}
\begin{aligned}
    \vecv (t=0) &= \vecv_0 \\
    \tensig (t=0) &= \tensig_0.
\end{aligned}
\end{equation}
The boundary value problem is complemented by the constitutive model \textit{TUBSsalt}
and will be solved using the commercial software FLAC3D \citep{FLAC3D2023}.

Only small deformations are considered in this work since no long-term analysis of the rock salt emplacement will be performed. Therefore, the kinematics are defined by
\begin{equation}\label{eq:kinematics}
    \dot{\teneps} = \frac{1}{2} \, \left( \operatorname{grad}\vecv + \operatorname{grad}\vecv^{T} \right),
\end{equation}
where $\dot{\teneps}$ is the strain rate tensor derived from the velocity vector $\vecv$.

\subsubsection{Constitutive model \textit{TUBSsalt}}\label{sec:TUBSsalt}

\begin{figure}[h]
    \centering
    \includegraphics[width=0.9\textwidth]{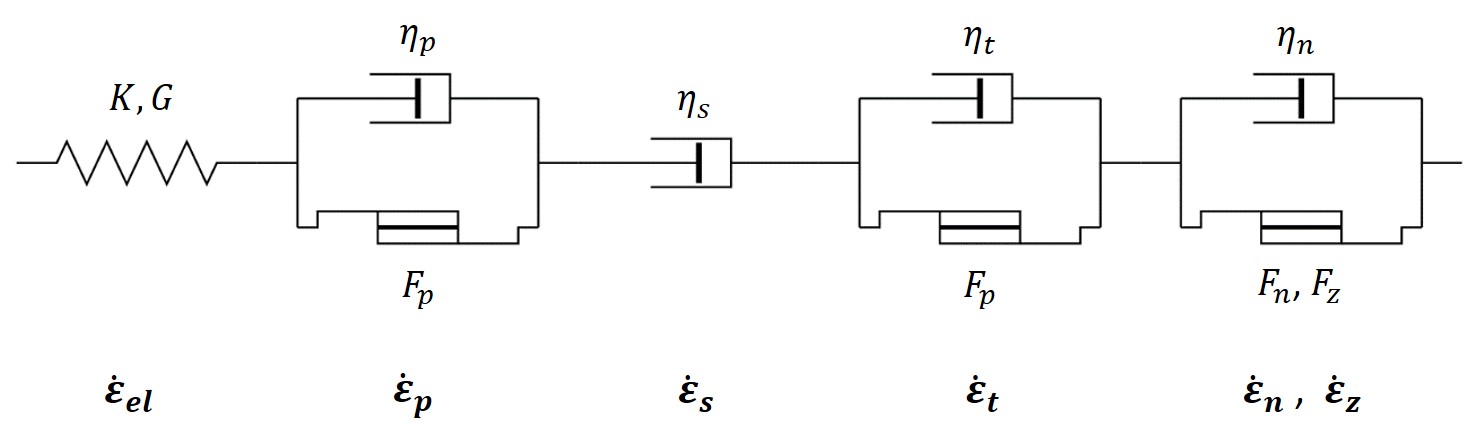}
    \caption{\textbf{Rheological model} \Revision{and corresponding strain components of the constitutive model \textit{TUBSsalt} for rock salt. While  $\dot{\teneps}_{el}$ is represented by a spring, $\dot{\teneps}_{p}$, $\dot{\teneps}_{s}$, $\dot{\teneps}_{t}$, $\dot{\teneps}_{n}$ as well as $\dot{\teneps}_{z}$ are modelled by hardening or softening sliders and viscous dampers.}}
    \label{fig:TUBSsalt}
\end{figure}

In this subsection, the constitutive model \textit{TUBSsalt} is briefly introduced. This model is based on the rheological model shown in \autoref{fig:TUBSsalt} and is proficient in characterizing various thermo-mechanical aspects of rock salt such as primary, secondary, and tertiary creep, recovery creep, shear, creep and tension failure, healing and the influence of temperature \citep{TUBSsalt2015, Vergleich2022TUBS}. 

Based on the rheological model, the total strain rate $\dot{\teneps}$ of the \textit{TUBSsalt} constitutive model is split additively into six components as 
\begin{equation}\label{eq:full_strain}
    \dot{\teneps} = \dot{\teneps}_{el} + \underbrace{\dot{\teneps}_p + \dot{\teneps}_s + \dot{\teneps}_t + \dot{\teneps}_n + \dot{\teneps}_z }_{\dot{\teneps}_{vp}}.
\end{equation}
Here, $\dot{\teneps}_{el}$ is the elastic strain rate and $\dot{\teneps}_{vp}$ the inelastic, visco-plastic strain rate. The latter consists of the primary creep rate $\dot{\teneps}_p$, the secondary creep rate $\dot{\teneps}_s$, the tertiary creep and healing rate $\dot{\teneps}_t$, the strain rate from creep and shear failure $\dot{\teneps}_n$ and the strain rate from tension failure $\dot{\teneps}_z$.
In \autoref{fig:TUBSsalt}, the elastic strain rate $\dot{\teneps}_{el}$ is represented by a spring, and the creep strain rates $\dot{\teneps}_p$, $\dot{\teneps}_s$ and $\dot{\teneps}_t$ as well as the failure strain rates $\dot{\teneps}_n$ and $\dot{\teneps}_z$ \Revision{are described by hardening or softening sliders with yield functions $F$ and viscous dampers, each characterised by an individual viscosity $\eta$}. 
For a visco-elasto-plastic material model, the stress increment depends on the elastic strain component and can be written as
\begin{equation}\label{eq:constitutive}
    \dot{\tensig} = \bm{D} \left(\tensig, \, \kappa \right) \cdot \left(\dot{\teneps}-\dot{\teneps}_{vp} \right),
\end{equation}
where $\kappa$ is a loading-history parameter depending on visco-plastic strain rate $\dot{\teneps}_{vp}$. Rock salt is considered to be an isotropic material, and thus $\bm{D}$ is the isotropic stiffness tensor, depending on bulk and shear moduli $K$ and $G$. Both of theses properties decrease as the level of the damage-induced dilatancy $\varepsilon_{v,d}$ \eqref{eq:dilatancy} increases. In the following, the specific forms of the strain components are introduced in more detail. \\

\noindent The \textbf{elastic strain} rate $\dot{\teneps}_{el}$ is calculated according to Hooke's law using the stiffness matrix $\bm{D}$ and the stress rate tensor $\dot{\tensig}$ as
\begin{equation}
    \dot{\teneps}_{el} = \bm{D}^{-1} \cdot \, \dot{\tensig}.
\end{equation}

\noindent \textbf{Primary and recovery creep} only occur after a load change. An increase in the stress deviator causes high primary creep rates, which decrease as the deformation progresses until no more primary deformations occur. There holds 
\begin{equation}
    F_p > 0 : \, \dot{\teneps}_p = \frac{F_p}{\eta_p^{*}} \cdot \frac{\partial\sigma_{eq}}{\partial\tensig}
\end{equation}
where $\dot{\teneps}_p$ represents the primary creep rate tensor, $F_p$ the yield function of primary creep, $\eta_p^{*}$ the current viscosity of primary creep and $\frac{\partial\sigma_{eq}}{\partial\tensig}$ denotes the derivatives of the equivalent stress $\sigma_{eq}$ with respect to the components of the stress tensor.
In case of a decrease of the stress deviator, the viscosity of primary creep $\eta_p^{*}$ is replaced by the viscosity for recovery creep $\eta_{p,rec}$.

\noindent \textbf{Secondary creep} is always active as soon as a stress deviator is present and leads to a constant creep rate for a constant stress state. It is therefore also referred to as stationary creep and defined as
\begin{equation}
     \dot{\teneps}_s = \frac{F_s \cdot q_s}{\eta_s} \cdot \frac{\partial\sigma_{eq}}{\partial \tensig},
\end{equation}
where $\dot{\teneps}_s$ represents the secondary creep rate tensor, $F_s$ the yield function of secondary creep, $\eta_s$ is the viscosity parameter for secondary creep and $q_s$ is the temperature coefficient for secondary creep.

\noindent \textbf{Tertiary creep} initiates as soon as the stress deviator exceeds the dilatancy criteria represented by the yield function $F_t$. Above this dilatancy strength, rock salt shows softening which is described by an accelerated reduction of the viscosity $\eta_t^{*}$ depending on the level of damage-induced dilatancy $\varepsilon_{v,d}$ \eqref{eq:dilatancy}: 
\begin{equation}
    F_t > 0 : \, \dot{\teneps}_t = \frac{F_t \cdot k}{\eta_t^{*} \cdot q_t} \cdot \frac{\partial Q_t}{\partial \tensig},
\end{equation}
where $\dot{\teneps}_t$ represents the tertiary creep rate tensor, $\eta_t^{*}$ is the current viscosity of tertiary creep, $k$ a coefficient for loading rate and stress state, $q_t$ a temperature coefficient and $\frac{\partial Q_t}{\partial\tensig}$ the directional derivatives of the potential function for tertiary creep $Q_t$ with respect to the components of the stress tensor. 
The increase in volume or dilatancy due to microcracks is taken into account in $\frac{\partial Q_t}{\partial\tensig}$.

\noindent The process of \textbf{healing} replaces the tertiary creep once a stress state falls below the dilatancy threshold and damage has already been determined. There holds
\begin{equation}
    F_t > -\sigma_z : \, \dot{\teneps}_t = \frac{F_t \cdot q_v}{\eta_v^{*}} \cdot \frac{\partial Q_v}{\partial \tensig},
\end{equation}
where $\sigma_z$ is the tensile strength, $q_v$ a temperature coefficient, $\eta_v^{*}$ the current viscosity of healing and $\frac{\partial Q_v}{\partial\tensig}$ the directional derivatives of the potential function for healing $Q_v$ with respect to the components of the stress tensor.

\noindent \textbf{Creep and shear failure:} Once the damage induced dilatancy $\varepsilon_{v,d}$ \eqref{eq:dilatancy} exceeds the failure volumetric strain $\varepsilon_{v,d,b,*}$, time-dependent failure deformations occur in addition to the deformations from the creep components:
\begin{equation}
    \varepsilon_{v,d} \geq \varepsilon_{v,d,b*}: \, \dot{\teneps}_n = \frac{F_n \cdot k}{\eta_n^{*} \cdot q_n} \cdot \frac{\partial \sigma_{eq}}{\partial \tensig},
\end{equation}
where $\dot{\teneps}_n$ is the strain rate tensor of the post-failure, $\eta_n^{*}$ the current viscosity parameter for healing and $q_n$ a temperature coefficient. 

\noindent \Revision{\textbf{Tension failure} occurs when tensile material strength $\sigma_{z,0}$ is exceeded and is denoted by the strain rate tensor $\dot{\teneps}_z$.} 

An important characteristic of the constitutive model is the damage-induced dilatancy $\varepsilon_{v,d}$, which is decisive for the evaluation of the geological integrity of the excavation damage zone of rock salt. As soon as the dilatancy strength of the material is exceeded, which is achieved when the yield function $F_t$ becomes greater than $0$, softening takes place as a result of crack formation, which in turn creates pathways for radionuclides. In a first step, the tensor of the damage-induced strain increment $\Delta \teneps_d$ can be determined using a timestep $\Delta t$ as
\begin{equation}\label{eq:damage}
    \Delta \teneps_d = \left( \dot{\teneps}_t + \dot{\teneps}_n \right) \cdot \Delta t.
\end{equation}
The increment of the damage-induced dilatancy $\varepsilon_{v,d}$ is equivalent to the first invariant of $\Delta \teneps_d$:
\begin{equation}\label{eq:dilatancy}
    \Delta \varepsilon_{v,d} = I_1 \left( \Delta \teneps_d \right) = \Delta \varepsilon_{d,xx} + \Delta \varepsilon_{d,yy} + \Delta \varepsilon_{d,zz}.
\end{equation}

For further details on quantities not explained in this section, such as the yield functions $F_p$, $F_t$ and $F_n$,  the potential functions for tertiary creep and healing $Q_t$ and $Q_v$, \Revision{as well as derivations of all equations of \textit{TUBSsalt} provided here}, the reader is referred to \cite{Vergleich2022TUBS}. In summary, the constitutive model \textit{TUBSsalt} comprises a total of $25$ material parameters, summarized in \autoref{appendixB}, \autoref{tab:parameter_config}.

\subsection{Numerical solution}\label{sec:model}

The constitutive model \textit{TUBSsalt} 
is implemented into the commercial software FLAC3D \citep{FLAC3D2023}, which is a program for three-dimensional engineering mechanics computations. FLAC3D relies on the finite difference method (FDM) to translate the continuum equations into ordinary differential equations for each element, which are then solved using an explicit central finite difference approach in time. The spatial discretization in FLAC3D is realized by meshing the continuum into hexahedral elements, with the option to use tetrahedral, wedge, and pyramid elements. It is recommended to use hexahedral elements, which are divided into tetrahedral elements to reduce the volumetric locking effect and thus achieve more accurate results.
The balance of the linear momentum, equation \eqref{eq:momentum}, 
is iterated to an equilibrium state, which is achieved when the unbalanced mechanical force for all the gridpoints in the model is negligibly low. 

Time-dependent phenomena such as creep require a timestep $\Delta t$ to solve the equations of the \textit{TUBSsalt} constitutive model. At the same time, for creep analysis, the state of equilibrium must be maintained, otherwise inertial effects may affect the solution. For this purpose, the unbalanced force is monitored in the model.
The finite volume scheme can be summarised for each time step as follows: After new strain rates are determined from nodal velocities, new stresses are calculated from strain rates and previous stresses using constitutive equations. By applying the balance of the linear momentum, new velocities and deformations are subsequently calculated from stresses and forces. More information on the FLAC3D solution algorithm can be found in \cite{FLAC3D2023}.

\section{Model calibration and sensitivity analysis}\label{sec:calibration}

In this section, we briefly state the calibration process of the mechanical model introduced in \autoref{sec:drift_model}. 
We distinguish between two model calibration stages. The first stage assumes that stress-strain data are available (\autoref{sec:regression}) and the constitutive model can be directly calibrated from common geomechanical laboratory tests. This approach is usually employed as the first calibration step to state some prior knowledge regarding the material parameters. The second stage uses in-situ monitoring data of the drift convergence (\autoref{sec:inverse}). This approach is used for model parameter recalibration and considers the real conditions of the deep repository. As part of the calibration process, a method for global sensitivity analysis is introduced in \autoref{sec:sensitivity_analysis}.

\subsection{Initial model calibration from mechanical testing data}\label{sec:regression}

Commonly, laboratory tests are used for the calibration of constitutive model parameters. In the case of rock salt and as shown in \autoref{tab:parameter_config}, short-term triaxial compression or extension tests, long-term creep tests and healing tests as well as indirect tensile tests need to be conducted under variation of temperature and confining stress to be able to cover all strain parts described in \autoref{sec:TUBSsalt}. While strength tests use a constant axial strain rate to apply a load on the mostly cylindrical specimens with a fixed confining pressure, the specimens in creep and healing tests have a predefined stress state so that the time-dependent behavior can be observed.

Common to the aforementioned mechanical tests is that strain states are considered, which yield well-defined stress states. Hence, stress-strain data can be obtained from the mechanical tests, and the calibration of the constitutive model is a regression problem that can be cast as the optimization problem
\begin{equation}\label{eq:regression}
    \boldsymbol{\theta}^* = \text{arg min}_{\boldsymbol{\theta}} 
    \sum_{i=1}^N \| \tensig^{i} - \tensig(\teneps^{i}; \tentheta) \|^2,
\end{equation}
with $N$ the number of stress-strain data pairs. Here, we consider the minimization of the least-squares error between measured stress $\tensig^{i}$ and predicted stress for the associated strain value $\teneps^{i}$. The constitutive model is parametrized in the material parameters $\tentheta$, and the semicolon denotes parametrization. This approach was used to determine and calibrate all \textit{TUBSsalt} parameters for Gorleben salt, resulting in a parameter set presented in \cite{IGG2016Gorleben}.

However, the samples used in laboratory tests are no longer in their original state and the material tests only give insight into the material behavior in localized regions from where the sample has been extracted. As a consequence, model predictions typically do not match with real-world monitoring data, necessitating re-calibration of the constitutive model. Therefore, model calibration based on monitoring data is addressed in the following Section. 

\subsection{Inverse problem formulation for the model re-calibration}\label{sec:inverse}

We now address the task of model calibration from monitoring data. The objective is to identify the optimal set of material model parameters \(\boldsymbol{\theta}\) that minimizes the discrepancy (least-squares error) between model predictions and the in-situ monitoring data. This process is framed as an inverse problem, solved with an optimization method. 

Let \(\mathbf{Y}_{\text{moni}}^T = [\vecy_{\text{moni}}^T(t_1),\, \vecy_{\text{moni}}^T(t_2),\, ...,\, \vecy_{\text{moni}}^T(t_n)]\) denote the vector of experimental observations at different time instances for a specific location with $n$ the total number of time instances. The corresponding model prediction is denoted by \(\vecs_i(\vectheta) = \vecs(t_i,\, \vectheta)\). Then, in analogy to equation \eqref{eq:regression}, material parameter are identified from
\begin{equation}\label{eq:calib_fullfield}
    \boldsymbol{\theta}^* = \text{arg min}_{\boldsymbol{\theta}} \ \sum_{i=1}^{n} \| \vecy_{\text{moni}}(t_i) - \vecs_i(\vectheta) \|^2.
\end{equation}
Note that in equations \eqref{eq:regression} and \eqref{eq:calib_fullfield} we omit a weighting matrix, that is often introduced to account for the size of measurement errors. In our setting, the error sizes did not have a relevant influence. The optimization problem \eqref{eq:calib_fullfield} is implemented using the SciPy Python library \citep{Virtanen2020scipy}, employing a global optimization algorithm. We opted for the differential evolution algorithm \citep{Storn1997diff_evolution}, but also experimented with genetic and dual annealing algorithms. However, differential evolution proved to be faster and more reliable in this context.

\subsection{Sensitivity analysis based on time-dependent Sobol' indices} \label{sec:sensitivity_analysis}

Mechanical models that are used in geomechanical contexts are often characterized by a large number of material parameters, and so is \textit{TUBSsalt}. However, not every parameter in a data set may be sensitive to specific monitoring data. To focus only on the important ones and to reduce the number of model evaluations when solving \eqref{eq:calib_fullfield}, we perform a sensitivity analysis. 

In this contribution, we compute sensitivities based on Sobol' indices \citep{Sobol2001globalSA,Saltelli2002SA,Saltelli2010varianceSA}. The \textbf{first-order Sobol' index} quantifies the contribution of each parameter \(\theta_j\) to the variance in the predicted solution \(\vecs_i(\vectheta)\), while averaging out the effects of other inputs. The first-order Sobol' index for each time point \( t_i \) is defined as:
\begin{equation}
  S1_i(\theta_j) = \frac{V_{\theta_j}(E_{\theta_{\sim j}}(\vecs_i(\boldsymbol{\theta}) | \theta_j))}{V(\vecs_i(\boldsymbol{\theta}))},
\end{equation}
where \( V(\vecs_i(\boldsymbol{\theta})) \) is the variance of the solution at time \( t_i \), and \( E_{\theta_{\sim j}}(\vecs_i(\boldsymbol{\theta}) | \theta_j) \) is the conditional expectation given \(\theta_j\).

Similarly, the \textbf{total-order Sobol' index} for each solution \(\vecs_i(\vectheta) = \vecs(t_i,\, \vectheta)\) and input parameter \(\theta_j\) at time \(t_i\) measures the total effect of \(\theta_j\) on the variance of the solution, including interactions with other inputs. The total-order Sobol' index is defined as:
\begin{equation}
  ST_i(\theta_j) = 1 - \frac{V_{\theta_{\sim j}}(E_{\theta_j}(\vecs_i(\boldsymbol{\theta}) | \theta_{\sim j}))}{V(\vecs_i(\boldsymbol{\theta}))},
\end{equation}
where \( E_{\theta_j}(\vecs_i(\boldsymbol{\theta}) | \theta_{\sim j}) \) is the conditional expectation given all parameters except \(\theta_j\).

Sobol' indices \(S1_i(\theta_j)\) and \(ST_i(\theta_j)\) allow us to quantify the sensitivity of the solution \(\vecs(t_i,\vectheta)\) at each time step \(t_i\) to the different parameters \(\boldsymbol{\theta}\). This analysis helps us understand how each parameter affects the system's behavior over time. To identify and potentially discard less influential parameters, we propose calculating global indicators: cumulated, time-averaged, and maximum Sobol' indices. These indicators offer insights into the overall influence of the parameters throughout the time-dependent solution and are defined as follows:
\begin{itemize}
    \item The \textbf{Normalized cumulated Sobol' indices} are calculated by summing the influence of each input parameter over all time points and then normalizing by the maximum integrated value across all input parameters:
    \begin{equation}
        \text{Norm.} \int{S1}_i = \frac{\sum_{i=1}^{n} S1_i}{\max\left(\sum_{i=1}^{n} S1_i, \sum_{i=1}^{n} ST_i\right)}, \quad
        \text{Norm.} \int{ST}_i = \frac{\sum_{i=1}^{n} ST_i}{\max\left(\sum_{i=1}^{n} S1_i, \sum_{i=1}^{n} ST_i\right)}.
    \end{equation}

    \item The \textbf{time-averaged Sobol' indices} provide an overall measure of the importance of each input parameter across all time points. They are computed as the mean of the Sobol' indices at each time step \(t_i\):
    \begin{equation}
        \overline{S1} = \frac{1}{n} \sum_{i=1}^{n} S1_i, \quad
        \overline{ST} = \frac{1}{n} \sum_{i=1}^{n} ST_i.
    \end{equation}
    
    \item The \textbf{maximum Sobol' indices} identify the time steps at which each input parameter has the highest influence. These are particularly useful for pinpointing critical moments in the time-dependent behavior of the model:
    \begin{equation}
        \max{S1}_i = \max_{t}(S1_i), \quad
        \max{ST}_i = \max_{t}(ST_i).
    \end{equation}
\end{itemize}
A general framework for time-dependent variance-based sensitivity analysis has been put forth in \cite{alexanderian2020variance}. Therein, generalized Sobol' indices for processes are defined via a decomposition of the covariance function of the process. If they are approximated with an unweighted quadrature on a uniform grid, a normalized version of the cumulated Sobol' indices introduced above is recovered. An even more general framework, presented in \cite{gamboa2014sensitivity}, considers sensitivity analysis of functional outputs, covering time-dependent responses as a special case. 

The \textbf{convergence of the Sobol' indices} estimation can be assessed using the maximum relative change between successive iterations. The relative change for each Sobol' index is calculated as the absolute difference between the current and previous indices, normalized by the previous indices plus a small constant to avoid division by zero. Specifically, the relative change for \(S1_i\) is given by:
\begin{equation}\label{eq:Sobol_indices_convergence}
\text{S1}_i\text{ relative change} = \frac{\left| \text{S1}_i{}_{\text{curr}} - \text{S1}_i{}_{\text{prev}} \right|}{\text{S1}_i{}_{\text{prev}} + 1 \times 10^{-10}},
\end{equation}
and the maximum value across all indices is monitored. The same applies to \(ST_i\). If the maximum change in both indices is below a predefined threshold (\(0.01\)), the process is deemed converged, indicating that the sampling effort is sufficient for accurate sensitivity analysis.

A bottleneck is that a sensitivity analysis using Sobol' indices with Saltelli's algorithm \cite{Saltelli2010varianceSA}, pick-and-freeze estimators \cite{gamboa2016statistical} or other methods requires a significant number of model evaluations. One way to bound the associated computational cost is by using surrogate models such as the Polynomial Chaos expansion \cite{sudret2008global} or GPs. In this contribution, GPs are used and presented in the next section.

\section{Gaussian processes as surrogate models for complex computer models}\label{sec:GPs}

The model $\vecs_i(\vectheta)$ represents the mechanical behavior of the deep geological repository introduced in \autoref{sec:drift_model}. Evaluating $\vecs_i(\vectheta)$ directly is computationally demanding, especially in many-query scenarios such as model calibration, optimization, or design. Surrogate modeling offers an efficient alternative to address the computational challenges associated with directly solving $\vecs_i(\vectheta)$. Among the available techniques, Proper Orthogonal Decomposition (POD) \citep{Agarwal2024POD} and Physics-Informed Neural Networks (PINNs) \citep{Anton2024PINNs} are particularly suitable for full-field simulations where capturing the entire spatial domain is crucial. However, GPs present a compelling option when the focus is on specific quantities of interest, such as an excavation convergence as in the present study, rather than the entire field. GPs are especially advantageous due to their ability to provide probabilistic predictions and handle smaller datasets effectively. In this study, we adopt GPs to develop a surrogate model, denoted as \(\hat{\vecs}_i(\boldsymbol{\theta})\), to approximate the complex behavior of $\vecs_i(\vectheta)$.

In scenarios where \(\vecs\) predicts a time-sequence output, we define a set of GPs, \(\{\hat{\vecs}_i\}_{i=1}^{n}\). Each \(\hat{\vecs}_i\) corresponds to a distinct time instance \(t_i\), as:
\begin{equation}
    \hat{\vecs}_i(\boldsymbol{\theta}) \sim \mathcal{GP}\left( m_i(\boldsymbol{\theta}),\, k_i(\boldsymbol{\theta}, \boldsymbol{\theta}') \right),
\end{equation}
where \(m_i(\boldsymbol{\theta})\) is the mean function and \(k_i(\boldsymbol{\theta}, \boldsymbol{\theta}')\) is the covariance (or kernel) function for the \(i\)-th GP. This function quantifies the similarity between two sets of parameter configurations, \(\boldsymbol{\theta}\) and \(\boldsymbol{\theta}'\), for the specific time point. Each GP \(\hat{\vecs}_i\) is independently trained on a subset of the data corresponding to its time point:
\begin{equation}
[\hat{\vecs}_i(\boldsymbol{\theta}_1),\, \hat{\vecs}_i(\boldsymbol{\theta}_2),\, \ldots,\, \hat{\vecs}_i(\boldsymbol{\theta}_N)]
\sim \mathcal\mathcal{N}\left( \mathbf{0}, \mathbf{K}_i \right),
\end{equation}
where \(\mathbf{K}_i\) is the covariance matrix for the \(i\)-th GP, computed using a kernel \(k_i\). Note that we could also introduce a priori correlation between GPs at different points in time; however, the independence assumption is very common. Here, for each GP, we use the Matérn kernel (with smoothness parameter $\nu=5/2$) defined as:
\begin{equation}
    k_i(\boldsymbol{\theta}, \boldsymbol{\theta}') = \sigma_i^2 \left(1 + \sqrt{5} r + \frac{5}{3} r^2 \right) \text{exp}(- \sqrt{5} r),
\end{equation}
where \(r(\boldsymbol{\theta}, \boldsymbol{\theta}') = \sqrt{\sum_{d=1}^D \left( \frac{\theta_d - \theta'_d}{l_{d,i}} \right)^2}\), with \(l_{d,i}\) and \(\sigma_i^2\) being the length-scales and variance parameter for the \(i\)-th GP.

The next crucial step is the determination of the GP-based surrogate model hyperparameters. These hyperparameters include the length-scales \(l_{d,i}\), variance \(\sigma_i^2\), and any other parameters specific to the chosen kernel function. The hyperparameters are typically optimized by maximizing the likelihood of the observed data under the GP model. This optimization can be formulated as:
\begin{equation}\label{eq:tune_hyper}
    \max_{{\sigma_i}, {l_{d,i}}} \ \text{log} \ \mathcal{L}({\sigma_i}, {l_{d,i}} | D_i),
\end{equation}
where \(\mathcal{L}\) is the likelihood of the training data \(D_i\) given the hyperparameters. This process is often carried out using gradient-based optimization techniques.

Upon training each \(\hat{\vecs}_i\) with a dataset \(D_i = \{ (\boldsymbol{\theta}_1, s_{i1}), \ldots, (\boldsymbol{\theta}_N, s_{iN}) \}\), where \(s_{ij}\) is the output from \(\vecs\) for input \(\boldsymbol{\theta}_j\) at time \(t_i\), each GP can make predictions for unseen parameter sets \(\boldsymbol{\theta}_*\) at its respective time instance. This ensemble of GPs allows for efficient evaluation of the time-sequenced outputs of \(\vecs(t_i, \vectheta)\) without relying on computationally expensive numerical solutions.

The mean prediction \(\mu_i(\boldsymbol{\theta}_*)\) and the standard deviation \(\sigma_i(\boldsymbol{\theta}_*)\) at each time point \(t_i\) are given by:
\begin{equation}\label{eq:GPs_mean_std}
    \mu_i(\boldsymbol{\theta}_*) = \mathbf{K}_*^T \mathbf{K}_i^{-1} \vecs, \quad
    \sigma_i^2(\boldsymbol{\theta}_*) = k_i(\boldsymbol{\theta}_*, \boldsymbol{\theta}_*) - \mathbf{K}_*^T \mathbf{K}_i^{-1} \mathbf{K}_*,
\end{equation}
where \(\mathbf{K}_*\) is the covariance vector between the training inputs and \(\boldsymbol{\theta}_*\), and \(\mathbf{K}_i\) is the covariance matrix of the training inputs.

These predictions not only provide the expected output of the model \(\vecs\) for an unseen parameter set but also quantify the uncertainty associated with these predictions. This feature of GPs is particularly useful in decision-making processes where uncertainty plays a crucial role.

It is important to note that before training the GP-based surrogate model, we perform feature scaling on both input parameters and outputs to achieve zero mean and unit variance. This step is crucial for enhancing the surrogate model's numerical stability, which is developed using the Scikit-learn library \citep{Pedregosa2011ScikitLearn}. This library offers a ready-to-use module to build surrogate models based on GPs for multivariate nonlinear regression problems. GP hyperparameters are optimized using the Limited-memory Broyden–Fletcher–Goldfarb–Shanno with bound constraints (L-BFGS-B) algorithm; see Scikit-learn library documentation for more details. This optimization process is key to maximizing the log-marginal likelihood defined in \eqref{eq:tune_hyper}, ensuring that the kernel parameters are finely tuned to best represent the underlying patterns of the data.

\section{Numerical analysis}\label{sec:results}

As described in \autoref{sec:drift_model}, \textit{TUBSalt} is a highly nonlinear time-dependent constitutive material model with $25$ material parameters. 
After a preliminary study, we train a GP-based surrogate model (\autoref{sec:GPs}) to capture the input-output relationship between selected parameters of the constitutive model \textit{TUBSsalt} and the drift convergence at different time instances, conduct a sensitivity analysis  and calibrate the constitutive model. This section is structured as follows: First, the high-fidelity model evaluation and dataset creation for training the GP-based surrogate model is presented in \autoref{sec:HF_model_numerical_results}. \autoref{sec:para_selection} deals with the training of the GP-based surrogate model, the accuracy evaluation with testing data and sensitivity analysis using Sobol' indices. Finally, in \autoref{sec:calibration_results}, the model parameters are calibrated using in-situ monitoring data.

\subsection{High-fidelity model simulation}
\label{sec:HF_model_numerical_results}

\noindent\textbf{Monitoring data:} The mechanical model described in \autoref{sec:drift_model} represents an open drift located in the northern main drift of Gorleben for which monitoring data is available. 
The first convergence measurement took place on October 20, 1999, one day after the excavation of the drift. This is important because it allows for the measurement of primary creep, which is most pronounced at the beginning of the excavation. On the day of the initial measurement, the horizontal distance (1-3) measured $8.94$ m, and the vertical distance (2-4) was $6.00$ m, as depicted in \autoref{fig:high_fidelity_model_simulation}\textbf{(a)}. The duration of the monitoring was approximately $14.3$ yr, with a significant variation of time intervals between measurements. During the first two months, measurements were taken every two to three days in order to capture the high creep rates. Subsequently, the measurement interval was increased from one week to one month, and from the end of 2002 onwards, measurements were taken approximately every six months as a constant creep rate had been achieved.
Since the measuring points of the convergence distances are anchored in the rock over an anchoring length, the deformations of the cavity are also measured at these fixed points in the rock salt. Those locations are defined as history points in the numerical model in FLAC3D, demonstrated in \autoref{fig:high_fidelity_model_simulation}\textbf{(b)}. The convergences 
$\vecy_{\text{moni}}^T(t_i) = \left[ \Delta u_x(t_i) ,\, \Delta u_z(t_i)\right]^T$ 
are measured as a vertical and horizontal distance between two points, therefore the displacements $u_x$ in horizontal and $u_z$ in vertical direction are obtained from the displacement vector $\vecu$ for the specific nodes.
\Revision{The uncertainty of the convergence measurements is around $\pm 0.5$ mm, while observations are on the order $\mathcal{O}(10^2)$. Consequently, we consider the monitoring data to be noise-free.}\\

\noindent\textbf{Model setup:} Since only one material is considered and the cross-section of the drift is almost symmetric, only the right symmetry half of the model needs to be discretized. Dirichlet boundary conditions are applied by fixing the left, right, and bottom sides of the system in their normal directions. We model the overburden with a load of $16.774$ MPa applied to the top of the model, which is a Neumann boundary condition. According to the geology \RevisionNew{of the site given in \cite{BGR2008Gorleben}}, the load results from the different heights and densities of the rock salt, cap rock, tertiary and quaternary layer. The densities were obtained from \cite{GRS2012VSG-286}.
The initial stress state is derived from the overburden load and displacements for time $t=0$ are set to zero. The initial stress state is assumed to be isotropic. The slight directional dependence of the monitoring data can be attributed to the differing dimensions of the drift cross-section in the $x$ and $z$ directions.
The absence of nearby geotechnical structures and the sufficient distance between the measurement site and other side drifts, or homogeneous areas allow the model to be simplified to a 2D model. The dimensions in the $x$ and $z$ direction for the model are given by $l_x=50$ m and $l_z=100$ m, so that an influence of the drift on the system boundaries can be excluded. Temperature measurements conducted in the close area of the drift show an almost constant value of $37~^{\circ}$C. Therefore, a constant temperature is assumed over the entire system as a simplification.
The mesh density of the model was optimized by performing a convergence study, ensuring that increasing the number of elements results in no significant changes in the curves of simulated horizontal and vertical convergences.
Prior to the time-dependent creep analysis, the model reaches an equilibrium state twice: first, after applying the initial and boundary conditions, and second, after setting the stresses in the zones inside the cross section to zero to simulate the excavation. The equilibrium state is reached as soon as the average ratio between out-of-balance and total forces falls below $1 \cdot 10^{-6}$. \\

\noindent \textbf{Preliminary parameter selection:} \textit{TUBSalt} has a total of $25$ material parameters. Out of these, $5$ parameters can be read directly from laboratory stress-strain data, namely $K_0$, $G_0$, $\varepsilon_{v,d,b}$, $\sigma_{z,0}$ and $\rho$. In the following, these are considered as well-calibrated and independent of the monitoring data. The remaining $20$ parameters serve as potential parameters for calibration. We then conducted a preliminary, empirical study with only two simulations: several damage-associated strain components were deactivated, and the result was compared to a reference solution obtained with the full set of strain components \eqref{eq:full_strain}. It has been found that the test case considered is dominated by primary and secondary creep mechanisms and damage-associated strain components are negligible. Details can be found in Appendix~\ref{appendixA}. As a result, in the following we focus on a total of $7$ \textit{TUBSsalt} parameters for primary and secondary creep: $\eta_p$, $\sigma_{0,eq,p}$, $p_p$ and $E_p$ for primary creep and $\eta_s$, $\sigma_{0,eq,s}$ and $p_s$ for secondary creep, and thus $\vectheta = \left[ \eta_p,\, \sigma_{0,eq,p},\, p_p,\, E_p,\, \eta_s,\, \sigma_{0,eq,s},\, p_s\right]$. The parameter ranges of those $7$ parameters, as well as the parameters that are considered as fixed, are summarized in Appendix~\ref{appendixB}, Table~\ref{tab:parameter_config}. \\

\noindent \textbf{Sampling of parameter values in Flac3D:} Following the approach in \autoref{sec:regression}, a parameter set for the constitutive model \textit{TUBSsalt} based on laboratory strength and creep tests has been determined in a project presented in \cite{IGG2016Gorleben} for the Gorleben site. These values serve as characteristic parameter values for Gorleben salt and also help define reasonable parameter ranges for those parameters that need calibration, see also \autoref{tab:parameter_config} in \autoref{appendixB}.
In order to create training and testing data by means of a multitude of simulations, the parameter values of the selected \textit{TUBSsalt} parameters are sampled in the given ranges in \autoref{tab:parameter_config} assuming a uniform distribution. For this purpose, a SciPy Quasi-Monte Carlo generator is used to generate a Sobol sequence by creating low-discrepancy, quasi-random numbers \citep{SobolSeq1967, QMC2019}. The loop for parameter sampling is implemented in the creep analysis of Flac3D, so that the saved equilibrium state of the model is recalled for every iteration. The creep analysis is performed for the same duration of the monitoring. To maintain the state of equilibrium, the average ratio between out-of-balance and total forces is limited to $5\cdot10^{-6}$. The timestep increases over time with a maximum of $\Delta t = 5.56$ h. With the given configuration, a single simulation run of the high-fidelity model requires approximately $4$ min.\\

\begin{figure}[b]
    \centering
    \includegraphics[width=0.90\textwidth]{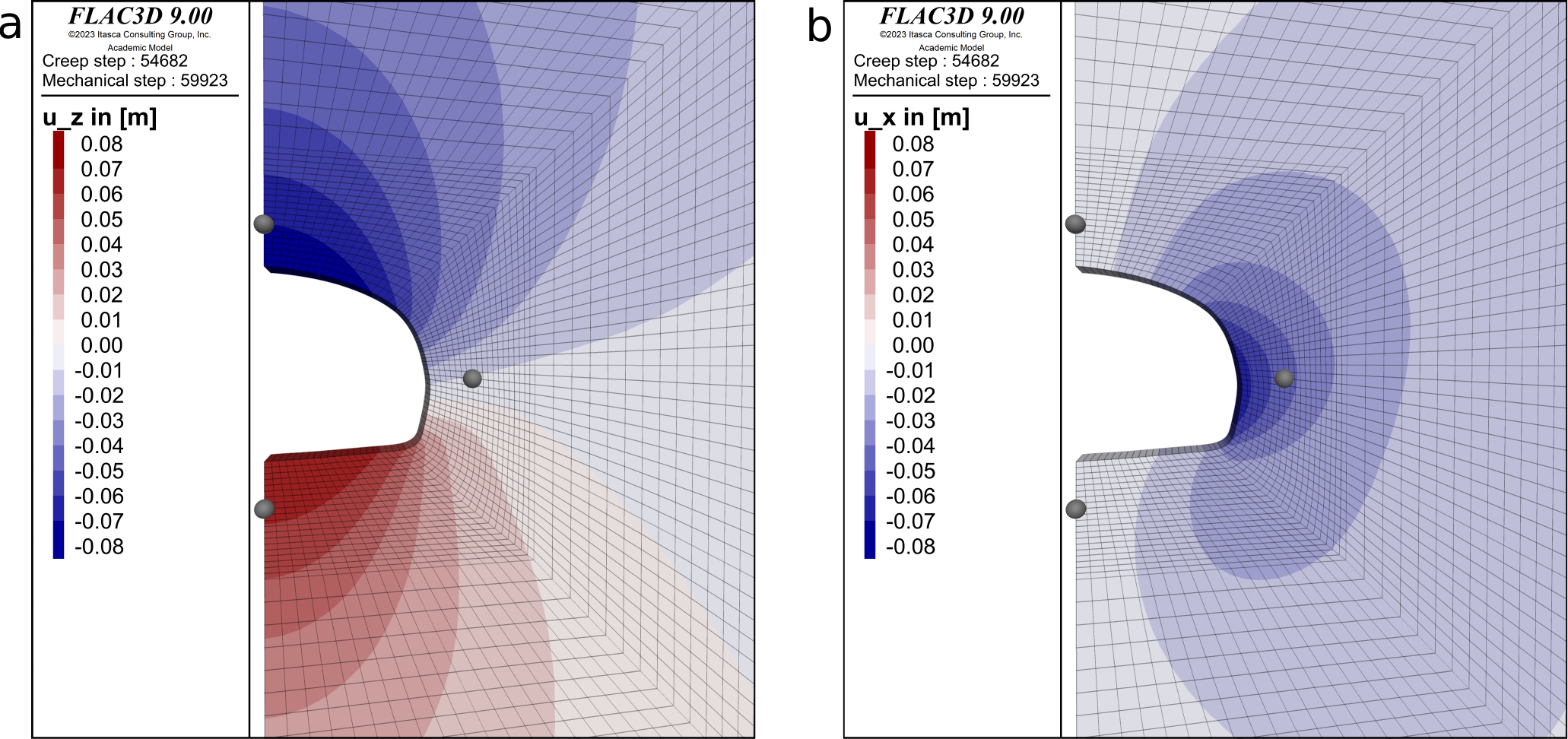}
    \caption{\textbf{Full-field simulation at $t = 14.5$ yr} corresponding to the geometry depicted in \autoref{fig:high_fidelity_model_simulation}\textbf{b}.
     \textbf{a.} Vertical \(u_z\) and \textbf{b.} horizontal displacements \(u_x\). History locations are marked by grey dots. The simulation uses \textit{TUBSsalt} parameters: $\eta_p = 8.0\cdot10^4$ MPa$\cdot$d, $E_p = 75$ MPa, $\sigma_{0,eq,p} = 30$ MPa, $p_p = 0.5$, $\eta_s = 3.0\cdot 10^7$ MPa$\cdot$d, $\sigma_{0,eq,s} = 30$ MPa and $p_s = 1.5$.}
    \label{fig:full-field-realization}
\end{figure}

\begin{figure}[h]
    \centering
    \includegraphics[width=0.9\textwidth]{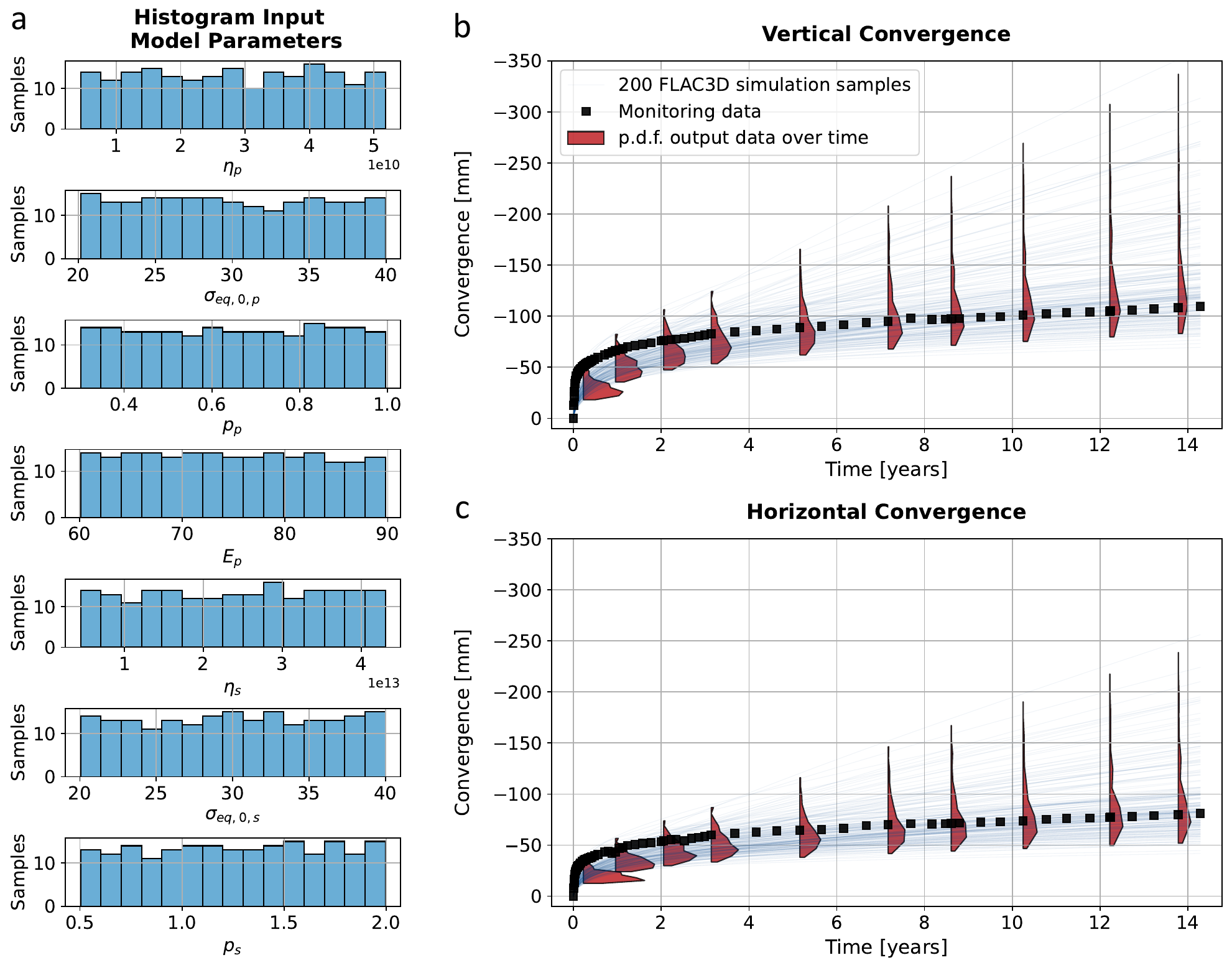}
    \caption{\textbf{Input-output simulation data at the monitoring location.} \Revision{\textbf{a.} Histograms for the input material parameters \(\eta_p\), \(E_p\), \(\eta_s\), \(p_s\), \(\sigma_{0,eq,p}\), \(p_p\), and \(\sigma_{0,eq,s}\) of \textit{TUBSsalt}. \textbf{b.} Vertical and \textbf{c.} horizontal convergence trajectories over time. \textbf{b.} and \textbf{c.} show the monitoring data (black squares) and $200$ model realizations (blue lines) from the FLAC3D simulations along with the obtained PDF at selected time instances (red area).}}
    \label{fig:simulation_samples_7param}
\end{figure}

\noindent \textbf{Processing the full-field simulation results to convergences:} After carrying out the simulations to create training and testing data, the displacements derived at the history locations are converted into convergences at the time instances for which monitoring data are available. To determine the vertical convergences, the amounts of the vertical displacements $u_z(t_i)$ at the top and bottom are summed up together. To determine the horizontal convergences, the amount of the horizontal displacements $u_x(t_i)$ is multiplied by two to account for the symmetry of the system. The displacements of an exemplary full-field simulation at $t = 14.5$ yr is depicted in \autoref{fig:full-field-realization}. In order to compare the simulated convergences with the monitoring data, the convergences of the simulations must be set to zero at the time of the initial monitoring measurement. Therefore, the convergences from the first day after excavation of the drift can not be analysed and will be neglected. Convergences are determined at all points in time at which monitoring is available. As a result, \Revision{\autoref{fig:simulation_samples_7param} shows the processed data of $200$ simulations. \autoref{fig:simulation_samples_7param}\textbf{a} depicts the histograms obtained after sampling the input material parameters, while \autoref{fig:simulation_samples_7param}\textbf{b,c} correspond to the vertical and horizontal convergences over time (blue lines) in comparison with the corresponding monitoring data (black squares) and the obtained probability density function (PDF) at selected time instances (red area). This figure illustrates how the uncertainty in material parameters propagates through the mechanical model implemented in FLAC3D, reflecting its nonlinearity. That is, uniformly sampled inputs result in heavily tailed outputs. The task of the GP-based surrogate is to represent this input-output relationship by learning the PDF of the convergences at different time instances as a function of the model parameters.}

\noindent In order to understand the individual and combined effect of the material parameters on the drift convergence, Sobol' indices-based global sensitivity analysis can be performed as outlined in \autoref{sec:sensitivity_analysis}.

\subsection{GP-based surrogate and sensitivity analysis}\label{sec:para_selection}
In this subsection, we focus on developing a surrogate model for predicting the temporal evolution of horizontal and vertical drift convergences based on the synthetic dataset described in the previous \autoref{sec:HF_model_numerical_results}.
The GP-based surrogate is constructed so that it predicts the convergences at those time instances $t_i$ for which monitoring data are available. As shown in \autoref{fig:simulation_samples_7param}, the original density of the measurement data is significantly higher at the beginning of the monitoring. To prevent distortion of both the time-dependent Sobol' indices and the subsequent calibration process, the available data points in the first $3.6$ years are filtered based on the time intervals in such a way that the measurement points are distributed as evenly as possible over time. As a result, our approach treats horizontal and vertical convergences at $40$ equidistant time instances as independent outputs, framing our problem a multivariate nonlinear regression with $7$ inputs (parameters of the \textit{TUBSsalt} material model for rock salt) and $80$ outputs (the drift convergence at $40$ different selected time instances). 
This finally leads to an ensemble of $80$ GPs, $40$ each for both vertical and horizontal convergences. Each GP takes the $7$ material parameters as input and outputs the value of the convergence at its specific time instance as output.  \\

\begin{figure}[b]
    \centering
    \includegraphics[width=0.9\textwidth]{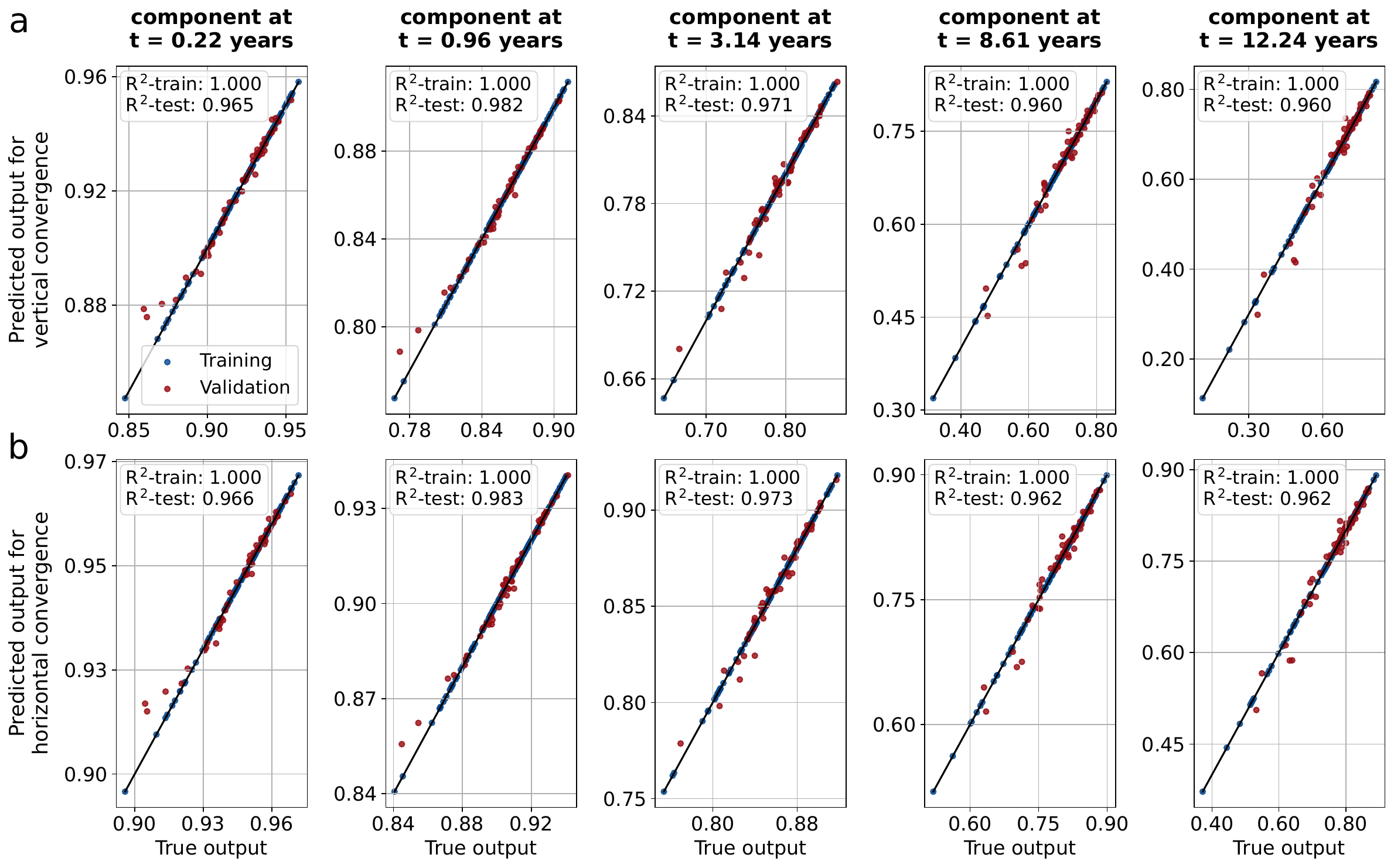}
    \caption{\textbf{Accuracy evaluation of the GP-based surrogate model.} The true versus predicted outputs for $5$ out of the $40$ GPs that constitute the surrogate model, for both \textbf{a.} vertical and \textbf{b.} horizontal convergence at different time instances. Each GP approximates the convergence at a different time instance as indicated in the title of each subplot. The blue dots represent training data, and the red dots represent testing data. The black diagonal line denotes the line of perfect prediction.}
    \label{fig:GPs_accuracy_7param}
\end{figure}

\noindent \textbf{GP-based surrogate model:} We split the dataset into training ($75\%$) and testing ($25\%$) datasets, which results in $150$ model realizations for training and $50$ for testing the accuracy of the surrogate model. The selected amount of training data results from an investigation of the relation between the amount of training data and the accuracy of the surrogate model prediction. On the one hand, we experienced that fewer model realizations lead to a significant decrease in the $R^2$ coefficient for the testing dataset. On the other hand, if more model realizations were used, only a marginal increase in the $R^2$ coefficient would be achieved compared to the computational cost required to train the surrogate model. Note that the number of model realizations can be further reduced and optimized using adaptive sampling strategies, see \cite{fuhg2021state} for a recent review. The generation of the GP-based surrogate model with $7$ parameters and $150$ training data samples takes $12.3$ s.

\autoref{fig:GPs_accuracy_7param} displays scatter plots comparing the true output versus predicted output for the GP-based surrogate model trained to predict vertical and horizontal convergence at different time instances. Each subplot represents a distinct time instance, in years, as indicated in their titles, respectively. The blue dots indicate training data, and the red dots indicate testing data. The black diagonal line represents perfect predictions. The high \(R^2\) scores for the training data across all components indicate that the model fits the training data extremely well. The testing \(R^2\) scores range from $0.960$ to $0.983$ for both vertical and horizontal convergence, indicating a very good generalization to unseen data. The consistently high \(R^2\) scores and tight clustering of data points confirm the model's robustness and reliability in predicting convergence accurately at the selected time instances.\\

\begin{figure}[b]
    \centering
    \includegraphics[width=0.9\textwidth]{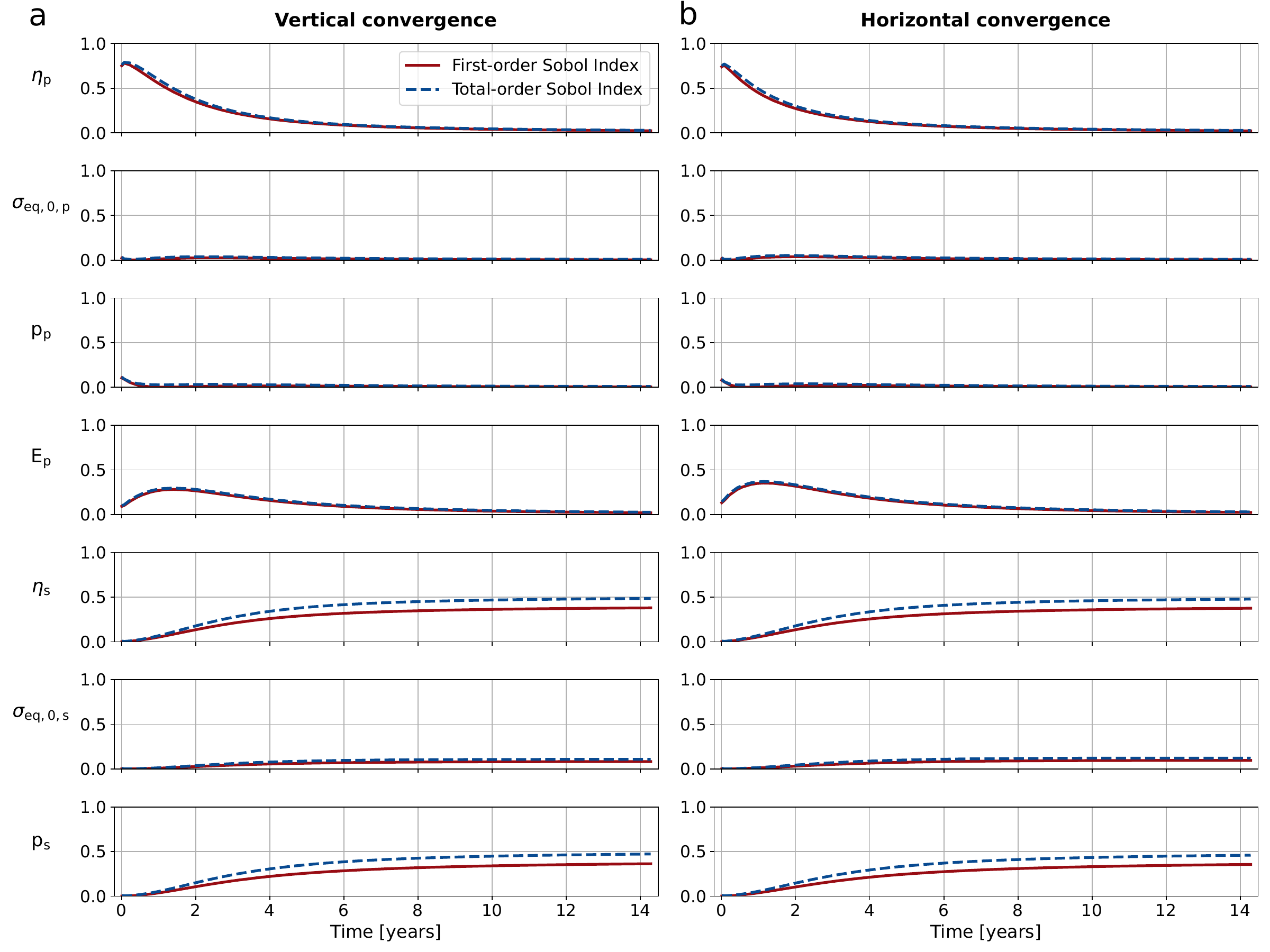}
    \caption{\textbf{Time-dependent global sensitivity analysis}. 
    First and total order Sobol' indices over time for \textbf{a.} vertical and \textbf{b.} horizontal convergences, showing the influence of the primary creep parameters (\(\eta_p\), \(\sigma_{0,eq,p}\), \(p_p\), \(E_p\)) and secondary creep parameters (\(\eta_s\), \(\sigma_{0,eq,s}\), \(p_s\)), presented from top to bottom, respectively.}
    \label{fig:SA_7_time}
\end{figure}

\noindent \textbf{Sensitivity analysis:} With the GP-based surrogate model at hand, we compute the first-order (S1) and the total-order (ST) Sobol' indices.
\autoref{fig:SA_7_time} displays the results of the time-dependent sensitivity analysis. From the figure, it is clear that the effect of each parameter varies over time. The parameters for primary creep $\eta_p$ and $E_p$ have a significant impact at the beginning of the simulation, but this effect diminishes as time progresses. Conversely, $\eta_s$ and $p_s$ initially have little influence but become more impactful over time as they are parameters of secondary creep. In particular, $\sigma_{0,eq,p}$, $p_p$ and $\sigma_{0,eq,s}$ have little to no influence on the output variation. Therefore, these material parameters can be fixed by taking the reference value from \cite{IGG2016Gorleben} and removed from further analysis.

\begin{figure}[h]
    \centering
    \includegraphics[width=0.9\textwidth]{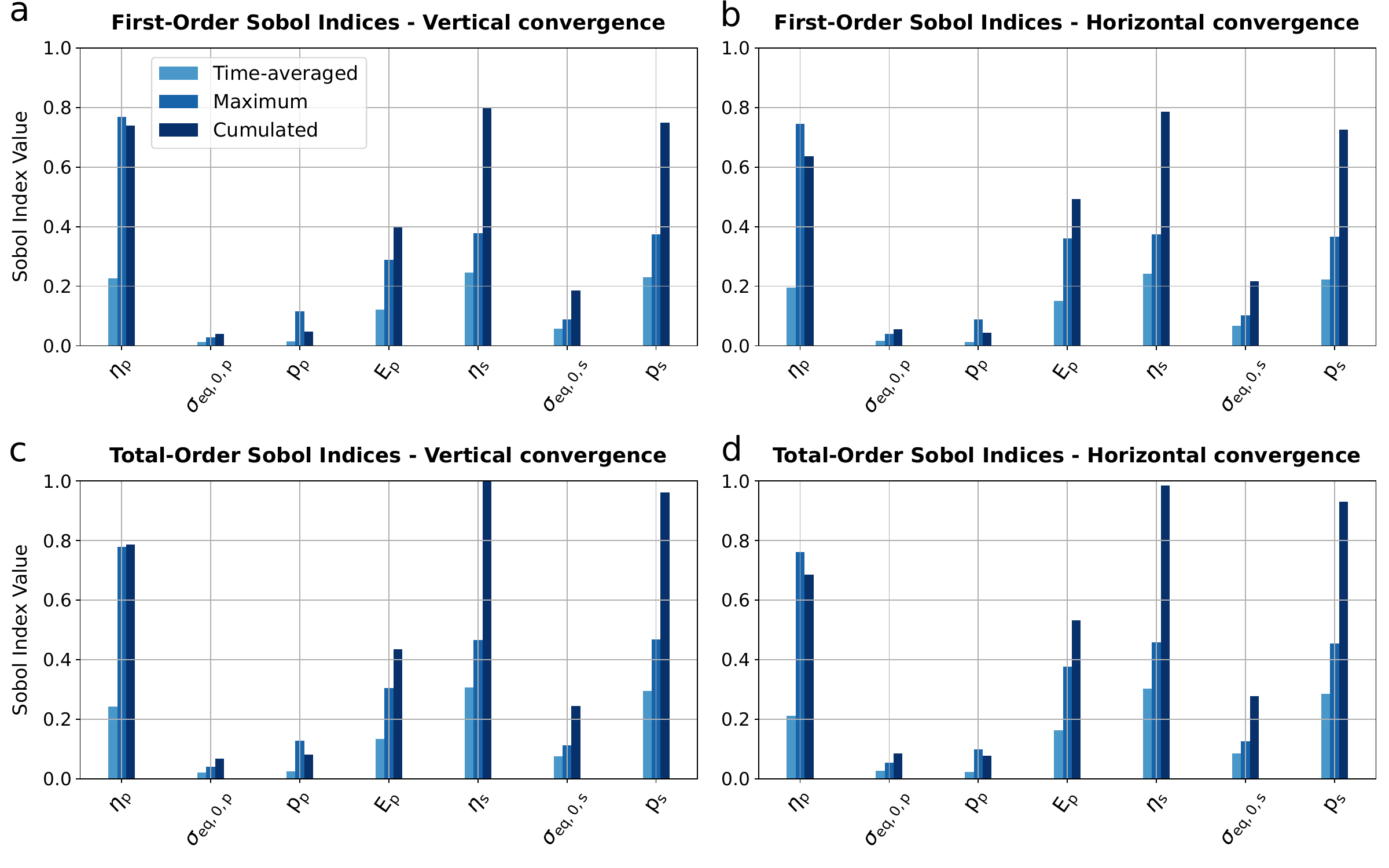}
    \caption{\textbf{Aggregated global sensitivity analysis.}
    Normalized first-order (\textbf{a. - b.}) and total-order (\textbf{c. - d.}) Sobol' indices for vertical (\textbf{a. - c.}) and horizontal (\textbf{b. - d.}) convergences, showing the influence of the primary creep parameters (\(\eta_p\), \(\sigma_{0,eq,p}\), \(p_p\), \(E_p\)) and secondary creep parameters (\(\eta_s\), \(\sigma_{0,eq,s}\), \(p_s\)), from left to right, respectively. The cumulated first-order and total Sobol' indices are normalized against their respective highest overall values for fair comparison with the other indicators.}
    \label{fig:SA_7_indicators}
\end{figure}

The indicators defined in \autoref{sec:GPs} are computed from the time series shown in \autoref{fig:SA_7_time} to quantify the effect of each parameter on the model output variation more precisely. \autoref{fig:SA_7_indicators} presents these indicators as bar plots for both the first-order (S1) and total-order (ST) Sobol' indices. These bar plots clearly indicate which parameters most significantly affect the model outputs. \autoref{fig:SA_7_indicators} complements our observation from \autoref{fig:SA_7_time}: the horizontal and vertical convergences were found to be sensitive only to the parameters $\vectheta=\left[\eta_p,\, E_p,\, \eta_s,\, p_s\right]$.

Convergence analysis of the Sobol' indices was performed as specified in \autoref{sec:sensitivity_analysis}.
The process begins with an initial set of $2^{7} = 128$ samples and iteratively increases the sample size by $2^{7}$ in each step until either a maximum of $2^{16} = 65536$ samples is reached or convergence is achieved. 
From \eqref{eq:Sobol_indices_convergence}, it was concluded that $2^{14} = 16384$ samples are enough to assure Sobol' indices analysis convergence according to criterion \eqref{eq:Sobol_indices_convergence}. In particular, the maximum changes for S1 and ST were $0.0093$ and $0.0071$, respectively. The simulation time required for $2^{14}$ surrogate model evaluations was $4.5$ min. 

\subsection{Model calibration from in-situ monitoring data}\label{sec:calibration_results}

\begin{figure}[h]
    \centering
    \includegraphics[width=0.9\textwidth]{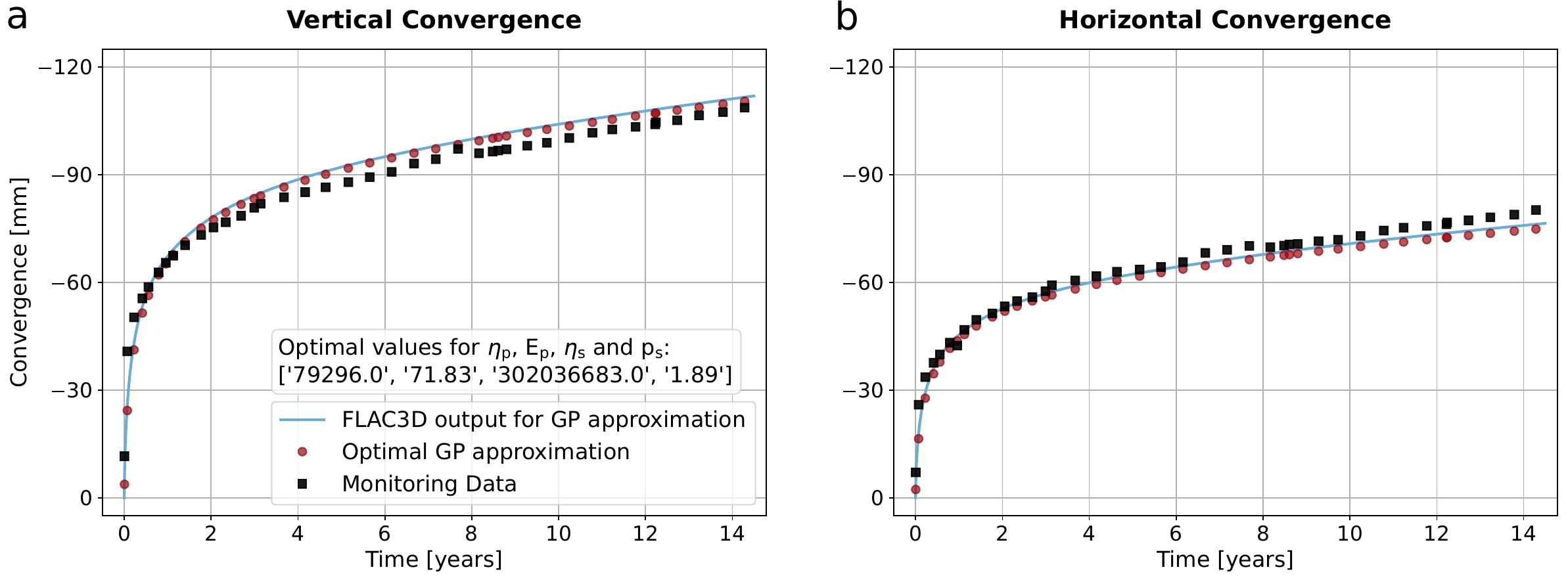}
    \caption{\textbf{Model calibration with monitoring data.}
    Comparison of monitoring data (black squares), GP-based optimal surrogate model prediction (red dots), and corresponding FLAC3D simulation results using the optimal parameter values (blue line) for \textbf{a.} the vertical and \textbf{b.} horizontal convergences.}
    \label{fig:GPs_based_opt}
\end{figure}

The final task is to calibrate the four sensitive model parameters comprised in $\vectheta$ using real drift convergence monitoring data. 
To achieve this, we embed the surrogate model trained in \autoref{sec:para_selection} into the optimization problem \eqref{eq:calib_fullfield} using only the mean values of the GP-based surrogate prediction as defined in \eqref{eq:GPs_mean_std}.
\autoref{fig:GPs_based_opt} displays the results obtained from the optimization problem. The results show that the simulation data generated from the high-fidelity model and the predictions from the GPs almost overlap, demonstrating the effectiveness of the surrogate model in calibration. It is crucial to highlight that model calibration is often a computationally demanding task, typically requiring multiple calls of the model. The global optimization of the remaining $4$ model parameters converged within $29$ iterations, which required $1805$ surrogate model evaluations. For comparison, training the GP-based surrogate model, used for both sensitivity analysis and calibration, only required $150$ model evaluations. The optimization took less than $1$ s to be completed on a standard laptop. This represents a dramatic reduction in the computational resources required to calibrate the deep-repository model, underscoring the efficiency of our surrogate modeling approach. The predicted parameter values agree very well with expert intuition so that the result can be considered reasonable.

\section{\RevisionNew{Conclusions}}\label{sec:conclusion}
\Revision{Deep geological disposal of hazardous materials requires robust numerical models to ensure long-term safety and stability. The calibration of such models with real-world monitoring data is essential for accurately reflecting in-situ conditions and enhancing repository management. This study contributes to advancing digital twinning for deep geological disposals by automating the calibration process using Gaussian Process (GP)-based surrogate models. The presented workflow combines sensitivity analysis, surrogate modeling, and optimization to enable efficient calibration of a mechanical model representing the behavior of an emplacement drift in rock salt formations located in the northern main drift of the Gorleben salt dome in Germany.} 

\Revision{The results demonstrate the efficiency and accuracy of the proposed approach. \RevisionNew{Initially, training of the GP-based surrogate model with $7$ input parameters, $150$ training data samples and $80$ outputs took $12.3$ s. The subsequent accuracy evaluation yielded $R^2$ scores between $0.960$ and $0.983$.} 
Afterwards, a sensitivity analysis using time-dependent Sobol' indices was performed with $16,384$ surrogate model calls within $4.5$ min to identify four relevant material parameters. Finally, the GP-based surrogate model was calibrated based on $14$ yr of convergence measurements, including the convergence of the global optimization in $29$ iterations, $1805$ model evaluations for gradient construction, and a duration of less than $1$ s. \RevisionNew{The surrogate model prediction provided both very good agreement with the monitoring data and valid values for parameters of the constitutive model \textit{TUBSsalt}.}}

\Revision{This approach reduces the computational burden associated with traditional high-fidelity models and enables rapid, iterative updates to model parameters as new monitoring data becomes available. By enhancing the scalability and adaptability of numerical models, this work lays the foundation for integrating advanced surrogate modeling techniques into the management of deep geological repositories.}

\Revision{While the workflow demonstrated high efficiency and accuracy for the presented mechanical model, further developments are required to extend its applicability. 
Specifically, first, the GP-based surrogate model calibration and optimization method must be extended to account for higher-dimensional monitoring data, which could include extensometer and permeability measurements. 
Second, an alternative formulation would be developing a time-dependent surrogate model, which enables forecasting capabilities together with uncertainty propagation. 
Third, efforts have to be undertaken to account for the multi-physics nature of deep geological disposal. Apart from the purely mechanical model investigated here, a variety of models has been developed in the past that describe, e.g., transport of radio-nuclides through fluid flow, heat generation from high-level radioactive waste, or hydration of a sealing structure.}

\begin{appendix}
\renewcommand{\thefigure}{A\arabic{figure}}
\setcounter{figure}{0} 

\section{Appendix - Preliminary parameter selection}\label{appendixA}

A preliminary parameter selection was performed in order to identify relevant strain components of \textit{TUBSsalt} given the considered monitoring data. For this purpose, several damage-associated strain components were deactivated, in particular tertiary creep $\dot{\varepsilon}_t$, creep and shear failure $\dot{\varepsilon}_n$ and tension failure $\dot{\varepsilon}_z$. A comparison between $274$ gridpoint displacements at $40$ time instances is given in \autoref{fig:comparison_displacements} below, in which the vertical (\autoref{fig:comparison_displacements}\textbf{a}) and horizontal (\autoref{fig:comparison_displacements}\textbf{b}) displacements were simulated once with and once without damage strains. From the comparison, it can be observed that no deviation from the diagonal line is visible, which is also confirmed by the $R^2$ value close to $1.0$. Thus, the considered test case is dominated by creep mechanisms and softening, and post-failure strains are considered negligible.

\begin{figure}[h]
    \centering
    \includegraphics[width=0.9\linewidth]{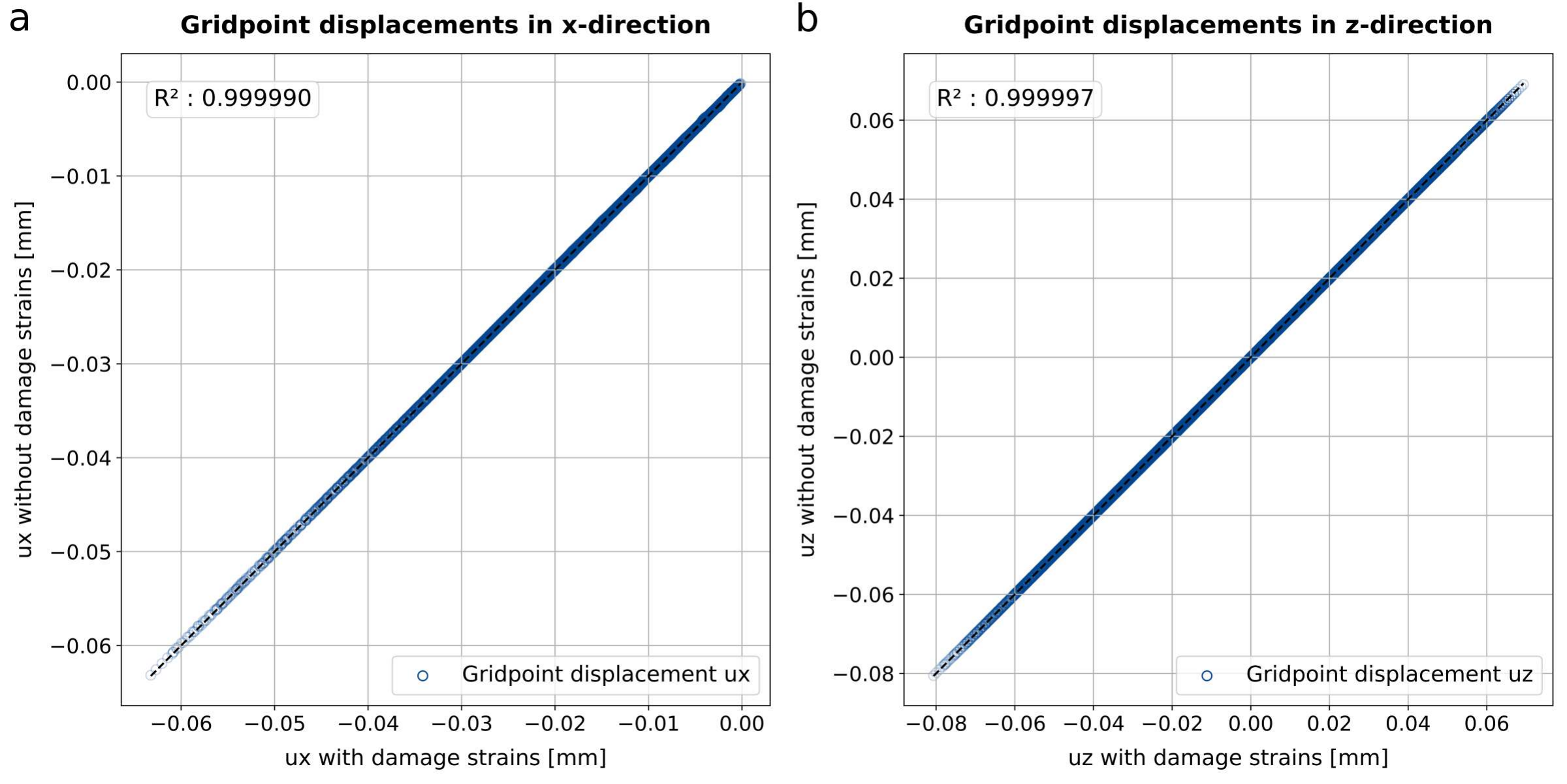}
    \caption{\textbf{Comparison of displacements with and without damage-associated strains} by evaluating $274$ gridpoints at $40$ time instances for \textbf{a.} vertical and \textbf{b.} horizontal displacements. The simulations use \textit{TUBSsalt} parameters: $\eta_p = 8.0\cdot10^4$ MPa$\cdot$d, $E_p = 75$ MPa, $\sigma_{0,eq,p} = 30$ MPa, $p_p = 0.5$, $\eta_s = 3.0\cdot 10^7$ MPa$\cdot$d, $\sigma_{0,eq,s} = 30$ MPa and $p_s = 1.5$.}
    \label{fig:comparison_displacements}
\end{figure}

\clearpage
\section{Appendix - Material parameter}\label{appendixB}
\renewcommand{\thetable}{B\arabic{table}}
\setcounter{table}{0}  

\autoref{tab:parameter_config} contains parameter values and ranges for the constitutive model \textit{TUBSsalt}. As reference serves a parameter set determined for Gorleben salt in \cite{IGG2016Gorleben}. However, it should be noted that the parameter set has been identified using data from various salt formations of Gorleben, and not only from the homogeneous \textit{z2HS2} area. As no parameters for healing are given in \cite{IGG2016Gorleben}, $\eta_v$ and $m_v$ are obtained from \cite{Vergleich2022TUBS}.

\definecolor{darkblue}{RGB}{0, 0, 200}
\definecolor{darkred}{RGB}{150, 0, 0}
\begin{table}[h]
    \centering
    \caption{\textbf{\textit{TUBSsalt} material parameter} under consideration of reference values from \cite{IGG2016Gorleben}. The parameters highlighted in red are directly read from experimental data. For the parameters highlighted in blue, a range of values is indicated in order to perform the sensitivity analysis.}
    \begin{tabularx}{\textwidth}{l p{5cm} p{4.5cm} X}
         \hline
         \textbf{Symbol} & \textbf{Description} & \textbf{Value or range} & \textbf{Unit} \\ \hline \hline
         \multicolumn{4}{l}{\textbf{Elastic Strain}} \\ \hline
         \rowcolor{darkred!5} $K_0$ & Initial bulk modulus & $22000$ & MPa \\
         \rowcolor{darkred!5} $G_0$ & Initial shear modulus & $14500$ & MPa \\
         $p_{el}$ & Damage exponent & $2.2$ & - \\ 
         \hline
         \multicolumn{4}{l}{\textbf{Primary Creep}} \\ \hline
         \rowcolor{darkblue!5} $\eta_p$ & Viscosity of primary creep & $5\cdot10^4 - 60\cdot10^4$ & MPa$\cdot$d \\
         \rowcolor{darkblue!5} $E_p$ & Hardening modulus & $60 - 90$ & MPa \\
         \rowcolor{darkblue!5} $\sigma_{0,eq,p}$ & Start of slope change & $20 - 40$  (later set to $30$) & MPa \\
         \rowcolor{darkblue!5} $p_p$ & Curvature parameter & $0.3 - 1.0$  (later set to $0.5$) & - \\
         \hline
         \multicolumn{4}{l}{\textbf{Secondary Creep}} \\ \hline
         \rowcolor{darkblue!5} $\eta_s$ & Viscosity of secondary creep & $5\cdot10^7 - 50\cdot10^7$ & MPa$\cdot$d \\
         \rowcolor{darkblue!5} $\sigma_{0,eq,s}$ & Start of slope change & $20 - 40$  (later set to $30$) & MPa \\
         \rowcolor{darkblue!5} $p_s$ & Curvature parameter & $0.5 - 2.0$ & - \\
         \hline
         \multicolumn{4}{l}{\textbf{Tertiary Creep}} \\ \hline
         $\eta_t$ & Viscosity of tertiary creep & $27.5$ & MPa$\cdot$d \\
         $t_0$ & Initial slope & $60$ & ° \\
         $t_1$ & Maximum yield stress & $27.5$ & MPa \\
         $\psi$ & Angle of dilatancy & $22.5$ & ° \\
         $m_t$ & Damage coefficient & $0.5$ & - \\
         \rowcolor{darkred!5} $\varepsilon_{v,d,b}$ & Volumetric strain at failure & $0.025$ & - \\
         \hline
         \multicolumn{4}{l}{\textbf{Healing}} \\ \hline
         $\eta_v$ & Viscosity of healing & $2.7\cdot10^6$ & MPa$\cdot$d \\
         $m_v$ & Healing coefficient & $0.55$ & - \\
         \hline
         \multicolumn{4}{l}{\textbf{Creep, shear and tension failure}} \\ \hline
         $\eta_n$ & Viscosity after failure & $12000$ & MPa$\cdot$d \\
         $n_0$ & Initial slope & $60$ & ° \\
         $n_1$ & Maximum yield stress & $27.5$ & MPa \\
         $m_n$ & Post failure coefficient & $2.2$ & - \\
         \rowcolor{darkred!5} $\sigma_{z,0}$ & Initial tensile strength & $1.5$ & MPa \\
         \hline
         \multicolumn{4}{l}{\textbf{Others}} \\ \hline
         \rowcolor{darkred!5} $\rho$ & Density & $2200\cdot10^{-6}$ & Gg/m$^3$ \\
         $Q$ & Activation energy & $22000$ & J/mol \\
         \hline
    \end{tabularx}
    \label{tab:parameter_config}
\end{table}
\end{appendix}

\clearpage

\begin{Backmatter}
\paragraph{Acknowledgments}
We are grateful for the provision and preparation of monitoring data from the Bundesgesellschaft für Endlagerung mbH (BGE).

\paragraph{Funding Statement}
This research is funded by the German Federal Ministry for the Environment, Nature Conservation, Nuclear Safety and Consumer Protection (BMUV) and managed by project management agency Karlsruhe (PTKA) under grant number 02E12102.

\paragraph{Competing Interests}
None

\paragraph{Data Availability Statement}
\RevisionNew{All source code and simulation data used for the presented benchmark studies are published at \url{https://doi.org/10.5281/zenodo.15049774}\nocite{ZenodoRepo2025}}.
Restrictions apply to the availability of the monitoring data, which were used under license for this study. Monitoring data are available from the authors with the permission of the Bundesgesellschaft für Endlagerung mbH (BGE).

\paragraph{Ethical Standards}
The research meets all ethical guidelines, including adherence to the legal requirements of the study country.

\paragraph{Author Contributions}
Conceptualization: L.P., JH.UQ., U.F., A.H. and H.W.; Methodology: L.P., JH.UQ., U.F., H.Y. and H.W.; Software: L.P., JH.UQ., U.F and A.H.; Data curation: L.P., U.F. and A.H.; Data visualisation: L.P., JH.UQ. and U.F.; Formal analysis: L.P., JH.UQ. and U.F.; Writing original draft: L.P., JH.UQ., U.F., A.H., H.Y., H.W., U.R. and J.S.; Funding acquisition: J.S., U.R. and H.W.; All authors approved the final submitted draft.

\bibliographystyle{unsrtnat}
\bibliography{literature}

\begin{thebibliography}{70}
\providecommand{\natexlab}[1]{#1}
\providecommand{\url}[1]{\texttt{#1}}
\expandafter\ifx\csname urlstyle\endcsname\relax
  \providecommand{\doi}[1]{doi: #1}\else
  \providecommand{\doi}{doi: \begingroup \urlstyle{rm}\Url}\fi

\bibitem[Pitz et~al.(2023)Pitz, Grunwald, Graupner, Kurgyis, Radeisen, Ma{\ss}mann, Ziefle, Thiedau, and Nagel]{Pitz2023BenchmarkingTH2M}
M.~Pitz, N.~Grunwald, B.~Graupner, K.~Kurgyis, E.~Radeisen, J.~Ma{\ss}mann, G.~Ziefle, J.~Thiedau, and T.~Nagel.
\newblock Benchmarking a new th2m implementation in ogs-6 with regard to processes relevant for nuclear waste disposal.
\newblock \emph{Environmental Earth Sciences}, 82\penalty0 (13):\penalty0 319, 2023.

\bibitem[Claret et~al.(2024)Claret, Prasianakis, Baksay, Lukin, Pepin, Ahusborde, Amaziane, B{\'a}tor, Becker, Bedn{\'a}r, et~al.]{Claret2024eurad}
F.~Claret, N.I. Prasianakis, A.~Baksay, D.~Lukin, G.~Pepin, E.~Ahusborde, B.~Amaziane, G.~B{\'a}tor, D.~Becker, A.~Bedn{\'a}r, et~al.
\newblock Eurad state-of-the-art report: development and improvement of numerical methods and tools for modeling coupled processes in the field of nuclear waste disposal.
\newblock \emph{Frontiers in Nuclear Engineering}, 3:\penalty0 1437714, 2024.

\bibitem[Wojnarowicz et~al.(2024)Wojnarowicz, Madaschi, and Laloui]{Wojnarowicz2024OptRepo}
M.~Wojnarowicz, A.~Madaschi, and L.~Laloui.
\newblock A methodology to optimize complex models in the context of nuclear waste repositories.
\newblock \emph{Computers and Geotechnics}, 173:\penalty0 106579, 2024.

\bibitem[Kurgyis et~al.(2024)Kurgyis, Achtziger-Zupan{\v{c}}i{\v{c}}, Bjorge, Boxberg, Broggi, Buchwald, Ernst, Fl{\"u}gge, Ganopolski, Graf, et~al.]{Kurgyis2024UncertaintiesRepo}
K.~Kurgyis, P.~Achtziger-Zupan{\v{c}}i{\v{c}}, M.~Bjorge, M.S. Boxberg, M.~Broggi, J.~Buchwald, O.G. Ernst, J.~Fl{\"u}gge, A.~Ganopolski, T.~Graf, et~al.
\newblock Uncertainties and robustness with regard to the safety of a repository for high-level radioactive waste: introduction of a research initiative.
\newblock \emph{Environmental Earth Sciences}, 83\penalty0 (2):\penalty0 82, 2024.

\bibitem[Myren and Lawrence(2021{\natexlab{a}})]{Myren2021comparisonGPsNNs}
S.~Myren and E.~Lawrence.
\newblock A comparison of gaussian processes and neural networks for computer model emulation and calibration.
\newblock \emph{Statistical Analysis and Data Mining: The ASA Data Science Journal}, 14\penalty0 (6):\penalty0 606--623, 2021{\natexlab{a}}.

\bibitem[Radaideh and Kozlowski(2020)]{Radaideh2020DeepGPs}
M.I. Radaideh and T.~Kozlowski.
\newblock Surrogate modeling of advanced computer simulations using deep gaussian processes.
\newblock \emph{Reliability Engineering \& System Safety}, 195:\penalty0 106731, 2020.

\bibitem[Sung and Tuo(2024)]{Sung2024RevModelCalibration}
C.-L. Sung and R.~Tuo.
\newblock A review on computer model calibration.
\newblock \emph{Wiley Interdisciplinary Reviews: Computational Statistics}, 16\penalty0 (1):\penalty0 e1645, 2024.

\bibitem[StandAG, 2017()]{StandAG2017}
StandAG, 2017.
\newblock \emph{Site Selection Act of 5 May 2017 (Federal Law Gazette I p. 1074), last modified by Article 8 of the Act of 22 March 2023 (Federal Law Gazette 2023 I No. 88)}.

\bibitem[Bollingfehr et~al.(2017)Bollingfehr, Buhmann, Dörr, Filbert, Gehrke, Heemann, Keller, Krone, Lommerzheim, Mönig, Mrugalla, Müller-Hoeppe, Rübel, Weber, and Wolf]{Bollingfehr2017DesignSalt}
W.~Bollingfehr, D.~Buhmann, S.~Dörr, W.~Filbert, A.~Gehrke, U.~Heemann, S.~Keller, J.~Krone, A.~Lommerzheim, J.~Mönig, S.~Mrugalla, N.~Müller-Hoeppe, A.~Rübel, J.R. Weber, and J.~Wolf.
\newblock Evaluation of methods and tools to develop safety concepts and to demonstrate safety for an hlw repository in salt.
\newblock \emph{Final report, TEC-03-2017-AB}, 2017.

\bibitem[Langer(1985)]{Langer1985}
M.~Langer.
\newblock {Hohlraumbau im Salzgebirge - {\"U}berblick {\"u}ber den Stand der Wissenschaft und Technik - Teil A Geologische und mechanische Grundlagen}.
\newblock \emph{Taschenbuch für den Tunnelbau 1985}, pages 287--300, 1985.

\bibitem[Wittke(2014)]{Wittke2014}
W.~Wittke.
\newblock \emph{Laboratory Tests}, chapter~14, pages 403--450.
\newblock John Wiley \& Sons, Ltd, 2014.
\newblock ISBN 9783433604281.

\bibitem[Fecker(2018)]{Fecker2018}
E.~Fecker.
\newblock \emph{Baugeologie}.
\newblock Springer Spektrum Berlin, Heidelberg, 3 edition, 2018.

\bibitem[Schulze et~al.(2007)Schulze, Heemann, Zetsche, Hampel, Pudewills, Günther, Minkley, Salzer, Hou, Wolters, Rokahr, and Zapf]{Saltmech2007}
O.~Schulze, U.~Heemann, F.~Zetsche, A.~Hampel, A.~Pudewills, R.-M. Günther, W.~Minkley, K.~Salzer, Z.~Hou, R.~Wolters, R.~Rokahr, and D.~Zapf.
\newblock Comparison of advanced constitutive models for the mechanical behavior of rock salt – results from a joint research project – i. modeling of deformation processes and benchmark calculations.
\newblock \emph{Proceedings of the 6th Conference on the Mechanical Behavior of Salt, Hannover, Germany}, page 77–88, 2007.

\bibitem[Hampel et~al.(2013)Hampel, Argüello, Hansen, Günther, Salzer, Minkley, Lux, Herchen, Düsterloh, Pudewills, Yildirim, Staudtmeister, Rokahr, Zapf, Gährken, Missal, and Stahlmann]{ARMA2013}
A.~Hampel, J.G. Argüello, F.D. Hansen, R.-M. Günther, K.~Salzer, W.~Minkley, K.-H. Lux, K.~Herchen, U.~Düsterloh, A.~Pudewills, S.~Yildirim, K.~Staudtmeister, R.~Rokahr, D.~Zapf, A.~Gährken, C.~Missal, and J.~Stahlmann.
\newblock Benchmark calculations of the thermo-mechanical behavior of rock salt – results from a us-german joint project.
\newblock \emph{Proceedings of the 47th US Rock Mechanics Symposium (ARMA 13-456), Salt Lake City, USA}, 2013.

\bibitem[Hampel et~al.(2022{\natexlab{a}})Hampel, Lüdeling, Günther, Sun-Kurczinski, Wolters, Düsterloh, Lux, Yildirim, Zapf, Wacker, Epkenhans, Stahlmann, and Reedlunn]{SaltMech2022}
A.~Hampel, C.~Lüdeling, R.-M. Günther, J.Q. Sun-Kurczinski, R.~Wolters, U.~Düsterloh, K.-H. Lux, S.~Yildirim, D.~Zapf, S.~Wacker, I.~Epkenhans, J.~Stahlmann, and B.~Reedlunn.
\newblock Weimos: Simulations of two geomechanical scenarios in rock salt resembling structures at wipp.
\newblock \emph{Proceedings of the 10th Conference on the Mechanical Behavior of Salt, Utrecht, Netherlands}, pages 421--435, 2022{\natexlab{a}}.

\bibitem[Hampel et~al.(2010)Hampel, G{\"u}nther, Salzer, Minkley, Leuger, Zapf, Rokahr, Herchen, Wolters, and D{\"u}sterloh]{Vergleich2010}
A.~Hampel, R.-M. G{\"u}nther, K.~Salzer, W.~Minkley, B.~Leuger, D.~Zapf, R.~Rokahr, K.~Herchen, R.~Wolters, and U.~D{\"u}sterloh.
\newblock {Vergleich aktueller Stoffgesetze und Vorgehensweisen anhand von 3D-Modellberechnungen zum mechanischen Langzeitverhalten eines realen Untertagebauwerks im Steinsalz - Synthesebericht}.
\newblock \emph{Research report, Federal Ministry of Education and Research (BMBF), Mainz, Germany}, 2010.

\bibitem[Hampel et~al.(2016)Hampel, Herchen, Lux, G{\"u}nther, Salzer, Winkley, Pudewills, Yildirim, Rokahr, Gährken, Missal, and Stahlmann]{Vergleich2016}
A.~Hampel, K.~Herchen, K.-H. Lux, R.-M. G{\"u}nther, K.~Salzer, W.~Winkley, A.~Pudewills, S.~Yildirim, R.~Rokahr, A.~Gährken, C.~Missal, and J.~Stahlmann.
\newblock {Vergleich aktueller Stoffgesetze und Vorgehensweisen anhand von Modellberechnungen zum thermo-mechanischen Verhalten und zur Verheilung von Steinsalz - Synthesebericht}.
\newblock \emph{Research report, Federal Ministry for Economic Affairs and Energy (BMWi), Berlin, Germany}, 2016.

\bibitem[Hampel et~al.(2022{\natexlab{b}})Hampel, Lüdeling, Günther, Salzer, Yildirim, Zapf, Epkenhans, Wacker, Gährken, Stahlmann, Sun-Kurczinski, Wolters, Herchen, and Lux]{Vergleich2022}
A.~Hampel, C.~Lüdeling, R.-M. Günther, K.~Salzer, S.~Yildirim, D.~Zapf, I.~Epkenhans, S.~Wacker, A.~Gährken, J.~Stahlmann, J.Q. Sun-Kurczinski, R.~Wolters, K.~Herchen, and K.-H. Lux.
\newblock {Weiterentwicklung und Qualifizierung der gebirgsmechanischen Modellierung für die HAW-Endlagerung im Steinsalz (WEIMOS) - Synthesebericht}.
\newblock \emph{Research report, Federal Ministry for Economic Affairs and Energy (BMWi), Berlin, Germany}, 2022{\natexlab{b}}.

\bibitem[Gährken et~al.(2015)Gährken, Missal, and Stahlmann]{TUBSsalt2015}
A.~Gährken, C.~Missal, and J.~Stahlmann.
\newblock A thermal-mechanical constitutive model to describe deformation, damage and healing of rock salt.
\newblock \emph{Proceedings of the 8th Conference on the Mechanical Behavior of Salt, Rapid City, USA}, pages 331--338, 2015.

\bibitem[Kennedy and O'Hagan(2000)]{Kennedy2000PredOut}
M.C. Kennedy and A.~O'Hagan.
\newblock Predicting the output from a complex computer code when fast approximations are available.
\newblock \emph{Biometrika}, 87\penalty0 (1):\penalty0 1--13, 2000.

\bibitem[Kennedy and O'Hagan(2001)]{Kennedy2001BayesianCal}
M.C. Kennedy and A.~O'Hagan.
\newblock Bayesian calibration of computer models.
\newblock \emph{Journal of the Royal Statistical Society: Series B (Statistical Methodology)}, 63\penalty0 (3):\penalty0 425--464, 2001.

\bibitem[Gu and Wang(2018)]{gu2018scaledGPS}
M.~Gu and L.~Wang.
\newblock Scaled gaussian stochastic process for computer model calibration and prediction.
\newblock \emph{SIAM/ASA Journal on Uncertainty Quantification}, 6\penalty0 (4):\penalty0 1555--1583, 2018.

\bibitem[Teckentrup(2020)]{teckentrup2020GPsConvergence}
A.~L. Teckentrup.
\newblock Convergence of gaussian process regression with estimated hyper-parameters and applications in bayesian inverse problems.
\newblock \emph{SIAM/ASA Journal on Uncertainty Quantification}, 8\penalty0 (4):\penalty0 1310--1337, 2020.

\bibitem[Myren and Lawrence(2021{\natexlab{b}})]{myren2021comparisonGPs_NNs}
S.~Myren and E.~Lawrence.
\newblock A comparison of gaussian processes and neural networks for computer model emulation and calibration.
\newblock \emph{Statistical Analysis and Data Mining: The ASA Data Science Journal}, 14\penalty0 (6):\penalty0 606--623, 2021{\natexlab{b}}.

\bibitem[Gramacy(2020)]{gramacy2020surrogates}
R.B. Gramacy.
\newblock \emph{Surrogates: Gaussian process modeling, design, and optimization for the applied sciences}.
\newblock Chapman and Hall/CRC, 2020.

\bibitem[Schulz et~al.(2018)Schulz, Speekenbrink, and Krause]{schulz2018GPstutorial}
E.~Schulz, M.~Speekenbrink, and A.~Krause.
\newblock A tutorial on gaussian process regression: Modelling, exploring, and exploiting functions.
\newblock \emph{Journal of Mathematical Psychology}, 85:\penalty0 1--16, 2018.

\bibitem[Wu et~al.(2018)Wu, Kozlowski, Meidani, and Shirvan]{wu2018invUQGPs}
X.~Wu, T.~Kozlowski, H.~Meidani, and K.~Shirvan.
\newblock Inverse uncertainty quantification using the modular bayesian approach based on gaussian process, part 2: Application to trace.
\newblock \emph{Nuclear Engineering and Design}, 335:\penalty0 417--431, 2018.

\bibitem[Mahdaviara et~al.(2021)Mahdaviara, Rostami, Keivanimehr, and Shahbazi]{mahdaviara2021PermeabilityGPs}
M.~Mahdaviara, A.~Rostami, F.~Keivanimehr, and K.~Shahbazi.
\newblock Accurate determination of permeability in carbonate reservoirs using gaussian process regression.
\newblock \emph{Journal of Petroleum Science and Engineering}, 196:\penalty0 107807, 2021.

\bibitem[Li et~al.(2023)Li, Hariri-Ardebili, Deng, Wei, and Cao]{li2023GPsDams}
Y.~Li, M.~A. Hariri-Ardebili, T.~Deng, Q.~Wei, and M.~Cao.
\newblock A surrogate-assisted stochastic optimization inversion algorithm: Parameter identification of dams.
\newblock \emph{Advanced Engineering Informatics}, 55:\penalty0 101853, 2023.

\bibitem[Veasna et~al.(2023)Veasna, Feng, Zhang, and Knezevic]{veasna2023GPsPlasticity}
K.~Veasna, Z.~Feng, Q.~Zhang, and M.~Knezevic.
\newblock Machine learning-based multi-objective optimization for efficient identification of crystal plasticity model parameters.
\newblock \emph{Computer Methods in Applied Mechanics and Engineering}, 403:\penalty0 115740, 2023.

\bibitem[Oakley and O'Hagan(2004)]{Oakley2004ProbSA}
J.E. Oakley and A.~O'Hagan.
\newblock Probabilistic sensitivity analysis of complex models: a bayesian approach.
\newblock \emph{Journal of the Royal Statistical Society Series B: Statistical Methodology}, 66\penalty0 (3):\penalty0 751--769, 2004.

\bibitem[Marrel et~al.(2009)Marrel, Iooss, Laurent, and Roustant]{marrel2009calculations}
A.~Marrel, B.~Iooss, B.~Laurent, and O.~Roustant.
\newblock Calculations of sobol indices for the gaussian process metamodel.
\newblock \emph{Reliability Engineering \& System Safety}, 94\penalty0 (3):\penalty0 742--751, 2009.

\bibitem[Srivastava et~al.(2017)Srivastava, Subramaniyan, and Wang]{Srivastava2017SA}
A.~Srivastava, A.K. Subramaniyan, and L.~Wang.
\newblock Analytical global sensitivity analysis with gaussian processes.
\newblock \emph{AI EDAM}, 31\penalty0 (3):\penalty0 235--250, 2017.

\bibitem[Kejzlar et~al.(2020)Kejzlar, Neufcourt, Nazarewicz, and Reinhard]{Kejzlar2020NuclearPhysics}
V.~Kejzlar, L.~Neufcourt, W.~Nazarewicz, and P.-G. Reinhard.
\newblock Statistical aspects of nuclear mass models.
\newblock \emph{Journal of Physics G: Nuclear and Particle Physics}, 47\penalty0 (9):\penalty0 094001, 2020.

\bibitem[Cheng et~al.(2021)Cheng, Konomi, Matthews, Karagiannis, and Kang]{Cheng2021intersatellite}
S.~Cheng, B.A. Konomi, J.L. Matthews, G.~Karagiannis, and E.~L. Kang.
\newblock Hierarchical bayesian nearest neighbor co-kriging gaussian process models; an application to intersatellite calibration.
\newblock \emph{Spatial Statistics}, 44:\penalty0 100516, 2021.

\bibitem[Thelen et~al.(2022)Thelen, Zhang, Fink, Lu, Ghosh, Youn, Todd, Mahadevan, Hu, and Hu]{Thelen2022DigitalTwinI}
A.~Thelen, X.~Zhang, O.~Fink, Y.~Lu, S.~Ghosh, B.D. Youn, M.D. Todd, S.~Mahadevan, C.~Hu, and Z.~Hu.
\newblock A comprehensive review of digital twin—part 1: modeling and twinning enabling technologies.
\newblock \emph{Structural and Multidisciplinary Optimization}, 65\penalty0 (12):\penalty0 354, 2022.

\bibitem[Thelen et~al.(2023)Thelen, Zhang, Fink, Lu, Ghosh, Youn, Todd, Mahadevan, Hu, and Hu]{Thelen2023DigitalTwinII}
A.~Thelen, X.~Zhang, O.~Fink, Y.~Lu, S.~Ghosh, B.~D. Youn, M.~D. Todd, S.~Mahadevan, C.~Hu, and Z.~Hu.
\newblock A comprehensive review of digital twin—part 2: roles of uncertainty quantification and optimization, a battery digital twin, and perspectives.
\newblock \emph{Structural and multidisciplinary optimization}, 66\penalty0 (1):\penalty0 1, 2023.

\bibitem[Sacks et~al.(1989)Sacks, Schiller, and Welch]{sacks1989designs}
J.~Sacks, S.B. Schiller, and W.J. Welch.
\newblock Designs for computer experiments.
\newblock \emph{Technometrics}, 31\penalty0 (1):\penalty0 41--47, 1989.

\bibitem[Bayarri et~al.(2007{\natexlab{a}})Bayarri, Berger, Paulo, Sacks, Cafeo, Cavendish, Lin, and Tu]{Bayarri2007CompModels}
M.J. Bayarri, J.O. Berger, R.~Paulo, J.~Sacks, J.A. Cafeo, J.~Cavendish, C.-H. Lin, and J.~Tu.
\newblock A framework for validation of computer models.
\newblock \emph{Technometrics}, 49\penalty0 (2):\penalty0 138--154, 2007{\natexlab{a}}.

\bibitem[Higdon et~al.(2004)Higdon, Kennedy, Cavendish, Cafeo, and Ryne]{Higdon2004DataCompModels}
D.~Higdon, M.~Kennedy, J.C. Cavendish, J.A. Cafeo, and R.D. Ryne.
\newblock Combining field data and computer simulations for calibration and prediction.
\newblock \emph{SIAM Journal on Scientific Computing}, 26\penalty0 (2):\penalty0 448--466, 2004.

\bibitem[Higdon et~al.(2008)Higdon, Gattiker, Williams, and Rightley]{Higdon2008HDOut}
D.~Higdon, J.~Gattiker, B.~Williams, and M.~Rightley.
\newblock Computer model calibration using high-dimensional output.
\newblock \emph{Journal of the American Statistical Association}, 103\penalty0 (482):\penalty0 570--583, 2008.

\bibitem[Bayarri et~al.(2007{\natexlab{b}})Bayarri, Berger, Cafeo, Garcia-Donato, Liu, Palomo, Parthasarathy, Paulo, Sacks, and Walsh]{Bayarri2007FuncOut}
M.J. Bayarri, J.O. Berger, J.~Cafeo, G.~Garcia-Donato, F.~Liu, J.~Palomo, R.J. Parthasarathy, R.~Paulo, J.~Sacks, and D.~Walsh.
\newblock Computer model validation with functional output.
\newblock \emph{The Annals of Statistics}, pages 1874--1906, 2007{\natexlab{b}}.

\bibitem[Santner et~al.(2003)Santner, Williams, Notz, and Williams]{Santner2003ExpDesign}
T.J. Santner, B.J. Williams, W.I. Notz, and B.J. Williams.
\newblock \emph{The design and analysis of computer experiments}, volume~1.
\newblock Springer, 2003.

\bibitem[Fang et~al.(2005)Fang, Li, and Sudjianto]{Fang2005CompExpDesign}
K.-T. Fang, R.~Li, and A.~Sudjianto.
\newblock \emph{Design and modeling for computer experiments}.
\newblock Chapman and Hall/CRC, 2005.

\bibitem[Loeppky et~al.(2009)Loeppky, Sacks, and Welch]{loeppky2009SampleSize}
J.L. Loeppky, J.~Sacks, and W.J. Welch.
\newblock Choosing the sample size of a computer experiment: A practical guide.
\newblock \emph{Technometrics}, 51\penalty0 (4):\penalty0 366--376, 2009.

\bibitem[Baker et~al.(2022)Baker, Barbillon, Fadikar, Gramacy, Herbei, Higdon, Huang, Johnson, Ma, Mondal, et~al.]{Baker2022StoCompModelsg}
E.~Baker, P.~Barbillon, A.~Fadikar, R.B. Gramacy, R.~Herbei, D.~Higdon, J.~Huang, L.R. Johnson, P.~Ma, A.~Mondal, et~al.
\newblock Analyzing stochastic computer models: A review with opportunities.
\newblock \emph{Statistical Science}, 37\penalty0 (1):\penalty0 64--89, 2022.

\bibitem[O’Hagan(2006)]{OHagan2006BayesianCompCode}
A.~O’Hagan.
\newblock Bayesian analysis of computer code outputs: A tutorial.
\newblock \emph{Reliability Engineering \& System Safety}, 91\penalty0 (10-11):\penalty0 1290--1300, 2006.

\bibitem[Hunsche et~al.(2003)Hunsche, Schulze, Walter, and Plischke]{Hunsche2003Gorleben}
U.~Hunsche, O.~Schulze, F.~Walter, and I.~Plischke.
\newblock {Projekt Gorleben: Thermomechanisches Verhalten von Salzgestein - Abschlussbericht}.
\newblock Technical report, Bundesanstalt für Geowissenschaften und Rohstoffe (BGR), Hannover, 2003.

\bibitem[Bornemann et~al.(2008)Bornemann, Behlau, Fischbeck, Hammer, Jaritz, Keller, Mingerzahn, and Schramm]{BGR2008Gorleben}
O.~Bornemann, J.~Behlau, R.~Fischbeck, J.~Hammer, W.~Jaritz, S.~Keller, G.~Mingerzahn, and M.~Schramm.
\newblock {Description of the Gorleben Site Part 3: Results of the geological surface and underground exploration of the salt formation}.
\newblock \emph{Bundesanstalt für Geowissenschaften und Rohstoffe (BGR), Hannover}, 2008.

\bibitem[Anand and Govindjee(2020)]{Anand2020CMM_solids_Book}
L.~Anand and S.~Govindjee.
\newblock \emph{Continuum mechanics of solids}.
\newblock Oxford University Press, 2020.

\bibitem[Itasca Consultants GmbH()]{FLAC3D2023}
Itasca Consultants GmbH.
\newblock \emph{Itasca Software 9.0 documentation - FLAC Theory and Backround}, 2023.
\newblock URL \url{https://docs.itascacg.com/itasca900/flac3d/docproject/source/theory/theory.html?node2293}.

\bibitem[Epkenhans et~al.(2022)Epkenhans, Wacker, and Stahlmann]{Vergleich2022TUBS}
I.~Epkenhans, S.~Wacker, and J.~Stahlmann.
\newblock {Weiterentwicklung und Qualifizierung der gebirgsmechanischen Modellierung für die HAW-Endlagerung im Steinsalz (WEIMOS): (Verbundprojekt: Teilprojekt D) Endbericht des Teilprojekts}.
\newblock \emph{Research report. Technische Universität Braunschweig, Braunschweig, Germany}, 2022.

\bibitem[Stahlmann et~al.(2016)Stahlmann, Missal, and Gährken]{IGG2016Gorleben}
J.~Stahlmann, C.~Missal, and A.~Gährken.
\newblock {Geomechanische Modellberechnungen zur Offenhaltungsphase des Bergwerkes Gorleben}.
\newblock unpublished, 2016.

\bibitem[Virtanen et~al.(2020)Virtanen, Gommers, Oliphant, Haberland, Reddy, Cournapeau, Burovski, Peterson, Weckesser, Bright, et~al.]{Virtanen2020scipy}
P.~Virtanen, R.~Gommers, T.~E. Oliphant, M.~Haberland, T.~Reddy, D.~Cournapeau, E.~Burovski, P.~Peterson, W.~Weckesser, J.~Bright, et~al.
\newblock {SciPy 1.0}: fundamental algorithms for scientific computing in {Python}.
\newblock \emph{Nature methods}, 17\penalty0 (3):\penalty0 261--272, 2020.

\bibitem[Storn and Price(1997)]{Storn1997diff_evolution}
R.~Storn and K.~Price.
\newblock Differential evolution--a simple and efficient heuristic for global optimization over continuous spaces.
\newblock \emph{Journal of global optimization}, 11:\penalty0 341--359, 1997.

\bibitem[Sobol(2001)]{Sobol2001globalSA}
I.~M. Sobol.
\newblock Global sensitivity indices for nonlinear mathematical models and their monte carlo estimates.
\newblock \emph{Mathematics and computers in simulation}, 55\penalty0 (1-3):\penalty0 271--280, 2001.

\bibitem[Saltelli(2002)]{Saltelli2002SA}
A.~Saltelli.
\newblock Making best use of model evaluations to compute sensitivity indices.
\newblock \emph{Computer physics communications}, 145\penalty0 (2):\penalty0 280--297, 2002.

\bibitem[Saltelli et~al.(2010)Saltelli, Annoni, Azzini, Campolongo, Ratto, and Tarantola]{Saltelli2010varianceSA}
A.~Saltelli, P.~Annoni, I.~Azzini, F.~Campolongo, M.~Ratto, and S.~Tarantola.
\newblock Variance based sensitivity analysis of model output. design and estimator for the total sensitivity index.
\newblock \emph{Computer physics communications}, 181\penalty0 (2):\penalty0 259--270, 2010.

\bibitem[Alexanderian et~al.(2020)Alexanderian, Gremaud, and Smith]{alexanderian2020variance}
A.~Alexanderian, P.A. Gremaud, and R.C. Smith.
\newblock Variance-based sensitivity analysis for time-dependent processes.
\newblock \emph{Reliability Engineering \& System Safety}, 196:\penalty0 106722, 2020.

\bibitem[Gamboa et~al.(2014)Gamboa, Janon, Klein, and Lagnoux]{gamboa2014sensitivity}
F.~Gamboa, A.~Janon, T.~Klein, and A.~Lagnoux.
\newblock Sensitivity analysis for multidimensional and functional outputs.
\newblock 2014.

\bibitem[Gamboa et~al.(2016)Gamboa, Janon, Klein, Lagnoux, and Prieur]{gamboa2016statistical}
F.~Gamboa, A.~Janon, T.~Klein, A.~Lagnoux, and C.~Prieur.
\newblock Statistical inference for sobol pick-freeze monte carlo method.
\newblock \emph{Statistics}, 50\penalty0 (4):\penalty0 881--902, 2016.

\bibitem[Sudret(2008)]{sudret2008global}
B.~Sudret.
\newblock Global sensitivity analysis using polynomial chaos expansions.
\newblock \emph{Reliability engineering \& system safety}, 93\penalty0 (7):\penalty0 964--979, 2008.

\bibitem[Agarwal et~al.(2024)Agarwal, Urrea-Quintero, Wessels, and Wick]{Agarwal2024POD}
G.~Agarwal, J.-H. Urrea-Quintero, H.~Wessels, and T.~Wick.
\newblock Parameter identification and uncertainty propagation of hydrogel coupled diffusion-deformation using pod-based reduced-order modeling.
\newblock \emph{Computational Mechanics}, pages 1--31, 2024.

\bibitem[Anton et~al.(2024)Anton, Tr{\"o}ger, Wessels, R{\"o}mer, Henkes, and Hartmann]{Anton2024PINNs}
D.~Anton, J.-A. Tr{\"o}ger, H.~Wessels, U.~R{\"o}mer, A.~Henkes, and S.~Hartmann.
\newblock Deterministic and statistical calibration of constitutive models from full-field data with parametric physics-informed neural networks.
\newblock \emph{arXiv preprint arXiv:2405.18311}, 2024.

\bibitem[Pedregosa et~al.(2011)Pedregosa, Varoquaux, Gramfort, Michel, Thirion, Grisel, Blondel, Prettenhofer, Weiss, Dubourg, Vanderplas, Passos, Cournapeau, Brucher, Perrot, and Duchesnay]{Pedregosa2011ScikitLearn}
F.~Pedregosa, G.~Varoquaux, A.~Gramfort, V.~Michel, B.~Thirion, O.~Grisel, M.~Blondel, P.~Prettenhofer, R.~Weiss, V.~Dubourg, J.~Vanderplas, A.~Passos, D.~Cournapeau, M.~Brucher, M.~Perrot, and E.~Duchesnay.
\newblock Scikit-learn: Machine learning in {P}ython.
\newblock \emph{Journal of Machine Learning Research}, 12:\penalty0 2825--2830, 2011.

\bibitem[Kock et~al.(2012)Kock, Eickemeier, Frieling, Heusermann, Knauth, Minkley, Navarro, Nipp, and Vogel]{GRS2012VSG-286}
I.~Kock, R.~Eickemeier, G.~Frieling, S.~Heusermann, M.~Knauth, W.~Minkley, M.~Navarro, H.-K. Nipp, and P.~Vogel.
\newblock {Vorläufige Sicherheitsanalyse für den Standort Gorleben - Bericht zum Arbeitspaket 9.1: Integritätsanalyse der geologischen Barriere}.
\newblock \emph{Gesellschaft für Anlagen- und Reaktorsicherheit (GRS) mbH}, 2012.

\bibitem[Sobol(1967)]{SobolSeq1967}
I.~M. Sobol.
\newblock On the distribution of points in a cube and the accurate evaluation of integrals.
\newblock \emph{Zhurnal Vychislitel’noi Matematiki i Matematicheskoi Fiziki 7, no. 4: 784-802}, 1967.

\bibitem[Owen(2019)]{QMC2019}
A.~B. Owen.
\newblock \emph{Monte Carlo Book: the Quasi-Monte Carlo parts}.
\newblock Stanford University, 2019.

\bibitem[Fuhg et~al.(2021)Fuhg, Fau, and Nackenhorst]{fuhg2021state}
J.N. Fuhg, A.~Fau, and U.~Nackenhorst.
\newblock State-of-the-art and comparative review of adaptive sampling methods for kriging.
\newblock \emph{Archives of Computational Methods in Engineering}, 28:\penalty0 2689--2747, 2021.

\bibitem[Paul et~al.(2025)Paul, Urrea-Quintero, and Fiaz]{ZenodoRepo2025}
L.~Paul, J.-H. Urrea-Quintero, and U.~Fiaz.
\newblock Code and data repository: Gaussian processes enabled model calibration in the context of deep geological disposal.
\newblock \emph{Zenodo}, 2025.
\newblock URL \url{https://doi.org/10.5281/zenodo.15049774}.

\end{thebibliography}

\end{Backmatter}

\end{document}